\documentclass[twocolumn]{aastex631}

\usepackage{xspace}
\usepackage{color, glossaries}

\hypersetup{
   pdftitle={UNITY1.5 and Union3}, 
   pdfauthor={David Rubin},
   colorlinks=true,        
   linkcolor=blue,         
   citecolor=blue,         
   urlcolor=blue           
}

\newcommand{\ang}{\AA\xspace}
\newcommand{\PD}[2]{\frac{\partial #1}{\partial #2}\xspace}
\newcommand{\PDnofrac}[2]{$\partial #1/ \partial #2$\xspace}
\newcommand{\DZP}{\ensuremath{\Delta \mathrm{sys}}\xspace}
\newcommand{\CALSPEC}{CALSPEC~03-2021\xspace}
\newcommand{\HST}{{\it HST}\xspace}
\newcommand{\bandpass}{passband\xspace}
\newcommand{\bandpasses}{passbands\xspace}
\newcommand{\zCMB}{\ensuremath{z_{\mathrm{CMB}}}\xspace}
\newcommand{\zhelio}{\ensuremath{z_{\mathrm{helio}}}\xspace}
\newcommand{\zstar}{\ensuremath{z_{\star}}\xspace}
\newcommand{\Steve}{{\it Steve}\xspace}
\newcommand{\BBC}{\texttt{BBC}\xspace}
\newcommand{\BEAMS}{\texttt{BEAMS}\xspace}
\newcommand{\RomanSpelled}{{\it Nancy Grace Roman Space Telescope}\xspace}

\newcommand{\NLandoltCALSPEC}{14\xspace}
\newcommand{\LandoltCALSPEC}{Landolt/CALSPEC-\NLandoltCALSPEC\xspace}
\newcommand{\LandoltCALSPECStars}{AGK+81~266, BD+26~2606, BD+75~325, Feige~110, Feige~34, G191-B2B, GD~153, GD~71, GRW+70~5824, HZ4, HZ44, LDS749B, P177D, P330E\xspace}

\newcommand{\NSmithCALSPEC}{9\xspace}
\newcommand{\SmithCALSPEC}{Smith/CALSPEC-\NSmithCALSPEC\xspace}
\newcommand{\SmithCALSPECStars}{BD+02~3375, BD+21~0607, BD+26~2606, BD+29~2091, BD+54~1216, BD+75~325, P177D, P330E, Feige~34\xspace}

\newcommand{\mBxc}{\ensuremath{\{m_B,\ x_1,\ c\}}\xspace}
\newcommand{\mBxcc}{\ensuremath{\{m_B,\ x_1,\ c_B,\ c_R\}}\xspace}
\newcommand{\mBxci}{\ensuremath{\{m_{B i},\ x_{1 i},\ c_i\}}\xspace}

\newcommand{\xonetruei}{\ensuremath{x_{1 i}^{\mathrm{true}}}\xspace}
\newcommand{\cBtruei}{\ensuremath{c_{B i}^{\mathrm{true}}}\xspace}
\newcommand{\cRtruei}{\ensuremath{c_{R i}^{\mathrm{true}}}\xspace}

\newcommand{\xonetrue}{\ensuremath{x_{1}^{\mathrm{true}}}\xspace}
\newcommand{\cBtrue}{\ensuremath{c^{\mathrm{true}}_B}\xspace}
\newcommand{\cRtrue}{\ensuremath{c^{\mathrm{true}}_R}\xspace}

\newcommand{\Phigh}{\ensuremath{P^{\mathrm{high}}}\xspace}
\newcommand{\Phigheff}{\ensuremath{P^{\mathrm{high}}_{\mathrm{eff}}}\xspace}

\newcommand{\betaRL}{\ensuremath{\beta_{RL}}\xspace}
\newcommand{\betaRH}{\ensuremath{\beta_{RH}}\xspace}

\newcommand{\Om}{\ensuremath{\Omega_m}\xspace}
\newcommand{\OL}{\ensuremath{\Omega_{\Lambda}}\xspace}
\newcommand{\LCDM}{$\Lambda$CDM\xspace}
\newcommand{\scriptM}{\ensuremath{\mathcal{M}_B}\xspace}
\newcommand{\zeff}{\ensuremath{z_{\mathrm{eff}}}\xspace}
\newcommand{\zprime}{\ensuremath{z^{\prime}}\xspace}

\newcommand{\NCrossCheckSNe}{250\xspace}
\newcommand{\NDatasets}{24\xspace}
\newcommand{\nPecVelEigen}{100\xspace}
\newcommand{\UThreeVsPP}{48\%\xspace}

\newcommand{\MegaGMinusAB}{0.0190\xspace}
\newcommand{\MegaRMinusAB}{0.0062\xspace}
\newcommand{\MegaIMinusAB}{$-0.0009$\xspace}
\newcommand{\MegaZMinusAB}{$-0.0126$\xspace}

\newcommand{\LCDMTension}{1.7--2.6$\sigma$\xspace}
\newcommand{\SNLSDepth}{$23.69^{+0.12}_{-0.14}$\xspace}

\newcommand{\HubbleCephChiInc}{26--33\xspace}
\newcommand{\wOwaChiDec}{7.3 (2.2$\sigma$, SNe+BAO+CMB) or 5.0 (1.7$\sigma$, SNe+BAO+CMB+$H_0^{\mathrm{TRGB}}$)\xspace}
\newcommand{\wOwaSigmaRange}{2.0$\sigma$ ($w_a$, SNe+BAO+CMB+$H_0^{\mathrm{TRGB}}$) to 2.6$\sigma$ ($w_0$, SNe+BAO+CMB)\xspace}
\newcommand{\BAOCMBChiDropwOwa}{2.1\xspace}

\newcommand{\wOwaTRGBFoMRange}{10\%--13\%\xspace}
\newcommand{\wOwaCephFoMRange}{33\%--48\%\xspace}

\newcommand{\foutlconstraint}{\ensuremath{0.031^{+0.010}_{-0.011}}\xspace}

\newcommand{\NSNefailingLCfits}{101\xspace}
\newcommand{\totalNSNe}{2087\xspace}

\newcommand{\CTNSNe}{18\xspace}
\newcommand{\CTzRange}{0.014--0.101\xspace}
\newcommand{\CfAOneNSNe}{10\xspace}
\newcommand{\CfAOnezRange}{0.016--0.125\xspace}
\newcommand{\KrisciunasNSNe}{9\xspace}
\newcommand{\KrisciunaszRange}{0.014--0.036\xspace}
\newcommand{\CfATwoNSNe}{14\xspace}
\newcommand{\CfATwozRange}{0.010--0.054\xspace}
\newcommand{\CfAThreeNSNe}{46\xspace}
\newcommand{\CfAThreezRange}{0.014--0.085\xspace}
\newcommand{\CfAFourNSNe}{31\xspace}
\newcommand{\CfAFourzRange}{0.011--0.070\xspace}
\newcommand{\CSPNSNe}{79\xspace}
\newcommand{\CSPzRange}{0.010--0.083\xspace}
\newcommand{\SCPNearbyNSNe}{7\xspace}
\newcommand{\SCPNearbyzRange}{0.039--0.156\xspace}
\newcommand{\LOSSNSNe}{101\xspace}
\newcommand{\LOSSzRange}{0.010--0.072\xspace}
\newcommand{\FoundationNSNe}{188\xspace}
\newcommand{\FoundationzRange}{0.010--0.109\xspace}
\newcommand{\CNZeroTwoNSNe}{100\xspace}
\newcommand{\CNZeroTwozRange}{0.010--0.029\xspace}
\newcommand{\LSQCSPNSNe}{21\xspace}
\newcommand{\LSQCSPzRange}{0.020--0.112\xspace}
\newcommand{\LSQLCONSNe}{59\xspace}
\newcommand{\LSQLCOzRange}{0.011--0.119\xspace}
\newcommand{\KnopNSNe}{11\xspace}
\newcommand{\KnopzRange}{0.357--0.863\xspace}
\newcommand{\SNLSNSNe}{235\xspace}
\newcommand{\SNLSzRange}{0.125--1.061\xspace}
\newcommand{\SDSSNSNe}{399\xspace}
\newcommand{\SDSSzRange}{0.037--0.550\xspace}
\newcommand{\ESSENCENSNe}{172\xspace}
\newcommand{\ESSENCEzRange}{0.145--0.805\xspace}
\newcommand{\PanSTARRSNSNe}{313\xspace}
\newcommand{\PanSTARRSzRange}{0.027--0.667\xspace}
\newcommand{\DESNSNe}{206\xspace}
\newcommand{\DESzRange}{0.077--0.849\xspace}
\newcommand{\HZTTonryNSNe}{5\xspace}
\newcommand{\HZTTonryzRange}{0.455--1.057\xspace}
\newcommand{\GOODSNSNe}{32\xspace}
\newcommand{\GOODSzRange}{0.461--1.550\xspace}
\newcommand{\AmanullahNSNe}{5\xspace}
\newcommand{\AmanullahzRange}{0.511--1.124\xspace}
\newcommand{\SCPHSTNSNe}{15\xspace}
\newcommand{\SCPHSTzRange}{0.621--1.713\xspace}
\newcommand{\MCTNSNe}{11\xspace}
\newcommand{\MCTzRange}{1.070--2.260\xspace}

\newcommand{\CTCalUncM}{0.03: $B$$R$$I$, 0.02: $V$\xspace}
\newcommand{\CTCalUncMWords}{0.03 mag in $B$$R$$I$ and 0.02 mag in $V$\xspace}
\newcommand{\CTCalUncA}{10\xspace}
\newcommand{\CfAOneCalUncM}{0.01\xspace}
\newcommand{\CfAOneCalUncA}{20\xspace}
\newcommand{\KrisciunasCalUncM}{0.05: $U$, 0.03: $B$$R$$I$, 0.02: $V$\xspace}
\newcommand{\KrisciunasCalUncMWords}{0.05 mag in $U$, 0.03 mag in $B$$R$$I$, and 0.02 mag in $V$\xspace}
\newcommand{\KrisciunasCalUncA}{10\xspace}
\newcommand{\CfATwoCalUncM}{0.01\xspace}
\newcommand{\CfATwoCalUncA}{20\xspace}
\newcommand{\CfAThreeCalUncM}{0.01\xspace}
\newcommand{\CfAThreeCalUncA}{20\xspace}
\newcommand{\CfAFourCalUncM}{0.01\xspace}
\newcommand{\CfAFourCalUncA}{20\xspace}
\newcommand{\CSPCalUncM}{0.01\xspace}
\newcommand{\CSPCalUncA}{10\xspace}
\newcommand{\SCPNearbyCalUncM}{0.01\xspace}
\newcommand{\SCPNearbyCalUncA}{20\xspace}
\newcommand{\LOSSCalUncM}{0.01\xspace}
\newcommand{\LOSSCalUncA}{20\xspace}

\newcommand{\CNZeroTwoCalUncM}{0.01\xspace}
\newcommand{\CNZeroTwoCalUncA}{20\xspace}
\newcommand{\LSQCSPCalUncM}{0.01\xspace}
\newcommand{\LSQCSPCalUncA}{10\xspace}
\newcommand{\LSQLCOCalUncM}{0.01\xspace}
\newcommand{\LSQLCOCalUncA}{20\xspace}
\newcommand{\KnopCalUncM}{0.02\xspace}
\newcommand{\KnopCalUncA}{10\xspace}
\newcommand{\SNLSCalUncM}{0.003\xspace}
\newcommand{\SNLSCalUncA}{10\xspace}
\newcommand{\SDSSCalUncM}{0.006: $u$$i$, 0.004: $g$, 0.003: $r$, 0.008: $z$\xspace}

\newcommand{\SDSSCalUncA}{10\xspace}
\newcommand{\ESSENCECalUncM}{0.01\xspace}
\newcommand{\ESSENCECalUncA}{10\xspace}
\newcommand{\PanSTARRSCalUncM}{0.002\xspace}
\newcommand{\PanSTARRSCalUncA}{10\xspace}
\newcommand{\DESCalUncM}{0.01\xspace}
\newcommand{\DESCalUncA}{10\xspace}
\newcommand{\HZTTonryCalUncM}{0.01\xspace}
\newcommand{\HZTTonryCalUncA}{10\xspace}

\newcommand{\AmanullahCalUncM}{0.01\xspace}
\newcommand{\AmanullahCalUncA}{10\xspace}
\newcommand{\SCPHSTCalUncA}{10\xspace}
\newcommand{\MCTCalUncM}{0.01\xspace}
\newcommand{\MCTCalUncA}{10\xspace}

\newcommand{\SCPHSTCalUncM}{0.01: ACS, 0.022: $F110W$, 0.023: $F160W$\xspace}

\shorttitle{Union3 and UNITY1.5}
\shortauthors{Rubin et al.}

\begin{document}

\title{Union Through UNITY: Cosmology with 2,000 SNe Using a Unified Bayesian Framework}
\newcommand{\uhawaii}{\affiliation{Department of Physics and Astronomy, University of Hawai`i at M{\=a}noa, Honolulu, Hawai`i 96822}}
\newcommand{\stsci}{\affiliation{Space Telescope Science Institute, 3700 San Martin Drive Baltimore, MD 21218, USA}}
\newcommand{\lbnl}{\affiliation{E.O. Lawrence Berkeley National Laboratory, 1 Cyclotron Rd., Berkeley, CA, 94720, USA}}
\newcommand{\ANUrsaa}{\affiliation{Research School of Astronomy and Astrophysics, The Australian National University, Canberra, ACT 2601, Australia}}
\newcommand{\ANUcga}{\affiliation{Centre for Gravitational Astrophysics, College of Science, The Australian National University, ACT 2601, Australia}}
\newcommand{\lancaster}{\affiliation{Physics Department, Lancaster University, Lancaster LA1 4YB, United Kingdom}}
\newcommand{\ucberkeley}{\affiliation{Department of Physics, University of California Berkeley, Berkeley, CA 94720, USA}}

\newcommand{\pilarone}{\affiliation{Instituto de F\'{\i}sica Fundamental, Consejo Superior de 
Investigaciones Cient\'{\i}ficas, E-28006, Madrid, Spain}}
\newcommand{\pilartwo}{\affiliation{Institute of Cosmos Sciences (UB--IEEC),  c/. Mart\'{\i} i Franqu\'es 1, E-08028, Barcelona, Spain}}

\newcommand{\fsu}{\affiliation{Department of Physics, Florida State University, 77 Chieftan Way, Tallahassee, FL 32306, USA}}

\newcommand{\usf}{\affiliation{Department of Physics and Astronomy, University of San Francisco, San Francisco, CA 94117-108, USA}}

\newcommand{\LPNHE}{\affiliation{LPNHE, CNRS/IN2P3 \& Sorbonne Universit\'e, 4 place Jussieu, 75005 Paris, France}}

\author[0000-0001-5402-4647]{David Rubin}
\uhawaii
\lbnl

\author{Greg Aldering}
\lbnl

\author{Marc Betoule}
\LPNHE

\author{Andy Fruchter}
\stsci

\author{Xiaosheng Huang}
\usf
\lbnl

\author{Alex G. Kim}
\lbnl

\author{Chris Lidman}
\ANUcga
\ANUrsaa

\author{Eric Linder}
\lbnl
\ucberkeley

\author{Saul Perlmutter}
\lbnl
\ucberkeley

\author{Pilar Ruiz-Lapuente}
\pilarone
\pilartwo

\author{Nao Suzuki}
\lbnl
\fsu

\newcommand{\secondhalfofsentence}{provided the first strong evidence that the expansion of the universe is accelerating \citep{riess98, perlmutter99}. By virtue of their combination of distance reach, precision, and prevalence, they continue to provide cosmological constraints complementary to other cosmological probes.\xspace}
\newcommand{\secondhalfofsetencenoref}{were instrumental in establishing the acceleration of the universe's expansion. By virtue of their combination of distance reach, precision, and prevalence, they continue to provide key cosmological constraints, complementing other cosmological probes.\xspace}

\begin{abstract}

Type Ia supernovae (SNe~Ia) \secondhalfofsetencenoref Individual SN surveys cover only over about a factor of two in redshift, so compilations of multiple SN datasets are strongly beneficial. We assemble an up-to-date ``Union'' compilation of \totalNSNe cosmologically useful SNe~Ia from \NDatasets datasets (``Union3''). We take care to put all SNe on the same distance scale and update the light-curve fitting with SALT3 to use the full rest-frame optical. Over the next few years, the number of cosmologically useful SNe~Ia will increase by more than a factor of ten, and keeping systematic uncertainties subdominant will be more challenging than ever. We discuss the importance of treating outliers, selection effects, light-curve shape and color populations and standardization relations, unexplained dispersion, and heterogeneous observations simultaneously. We present an updated Bayesian framework, called UNITY1.5 (Unified Nonlinear Inference for Type-Ia cosmologY), that incorporates significant improvements in our ability to model selection effects, standardization, and systematic uncertainties compared to earlier analyses. As an analysis byproduct, we also recover the posterior of the SN-only peculiar-velocity field, although we do not interpret it in this work. We compute updated cosmological constraints with Union3 and UNITY1.5,  finding weak \LCDMTension tension with $\Lambda$CDM and possible evidence for thawing dark energy ($w_0 > -1$, $w_a< 0$). We release our SN distances, light-curve fits, and UNITY1.5 framework to the community.
\end{abstract}



\section{Introduction} \label{sec:Introduction}

Type Ia supernovae (SNe~Ia) measure cosmological distances through their standardizable luminosities \citep{Pskovskii1967, Phillips1993}, which enable an inverse-square relation between their observed brightnesses and their (luminosity) distances. Their distance vs. redshift relation \secondhalfofsentence For example, they provide measurements of the physical nature of the acceleration \citep{Abbott2019, Brout2022Cosmology}, and extend the local distance ladder from Mpc distances into the smooth Hubble flow \citep[e.g.,][]{Riess2022}. For an excellent review, see \citet{Goobar2011}.

Generally, different supernova surveys trade area and depth differently and end up with most of their SNe spread only over about a factor of two in redshift. Combining supernova surveys thus gives much stronger cosmological constraints \citep[e.g.,][]{perlmutter97}. To this end, we create an up-to-date compilation of SNe. As we follow much the same dataset-inclusive philosophy of the ``Union'' compilations \citep{kowalski08, amanullah10, suzuki12}, and this is the second major revision to the original Union, we call this dataset ``Union3.'' However, as motivated below, our updated analysis framework is Bayesian and thus is quite different (improved) from the earlier frequentist Union analyses. Our compilation is of a similar size to (but 1/3 larger than) Pantheon+ \citep{Scolnic2022}, but we adopt different calibration paths (routing many surveys through \citealt{Landolt1992} or \citealt{Smith2002} stars for absolute calibration as motivated in Section~\ref{sec:PhotometryCompilation}), select SNe differently (Section~\ref{sec:ValidQuality}), and fit the full rest-frame optical wavelength range using the new SALT3 model (Section~\ref{sec:lcfit}). This makes our compilation a useful comparison (in addition to the comparison provided by our Bayesian analysis framework described below).

SNe~Ia require standardization, most commonly through their measured light-curve-shape \citep{Pskovskii1967, Phillips1993} and color \citep{Riess1996, Tripp1998} parameters, as well as host-galaxy properties \citep{kelly10, sullivan10}. (The inclusion of host-galaxy properties indicates that a single shape and color parameter do not capture the full range of relevant astrophysics affecting the measurements, e.g., \citealt{Boone2021B}.) These parameters have uncertainties that are a combination of measurement uncertainties and scatter unexplained by the light-curve model \citep[e.g.,][]{Riess1996}. Luminosity, the dependent variable, similarly has uncertainties that are a combination of measurement uncertainty and unexplained scatter.

Regression with significant uncertainties in both dependent and independent variables requires careful treatment.\footnote{
``Significant uncertainties'' here means uncertainties that are a meaningful fraction of the population width. For example, in the SALT2 model \citep{Guy2007}, light-curve shape $x_1$ has a population width of $\sim 1$ and so typical $x_1$ measurement uncertainties of $\sim 0.3$ are considered significant. The same is true of color $c$, which has a population width $\sim 0.1$~magnitudes and measurement uncertainties $\sim 0.04$ magnitudes. Furthermore, we note that ``uncertainties'' in this context includes any unexplained scatter around the model, so simply increasing light-curve signal-to-noise may still leave significant uncertainties. As we discuss, Bayesian Hierarchical Models naturally handle inference when such significant uncertainties are present.\\
Intrinsically correlated independent variables, e.g., light-curve shape earlier and later than maximum light \citep{hayden19} increase the challenge, as a relevant population width is the (now much smaller) direction perpendicular to the correlation \citep{minka1999linear}. In this example, this population width would be the width of the population distribution of (light-curve shape early) $-$ (light-curve shape late) which is smaller than the width of either light-curve shape population distribution and is more difficult to measure. Other examples include correlations between host-galaxy properties \citep{Rose2020} and correlations between light-curve parameters and host-galaxy properties \citep{Dixon2021}. Bayesian Hierarchical Models can handle these correlated cases as well.}
\citet{Gull1989} shows how modeling the stochastic process that generated the observed values of both the independent and dependent variables is key to addressing this case. Furthermore, \citet{Gull1989} shows that building this model requires including parameters for the ``true'' values of the independent variables and these values require informative priors. As applying incorrect informative priors on these parameters causes a bias, the safest solution (advocated for by \citealt{Gull1989}) is to marginalize over the parameters of the priors (``hyperparameters'') at the same time as the other parameters, i.e., to build a Bayesian Hierarchical Model. \citet{March2011} applied the \citet{Gull1989} model to SN cosmology, using redshift-independent informative priors on light-curve shape and color. However, applying informative priors that are driven by well-measured low-redshift SNe can cause a bias if the population of independent variables falls further from the prior as a function of redshift \citep{Wood-Vasey2007}, so the proper solution is to allow for the hyperparameters to change with redshift \citep{rubin15b, rubin16}.

Although correct standardization requires Bayesian Hierarchical Models, these models provide other advantages. The unexplained luminosity dispersion (remaining after standardization) can be treated as just another fit parameter \citep{March2011}, as opposed to frequentist regression where it must be treated separately. Indeed, even multidimensional parameterizations of how unexplained dispersion might impact shape and color \citep[e.g.,][]{Marriner2011, Kessler2013} can be included in the model. Outliers, e.g., non-Ia contamination, can be modeled simultaneously as another population \citep{Kunz2007}. Bayesian Hierarchical Models can even directly include selection effects, e.g., the tendency to find SNe with brighter apparent magnitudes \citep{rubin15b}. Of course, the standard terms in a cosmological analysis like calibration uncertainties can be included in any framework, including a Bayesian Hierarchical Model.

The Unified Nonlinear Inference for Type Ia cosmologY (UNITY) framework of \citet{rubin15b} addressed all of the preceding issues, unifying them into one model. UNITY found cosmology constraints consistent with the earlier Union frequentist analysis \citep{suzuki12} when run on the same data (Union2.1). Interestingly, the related Bayesian \Steve model \citep{Hinton2019} inferred different cosmological constraints when comparing results to those from the simulation-based \BBC framework (\BEAMS with Bias Corrections, \citealt{Kessler2017}) on simulated data designed to test both \Steve and \BBC.\footnote{\BEAMS stands for Bayesian Estimation Applied to Multiple Species \citep{Kunz2007}.} When measuring the dark energy equation of state parameter $w$ on simulated datasets, \citet{Hinton2019} found biases and scatter, both about half the size of the total uncertainties, between \Steve and \BBC. On real data, they found a large 0.07 offset on the equation-of-state parameter $w$ between \Steve and \BBC (roughly the size of the total uncertainty) which remains only partially expected from the simulation results. \Steve (like UNITY) includes additional parameters over \BBC, but the full source of this offset remains unexplained \citep{Brout2019Systematics}.\footnote{Part of the source of the offset is that the priors on \Steve seem to be tuned to favor the model of unexplained dispersion from \citet{guy10} over the model from \citet{Chotard2011}. See Section~\ref{sec:unexplaineddispersion} for UNITY's parameterization of unexplained dispersion. Another part of the offset may be related to the increased flexibility of \Steve (and UNITY) to model effects like changes in the light-curve shape and color population distributions with redshift. See Section~\ref{sec:ShapeAndColorPopulation} for UNITY's parameterization of light-curve color and shape populations.} Such large differences in inferred cosmology (on the same data) motivate a unified Bayesian approach to ensure robustness.

A final benefit of our analysis is our handling of analysis choices with consequences for the cosmological results. If there is an expected or desired answer, and the analysis results are computed and revealed while the analysis is developed, then unconscious bias can push the analysis results towards the expected answer (``experimenter bias'').\footnote{A related problem is that analyses can be tuned to find statistically significant results \citep{gelman2013garden}.} Analyses that hide results until the analysis is finalized are known as ``blinded analyses'' and have long been used in particle physics (e.g., \citealt{Roodman2003}) and cosmology (e.g., \citealt{Conley2006, kowalski08}). Blinded analyses seek to minimize conscious and unconscious choices that may push analyses towards expected results \citep{MacCoun2015}. Our cosmology analysis is blinded, in the sense that the cosmological constraints were hidden while the analysis was finalized. Even the choices of which cosmological models to present in Section~\ref{sec:CosmologyConstraints} and which external datasets to include were made while blinded. We performed our blinding by fitting separate absolute magnitudes for each sample (fixing the cosmology to flat \LCDM with $\Omega_m=0.3$) and removing these per-sample magnitude differences.

Our paper is organized as follows. Section~\ref{sec:PhotometryCompilation} presents our compilation of photometry from \NDatasets datasets and updates to their calibrations. Next, we uniformly fit the data, and apply selection cuts (Section~\ref{sec:UnionCompilation}), producing the updated Union3 compilation of SNe. Section~\ref{sec:UNITY} presents the updated version of the UNITY framework, which we refer to as UNITY1.5. This section outlines several improvements over the original framework as well as provides a general review of UNITY. Section~\ref{sec:CosmologyConstraints} shows the updated cosmological constraints both with SNe alone, and SNe with distance measures from Baryon Acoustic Oscillations (BAO, with and without Big Bang nucleosynthesis, BBN), the Cosmic Microwave Background (CMB), and the Hubble constant ($H_0$). Section~\ref{sec:CosmologyConstraints} also shows the decomposition of uncertainties for two cosmological parameters ($\Omega_m$ and $w_a$). Finally, Section~\ref{sec:Conclusions} summarizes and concludes. Some of the technical details are more suitable for appendices; Appendix~\ref{sec:scriptM} shows our parameterization of distance moduli and the absolute magnitude (similar to \citealt{perlmutter97}), Appendix~\ref{sec:Priors} shows our parameters and priors, Appendix~\ref{sec:colorpop} presents the details of our light-curve shape and color population model, Appendix~\ref{sec:MoreZeropoints} presents some zeropoints derived in this work, Appendix~\ref{sec:BAOCollection} shows the BAO constraints we include, and Appendix~\ref{sec:CMBCompression} derives a compressed Cosmic Microwave Background likelihood to speed up computation, and finally Appendix~\ref{sec:validLCfit} uses our simulated-data testing to validate our light-curve fitting and constrain deviations from Gaussian uncertainties.

\section{Data Compilation} \label{sec:PhotometryCompilation}

This section describes the compiled SN light curves and the work to place them on the same magnitude scale and thus the same distance scale.

\subsection{General Photometric Calibration Considerations}

As with all standardizable candles, any systematic error in relative apparent SN~Ia magnitudes becomes a systematic error in relative distances. Thus, relative photometric calibration is key to using SNe~Ia for cosmological analyses. At a minimum, this calibration must span magnitude (more distant SNe are fainter than nearby SNe), position on the sky (more distant SNe are generally found in dedicated deep survey fields while nearby SNe can be anywhere), and wavelength (more distant SNe are redshifted compared to nearby SNe). In addition, the \bandpasses for the survey must be measured (including the full optical path with any atmospheric absorption, not just scans of the filter transmission) giving the range of wavelengths that the survey actually observed at. In addition, if multiple surveys are combined together (as they are here), then the relative calibration between surveys must be verified, generally using reported magnitudes for field stars or SNe in common. We summarize each of these points for this work in turn.

\subsubsection{Calibration in Magnitude (Linearity Calibration)}

Generally speaking, we expect the group behind each published dataset to have measured and corrected the nonlinearity of the camera(s) used, enabling flux measurements that are linearly related to the photon rate on the detector(s). We also expect other aspects of the photometry that could affect the linearity (like mismatches between the model and true PSFs or inaccurate sky subtraction) to be well controlled. There is no absolutely linear magnitude scale to compare against, but different surveys can be compared using their quoted magnitudes for field stars to see if any survey stands out as nonlinear. For example, \citet{currie20} find that Pan-STARRS aperture photometry is more linear when compared to other surveys than Pan-STARRS PSF photometry.

\subsubsection{Calibration Over the Sky}

Again, different surveys can be compared using their field stars to check and improve survey uniformity. For example, \citet{Finkbeiner2016} calibrated the Sloan Digital Sky Survey \citep{York2000} against Pan-STARRS \citep{Chambers2016}. In general, for large surveys, spatial calibration is controlled to the point where it is a subdominant systematic uncertainty. As another example, the four SuperNova Legacy Survey (SNLS) fields are completely consistent with each other although disjoint on the sky \citep{betoule14}.

\subsubsection{Wavelength-to-Wavelength Calibration (Fundamental- or Absolute-Color Calibration)} \label{sec:FundamentalCalibration}

As noted, calibrating SNe together across different redshifts requires building a calibrated spectral-energy-distribution (SED) model for SNe Ia, and fitting it to multi-band light curves (or spectrophotometry). The relative calibration of the light curves as a function of wavelength is important, as is the relative calibration of the rest-frame model as a function of wavelength (which is trained on a combination of light curves and spectra).

Surveys can accomplish this calibration in different ways. In general, the best practice at present is to directly observe spectrophotometric standard stars like those from the \HST CALSPEC system \citep{Bohlin2014}, ideally interspersed with the survey to control any changes with time or telescope pointing. The CALSPEC system is calibrated to the non-local-thermodynamic-equilibrium (NLTE) models for the atmospheres of three primary hot DA white dwarfs, each close enough to have essentially zero extinction \citep{Bohlin2020}. Soon, artificial light sources referenced to calibrated detectors may have better known SEDs than stars on the sky \citep[e.g.,][]{Lombardo2017}.

Unfortunately, many surveys do not observe spectrophotometric standard stars, choosing instead to observe only stars from a catalog of magnitudes (generally \citealt{Landolt1992} stars for $UBVRI$ filters, \citealt{Smith2002} for $ugriz$ filters, or even just relying on magnitudes of field stars from overlapping wide-area surveys like SDSS). Most of these stars are not in CALSPEC, so this interjects another photometric system between the SN observations and spectrophotometric standards. In the future, depending on the calibration accuracy that can be achieved, comparing {\it Gaia} all-sky spectrophotometry \citep{Carrasco2021, GaiaCollaboration2022} to each survey's field stars may become the best calibration path for such surveys.

For surveys calibrated to the \citet{Landolt1992} or \citet{Smith2002} systems, we compute AB offsets by color transforming every CALSPEC star in those systems to the natural system of the SNe, computing synthetic AB photometry in the natural system passbands, and averaging over all the CALSPEC stars. As there is some internal star-to-star tension in this comparison \citep{Bohlin2015, currie20}, averaging over stars reduces uncertainties (instead of just taking one star like BD+17$^{\circ}$~4708 as the fundamental reference, e.g., \citealt{Kessler2009}). We call these systems the \LandoltCALSPEC system (as it is based on \NLandoltCALSPEC stars: \LandoltCALSPECStars) and the \SmithCALSPEC system (based on \NSmithCALSPEC stars: \SmithCALSPECStars). This process yields uncertainties of around 5~mmag per filter \citep{Bohlin2015}.

\subsubsection{Passbands} \label{sec:BandpassUnc}

The gold standard for \bandpass measurements (the response of the system as a function of wavelength for each filter) is to scan the response of the system wavelength-by-wavelength with a light source referenced to a calibrated detector and also monitor the atmospheric part of the transmission \citep{Stubbs2006}. As noted above, the same sort of system could be used for inter-filter calibration, but practically such systems as of yet have scattered light and other systematic uncertainties that prevent precise relative measurements at widely spaced wavelengths (e.g., requiring the ``tweak'' term of \citealt{Tonry2012}). Thus, we distinguish \bandpass measurements of filters (intra-filter response as a function of wavelength) from the inter-filter response as a function of wavelength discussed previously.

For surveys that have not measured their \bandpasses (or have implemented such measurements only after significant evolution with time, e.g., \citealt{hicken12}), observing stars of different SEDs is the only way to verify the given \bandpasses. There are two sets of star observations that one might consider using: SN field stars and standard stars. Unfortunately, many surveys do not even quote their standard-star measurements (\citealt{Betoule2013} being a welcome exception). This leaves SN field stars, which unfortunately tend to be redder than low-redshift SNe and which are not always observed in similar filters as overlapping well-calibrated all-sky surveys (e.g., $BVRI$ photometry must be transformed to Pan-STARRS $grizy$, \citealt{Scolnic2015}).

We pursue a compromise in this work for the low-redshift datasets with unmeasured or evolving \bandpasses: shifting the natural-system \bandpasses until the observed color terms match synthetic color terms. We synthesize our color terms using the complete set of stars from the stellar libraries of \citet{Pickles1998}, the STIS Next Generation Spectral Library
Version 2 (\url{https://archive.stsci.edu/prepds/stisngsl}), and the STIS Low Resolution Stellar Library \citep{Pal2023}. We average the derived \bandpass shifts between libraries, which generally agree to a few angstroms, except for a few broader \bandpasses. Table~\ref{tab:FilterShifts} shows our recovered mean \bandpass shifts, which we discuss in more detail in Section~\ref{sec:datasets}. We take the \citet{Bohlin2015} shifts of the \citet{Bessell2012} passbands as the approximation to the \citet{Landolt1992} natural system,\footnote{The Landolt system is heterogeneous and thus does not exactly have one natural system; see the summary in \citet{conley11}.} and derive our own shifts for the \citet{Smith2002} passbands using their observations of \NSmithCALSPEC CALSPEC spectrophotometric standard stars.

\startlongtable
\begin{deluxetable}{lr}
  \tablecaption{Bandpass shifts for each filter to bring the synthesized color terms in line with observed color terms. For the \citet{Bessell2012} filters, we take the shifts from \citet{Bohlin2015}; for the \citet{Smith2002} filters, we compute our own shifts using stars in common with CALSPEC.\label{tab:FilterShifts}}
 \tablehead{
 \colhead{Filter} & \colhead{Shift (\ang)}}
 \startdata
\citet{Bessell2012} $U$ & -8 \\
\citet{Bessell2012} $B$ & -20 \\
\citet{Bessell2012} $V$ & -20 \\
\citet{Bessell2012} $R$ & -31 \\
\citet{Bessell2012} $I$ & -27 \\
\hline 
\citet{Smith2002} $u$ & $7$\\
\citet{Smith2002} $g$ & $19$\\
\citet{Smith2002} $r$ & $7$\\
\citet{Smith2002} $i$ & $30$\\
\citet{Smith2002} $z$ & $-12$\\
\hline 
LOSS K1 $B$ & $-13$\\
LOSS K2 $B$ & $-28$\\
LOSS K3 $B$ & $-22$\\
LOSS K4 $B$ & $-21$\\
LOSS N1 $B$ & $-22$\\
LOSS N2 $B$ & $73$\\
LOSS K1 $V$ & $-18$\\
LOSS K2 $V$ & $-17$\\
LOSS K3 $V$ & $-18$\\
LOSS K4 $V$ & $-18$\\
LOSS N1 $V$ & $-18$\\
LOSS N2 $V$ & $-17$\\
LOSS K1 $R$ & $-46$\\
LOSS K2 $R$ & $-49$\\
LOSS K3 $R$ & $-49$\\
LOSS K4 $R$ & $-44$\\
LOSS N1 $R$ & $-47$\\
LOSS N2 $R$ & $-51$\\
LOSS K1 $I$ & $-180$\\
LOSS K2 $I$ & $-58$\\
LOSS K3 $I$ & $-56$\\
LOSS K4 $I$ & $-55$\\
LOSS N1 $I$ & $-133$\\
LOSS N2 $I$ & $-133$\\
\hline 
LCO $B$ & $-23$\\
LCO $V$ & $124$\\
LCO $g$ & $-65$\\
LCO $r$ & $-9$\\
LCO $i$ & $62$\\
\hline 
CfA1 FLWO $B$ thick & $12$\\
CfA1 FLWO $B$ thin & $11$\\
CfA1 FLWO $V$ thick & $-35$\\
CfA1 FLWO $V$ thin & $-9$\\
CfA1 FLWO $R$ thick & $-73$\\
CfA1 FLWO $R$ thin & $-63$\\
CfA1 FLWO $I$ thick & $27$\\
CfA1 FLWO $I$ thin & $-61$\\
\hline 
CfA2 4Sh1 Harris $U$ & $38$\\
CfA2 4Sh1 SAO $U$ & $5$\\
CfA2 4Sh3 Harris $U$ & $38$\\
CfA2 4Sh3 SAO $U$ & $5$\\
CfA2 AC Harris $U$ & $43$\\
CfA2 AC SAO $U$ & $24$\\
CfA2 4Sh1 Harris $B$ & $23$\\
CfA2 4Sh1 SAO $B$ & $11$\\
CfA2 4Sh3 Harris $B$ & $23$\\
CfA2 4Sh3 SAO $B$ & $11$\\
CfA2 AC Harris $B$ & $-8$\\
CfA2 AC SAO $B$ & $-14$\\
CfA2 4Sh1 Harris $V$ & $-13$\\
CfA2 4Sh1 SAO $V$ & $-14$\\
CfA2 4Sh3 Harris $V$ & $-13$\\
CfA2 4Sh3 SAO $V$ & $-14$\\
CfA2 AC Harris $V$ & $-30$\\
CfA2 AC SAO $V$ & $-19$\\
CfA2 4Sh1 Harris $R$ & $-57$\\
CfA2 4Sh1 SAO $R$ & $-42$\\
CfA2 4Sh3 Harris $R$ & $-57$\\
CfA2 4Sh3 SAO $R$ & $-42$\\
CfA2 AC Harris $R$ & $-62$\\
CfA2 AC SAO $R$ & $-39$\\
CfA2 4Sh1 Harris $I$ & $-88$\\
CfA2 4Sh1 SAO $I$ & $-157$\\
CfA2 4Sh3 Harris $I$ & $-88$\\
CfA2 4Sh3 SAO $I$ & $-157$\\
CfA2 AC Harris $I$ & $-228$\\
CfA2 AC SAO $I$ & $-157$\\
\hline 
CfA3/4 4Sh Harris $B$ & $51$\\
CfA3/4 KC $B1$ & $2$\\
CfA3/4 MC $B$ & $3$\\
CfA3/4 KC $B1$ & $-9$\\
CfA3/4 KC $B2$ & $-64$\\
CfA3/4 4Sh Harris $V$ & $4$\\
CfA3/4 KC $V1$ & $4$\\
CfA3/4 MC $V$ & $-26$\\
CfA3/4 KC $V1$ & $-3$\\
CfA3/4 KC $V2$ & $-3$\\
CfA3/4 4Sh Harris $R$ & $-27$\\
CfA3/4 4Sh Harris $I$ & $-92$\\
CfA3/4 KC $r1$ & $24$\\
CfA3/4 MC $r$ & $17$\\
CfA3/4 KC $r1$ & $30$\\
CfA3/4 KC $r2$ & $-11$\\
CfA3/4 KC $i1$ & $-30$\\
CfA3/4 MC $i$ & $-33$\\
CfA3/4 KC $i1$ & $-28$\\
CfA3/4 KC $i2$ & $-28$\\
\hline 
Swope $B$ & $0$\\
Swope $V3009$ & $-12$\\
Swope $V3014$ & $0$\\
Swope $V9844$ & $-3$\\
Swope $u$ & $-17$\\
Swope $g$ & $1$\\
Swope $r$ & $-11$\\
Swope $i$ & $-7$\\
\enddata
\end{deluxetable}

\newcommand{\synthsentence}{We synthesize magnitudes for the stars (plotted with red stars) using the \citet{Pickles1998} stellar library and magnitudes for the SNe~Ia (plotted with blue dots) using the \citet{hsiao07} SN~Ia template from $-10$ to +15 days, redshifted from $z=0.01$ to 0.10. We subtract the linear color terms from the synthesized magnitude differences, as shown at the top of each panel. Mean offsets of a few hundredths of a magnitude are visible between stars and SNe, and the stars are generally more tightly clustered around the linear relations.\xspace}

\begin{figure*}[h!tbp]
    \centering
    \includegraphics[width = \textwidth] {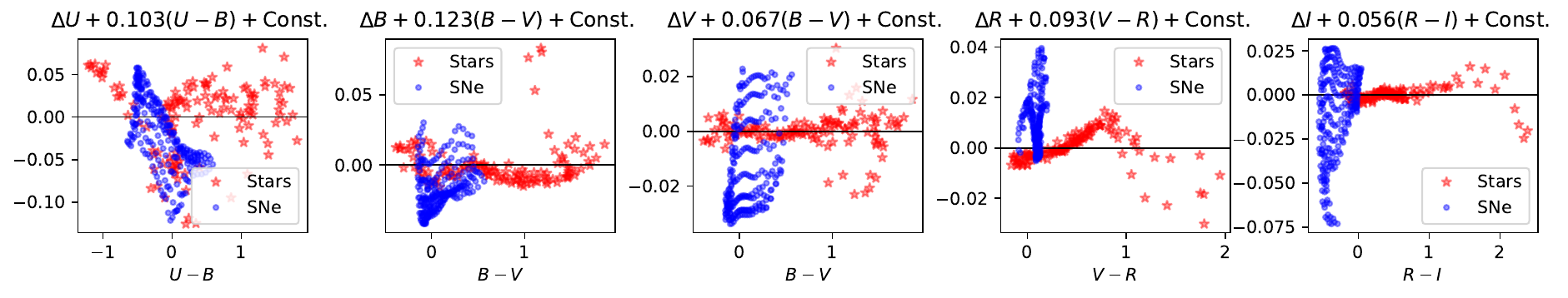}
    \caption{Residuals from a linear color transformation between \citet{Bessell2012} $UBVRI$ filters and the same filters shifted by 100\ang. Each panel shows $UBVRI$ from left to right. \synthsentence 
    \label{fig:ColorTransformError}}
\end{figure*}

We do not have any natural-system photometry for some SNe. Instead the SNe were linearly color transformed onto the \citet{Landolt1992} system using relations derived for stars in common with Landolt. Figure~\ref{fig:ColorTransformError} shows why this is worse than working in the natural system and gives a sense of the scale of the error (see also \citealt{Stritzinger2002}). We use the \citet{Bessell2012} \bandpasses with the \citet{Bohlin2015} \bandpass shifts as an approximation to the Landolt system with and without an additional 100\ang shift. (Comparing synthetic photometry between these \bandpasses gives color terms similar in size to many observed color terms.) We synthesize Vega magnitudes, which are within a few percent of the Landolt system \citep{Bessell2005}. \synthsentence 

\subsubsection{Relative Survey Calibration}

If the previous steps were done correctly, then surveys would only need inter-calibration as a crosscheck. Observations of field stars in common can be used to place SN surveys on the same calibration as long as the filters are sufficiently similar and the stellar SEDs are known \citep{Scolnic2015, currie20, Fragilistic2021}. As noted previously, surveys generally do not publish their standard-star measurements and the field stars are offset in color compared to low-redshift SNe. So calibration based only on field stars may increase or decrease the calibration precision. For example, \citet{Scolnic2015} found improved inter-survey agreement after inter-calibrating SN surveys against Pan-STARRS, but \citet{currie20} did not.

Another possibility is to check relative calibrations using SNe observed in common between surveys \citep[e.g.,][]{Stritzinger2002, hicken09, Scolnic2022}. After compiling our datasets and fitting light curves, we carry out such a test in Section~\ref{sec:DeltaZPNearby}.

\subsection{Datasets} \label{sec:datasets}

Table~\ref{tab:DatasetSummary} presents the datasets included in this work. For each dataset, we include the photometry reference(s), the system(s) it is calibrated on, our assumed calibration uncertainties, and the approximate survey depth, which we use as the central value of a broad $\pm 0.5$~magnitude prior for the selection-effect part of UNITY (discussed more in Section~\ref{sec:MalmquistBias}). We also give our estimated calibration uncertainties.

\begin{longrotatetable}
\begin{deluxetable*}{p{2 cm}p{3.2 cm} p{2.5 cm} p{1.6 cm} c ccc rr} 
  \label{tab:DatasetSummary}
\tablehead{
 \colhead{SN Sample} & \colhead{Reference} & \colhead{Calibration} & \multicolumn{2}{c}{Uncertainty} & \colhead{Depth} & \colhead{$\sigma_{\mathrm{Depth}}$} & \colhead{SNe} & \colhead{$z$ Range} & \colhead{\S}  \\
 &  &  &  \colhead{Mag} & \colhead{\ang} &  \colhead{$\pm 0.5$ Mag} & \colhead{$\pm 0.25$ Mag} & \colhead{Pass.} &  & 
 }
 \startdata
 C/T & \citet{hamuy96} & Landolt (Standard System) & \CTCalUncM & \CTCalUncA & $R=18.5$ & 1.0 & \CTNSNe & \CTzRange & \ref{subsec:CT} \\
 CfA1 & \citet{riess99} & Landolt  & \CfAOneCalUncM & \CfAOneCalUncA & $R=17.0$ & 1.0 & \CfAOneNSNe & \CfAOnezRange & \ref{subsec:CfA1} \\
Krisciunas & \citet{krisciunas01, krisciunas04, krisciunas04b, krisciunas06} & Landolt (Standard System) & \KrisciunasCalUncM & \KrisciunasCalUncA & $R=16.0$ & 1.0 & \KrisciunasNSNe & \KrisciunaszRange & \ref{subsec:Krisciunas} \\
 CfA2 & \citet{jha06} & Landolt & \CfATwoCalUncM & \CfATwoCalUncA & $R=16.5$ & 1.0 & \CfATwoNSNe & \CfATwozRange & \ref{subsec:CfA2} \\
 CfA3 & \citet{hicken09} & Landolt \& Smith & \CfAThreeCalUncM & \CfAThreeCalUncA & $R=17.0$ & 1.0 & \CfAThreeNSNe & \CfAThreezRange & \ref{subsec:CfA3} \\
 CfA4 & \citet{hicken12} & Landolt \& Smith & \CfAFourCalUncM & \CfAFourCalUncA & $R=17.5$ & 1.0 & \CfAFourNSNe & \CfAFourzRange & \ref{subsec:CfA4} \\
 CSP1 DR3 & \citet{krisciunas17} & Landolt \& Smith & \CSPCalUncM & \CSPCalUncA & $R=18.0$ & 1.0 & \CSPNSNe & \CSPzRange & \ref{subsec:CSP1} \\
 SCP & \cite{kowalski08} & Landolt & \SCPNearbyCalUncM & \SCPNearbyCalUncA & $R=19.5$ & 1.0 & \SCPNearbyNSNe & \SCPNearbyzRange & \ref{subsec:SCP99}\\
 LOSS & \cite{ganeshalingam10} \& \citet{stahl19} & Landolt, Pan-STARRS & \LOSSCalUncM & \LOSSCalUncA & $R=17.0$ & 1.0 & \LOSSNSNe & \LOSSzRange & \ref{subsec:LOSS} \\
 Foundation & \citet{foley18} & Pan-STARRS & \PanSTARRSCalUncM & \PanSTARRSCalUncA & $R=17.5$ & 1.0 & \FoundationNSNe & \FoundationzRange & \ref{subsec:Foundation} \\
 CNIa0.02 & \citet{Chen2020} & Pan-STARRS, Refcat2 & \CNZeroTwoCalUncM & \CNZeroTwoCalUncA & $R=16.0$ & 1.0 & \CNZeroTwoNSNe & \CNZeroTwozRange & \ref{subsec:CNIa0.02} \\
 LSQ + CSP & \citet{Walker2015} & Landolt \& Smith & \LSQCSPCalUncM & \LSQCSPCalUncA & $R=19.0$ & 1.0 & \LSQCSPNSNe & \LSQCSPzRange & \ref{subsec:LSQ} \\
 LSQ + LCO & \citet{Baltay2021} & APASS & \LSQLCOCalUncM & \LSQLCOCalUncA & $R=18.0$ & 1.0 & \LSQLCONSNe & \LSQLCOzRange & \ref{subsec:LSQLCOGT} \\
\hline
SCP & \citet{knop03} \& \citet{nobili09} & Landolt (Standard System), CALSPEC & \KnopCalUncM & \KnopCalUncA & $R=24.0$ & 0.5 & \KnopNSNe & \KnopzRange & \ref{subsec:SCP98} \\
SNLS & \citet{betoule14} & Pan-STARRS & \SNLSCalUncM & \SNLSCalUncA & $i=23.8$ & 0.5 & \SNLSNSNe & \SNLSzRange & \ref{subsec:SNLS} \\
SDSS & \citet{Sako2018} & SDSS PT & \SDSSCalUncM & \SDSSCalUncA & $r=22.0$ & 0.5 & \SDSSNSNe & \SDSSzRange & \ref{subsec:SDSS} \\
ESSENCE & \citet{narayan16} \& \citet{Krisciunas2005} & Landolt, CALSPEC & \ESSENCECalUncM &\ESSENCECalUncA & $R=23.5$ & 0.5 & \ESSENCENSNe & \ESSENCEzRange & \ref{subsec:ESSENCE} \\
Pan-STARRS & \citet{Jones2018} \& \citet{Scolnic2018} & Pan-STARRS & \PanSTARRSCalUncM & \PanSTARRSCalUncA & $r=22.0$ & 0.5 & \PanSTARRSNSNe & \PanSTARRSzRange & \ref{subsec:PS1} \\
DES & \citet{Brout2019Photometry} & CALSPEC & \DESCalUncM & \DESCalUncA & $i=23.0/24.0$ & 0.5 & \DESNSNe & \DESzRange & \ref{subsec:DES} \\
\hline
HZT & \citet{tonry03} & Landolt, SDSS, Persson & \HZTTonryCalUncM & \HZTTonryCalUncA & $I=23.8$ & 0.5 & \HZTTonryNSNe & \HZTTonryzRange & \ref{subsec:HZT03} \\
GOODS HZT & \citet{riess07} & CALSPEC & \SCPHSTCalUncM & \SCPHSTCalUncA & $z=25.7$ & 0.5 & \GOODSNSNe & \GOODSzRange & \ref{subsec:HZT07} \\
SCP & \citet{amanullah10} & SDSS to Landolt, CALSPEC, Persson & \AmanullahCalUncM & \AmanullahCalUncA & $I=23.8$ & 0.5 & \AmanullahNSNe & \AmanullahzRange & \ref{subsec:SCP01} \\
MCT & \citet{riess18} & CALSPEC & \MCTCalUncM & \MCTCalUncA & $F125W = 26.6$ & 0.25 & \MCTNSNe & \MCTzRange & \ref{subsec:MCT} \\
SCP & \citet{suzuki12} and \citet{rubin13} & CALSPEC, WFC3 & \SCPHSTCalUncM & \SCPHSTCalUncA & $z=25.7$ & 0.5 & \SCPHSTNSNe & \SCPHSTzRange & \ref{subsec:SCPHST} \\
 \enddata
\tablecomments{Summaries of each dataset. The ``Calibration'' column explains the calibration path to CALSPEC, which is generally Landolt, Smith, Pan-STARRS, or direct to CALSPEC. The ``Uncertainty'' columns show our estimated uncertainties from that path to the SNe, but there are additional correlated uncertainties for the indirect-to-CALSPEC paths (e.g., Appendix~\ref{sec:MoreZeropoints}). The depth is the central value of a prior for the median depth of the survey (50\% chance of a SN at this magnitude making it into our sample). SNe pass is the number of SNe passing selection cuts and $z$~Range gives their redshift range. When a SN is observed by more than one group, we assign it to the group where their data have the highest total S/N across the light curve; the SN counts are thus mutually exclusive.}
\end{deluxetable*}
\end{longrotatetable}

With the recent publication of large mid-redshift supernova samples \citep{guy10, betoule14, Sako2018, Jones2018, Scolnic2018, Brout2019Photometry}, it is worth reconsidering the selection of SNe in this redshift range compared to earlier Union compilations. Smaller samples in the same redshift range cannot contribute much statistical weight, and are too small for some of the data-quality checks (like fitting standardization coefficients) possible with larger samples. On the other hand, the calibration uncertainties in each sample have some level of independence, providing cross-checks of the larger samples and some robustness to calibration problems.

For the purposes of our work here, we do not include the following supernova samples that were in Union2.1: \citet{riess98}, \citet{perlmutter99}, \citet{Barris2004}, and \citet{amanullah08}. When fitting each sample with its own standardization parameters as a test, \citet{kowalski08} found the inferred color standardization parameter for \citet{riess98} and \citet{perlmutter99} data to be low, indicating that the uncertainties were underestimated (probably in $I$ band which had less weight than $R$ in the original fits which used color cuts and color priors). The \citet{Barris2004} sample does not have proper host-galaxy-only (``reference'' or ``template'') images, and so proper use of those data requires a light-curve model extending to very late phases, which SALT does not. Finally, the \citet{amanullah08} sample is small (5~SNe) and only measured in two filters.

We do include any sample (of any size) in this redshift range that has near-IR photometry, as this longer wavelength baseline improves SN distances and robustness to zeropoint uncertainties. The mid-to-high redshift supernova samples falling into this category are \citet{knop03} + \citet{nobili09}, \citet{tonry03}, \citet{narayan16} + \citet{Krisciunas2005}, \citet{amanullah10}, and \citealt{Jones2018} + \citet{Scolnic2018} + \citealt{Jones2022} (with additional \citealt{Brout2019Photometry} SNe having IR coverage too red for SALT3).

In contrast with mid-redshift where we are mildly selective, we include many nearby ($z \lesssim 0.1$) SNe, even those where we do not have good determinations of the natural system or the exact calibration path. Many nearby SNe are important, either for measuring nearby peculiar velocities or the local distance ladder, so it is better to be inclusive here. We investigate photometric calibration tensions between nearby SN datasets in Section~\ref{sec:DeltaZPNearby}.

The UNITY1.5 selection-effect model (discussed in Section~\ref{sec:MalmquistBias}) has no informative SN relative-rate model and thus benefits from a prior on the selection efficiency of each survey. The nearby SN samples are frequently too inhomogeneous to have good estimates of their selection efficiencies, so we evaluate approximate magnitude limits (where $\sim 50$\% of SNe would make it into the sample) as follows. We consider SALT $R$-band magnitudes at peak for each SN, as much of the photometric selection is based on unfiltered magnitudes which are similar to $R$ and much of the spectroscopic selection is based on the Si~II 6150\ang line. We histogram these values and fit these counts with a model of
\begin{equation}
    10^{0.6\, t \, m_R} \left[ 1 - \mathrm{erf} \left( \frac{m_R - m_0}{\sqrt{2} \sigma_m} \right) \right]\;. \label{eq:MagnitudeDistribution}
\end{equation}
The $10^{0.6\, m_R}$ term is the magnitude distribution for a constant volumetric rate: going one magnitude fainter increases the limiting distance by a factor $10^{0.2}$ (ignoring relativistic effects for this low redshift range) and increases the volume and thus the SN rate by $10^{0.6}$. However, some SNe are selected from known galaxy catalogs, and these are closer to uniform in distance (the distribution of magnitudes for this possibility is $10^{0.2 \, m_R}$); this is roughly true of the galaxies targeted by the Lick Observatory Supernova Search \citep{Leaman2011}. $t$ is thus a factor that interpolates between these possibilities, and we constrain $1/3 < t < 1$. $m_0$ is the limiting magnitude and $\sigma_m$ is the width of the selection function. Table~\ref{tab:DatasetSummary} displays these survey depths with some modest tweaks based on plots of colors and magnitudes against redshift (the $t$ values do not enter UNITY so we do not quote them). When using our estimated depths as priors, we assume uncertainties of $\pm 0.5$~magnitudes. Many of the posteriors on the depth for the low-redshift SN samples have uncertainties comparable to 0.5~magnitudes, so these priors provide meaningful constraints, especially for small samples. For large samples, which contribute most the weight in the cosmology fits, the priors do not have much impact.\footnote{Of course, using one prior width of $\pm 0.5$~magnitudes for all surveys is only an approximation, but it is one we make for a few reasons. 1) typical surveys have uncertainties from the number-versus-magnitude fits of roughly 0.5 magnitudes. Some small surveys, e.g., CfA1, have larger uncertainties, but these are also small surveys without much weight in the analysis. 2) Some larger surveys have smaller uncertainties from the number-versus-magnitude fits or selection depths estimated from simulations. But these can disagree with the UNITY1.5 model. For example, Figure~9 of \citep{Perrett2010} shows a discovery+spectroscopy+cuts 50\% efficiency a bit fainter than $i=24$ relative to the efficiency at brighter magnitudes. But the SNLS depth posterior from UNITY1.5 is \SNLSDepth, showing moderate tension (Figure~\ref{fig:MagntiudeLimits}). The UNITY1.5 predictive posteriors look reasonable for SNLS (Figure~\ref{fig:PPDvssamp}), so the origin of this discrepancy is not clear. We thus view it as prudent at this time to allow UNITY1.5 the additional freedom to match the data and do not tighten the priors for larger samples. 3) As a test, we shrink the prior to $\pm 0.25$~magnitudes (comparable to the 20\% systematic uncertainty in S/N assumed by \citealt{Brout2022Cosmology}) and note that this does not have much impact on our cosmological parameter values or their uncertainties (for example, the mean changes by less than $0.1\sigma $ for $w_0$-$w_a$ and the uncertainties change by less than 1\%). So the exact assumptions here do not matter too much for UNITY1.5.}

\subsubsection{Cal\'{a}n/Tololo Supernova Survey} \label{subsec:CT}

\citet{hamuy96} presented $BVRI$ light curves of 29 SNe, mostly from the Cal\'{a}n/Tololo Supernova Survey. These data are color-transformed to the Landolt system and no estimates of the natural-system \bandpasses are provided. Figure~\ref{fig:ColorTransformError} shows the size of the error that can be introduced, as our 100\ang shifts give synthetic color terms similar to the observed color terms \citep{Hamuy1993}. We use the color-transformed photometry, assuming systematic uncertainties of \CTCalUncMWords.

\subsubsection{CfA1} \label{subsec:CfA1}

\citet{riess99} presents $BVRI$ light curves for 22 SNe~Ia. These data are color transformed into the standard system, but estimates of the natural-system \bandpasses are provided. Thus, we undo the color corrections that are applied to the supernova data using the quoted color terms. (When a color needed for undoing the color terms is missing, we do not undo the correction to the standard system.) For the \citet{riess99} data, we assume all of the data were taken at the FLWO 1.2 meter, as those are the only \bandpasses provided ($\sim 95\%$ of the data were taken there). The mean color term in $R$ band is slightly larger than that originally reported (A. Riess, private communication in \citealt{currie20}); our complete set of color terms is $B - b = 0.04\, (B - V)$, $V - v = -0.03\, (B - V)$, $R - r = -0.1075\, (V - R)$, and $I - i = 0.06\, (V - I)$, where lower case are natural-system and upper case are standard-system magnitudes.

\subsubsection{SN Set of Krisciunas et al.} \label{subsec:Krisciunas}

\citet{krisciunas01}, \citet{krisciunas04}, \citet{krisciunas04b} and \citet{krisciunas06} presented the optical light-curves of 21 low-redshift SNe Ia. From these, we select 19 SNe (excluding the peculiar SN~1991bg and the somewhat peculiar SN~1991T which would fail our redshift cut in any case). The SN data are color-transformed onto the standard \citet{Landolt1992} system, so we use large uncertainties of \KrisciunasCalUncMWords as motivated in Section~\ref{sec:BandpassUnc}.

\subsubsection{CfA2} \label{subsec:CfA2}

For the 44 CfA2 SNe presented by \citet{jha06}, we use the natural-system photometry provided to us by the authors, rather than the color-transformed system as with previous Union compilations. We use the \bandpasses provided in \citet{jha06}, converted to photonic \bandpasses, and multiplied by the atmospheric transmission (noting that the $U$-band \bandpass already includes the atmosphere). We use the atmospheric transmission of KPNO from IRAF \citep{Tody1986}, although it is from a lower altitude than FLWO. We then apply the \bandpass shifts in Table~\ref{tab:FilterShifts} to bring our synthetic color terms into alignment with the observed color terms. These shifts are most significant in the $I$ bands, probably indicating a somewhat faster CCD falloff in the red than was assumed.

\subsubsection{CfA3} \label{subsec:CfA3}

\citet{hicken09} present natural-system light curves of 185 low-redshift SNe taken with three different cameras (4Shooter, Minicam, and Keplercam) on the FLWO 1.2m.

The 4Shooter data were taken with $UBVRI$ filters (as were the last SNe with CfA2), so we use those \bandpasses as initial estimates before applying \bandpass shifts to match synthetic and observed color terms.

The Minicam and Keplercam data were taken with $UBVri$ filters. No quoted \bandpasses exist for Minicam, so we use Keplercam filters as our initial estimates before applying \bandpass shifts. \citet{hicken09} quotes nominal \bandpasses for the $BVri$ filters, while \citet{hicken12} present $BV$ \bandpasses measured with a monochromatic flat-field illumination system. We use the \citet{hicken12} $BV$ and \citet{hicken09} $ri$ as initial estimates for Kepler and Minicam \bandpasses before applying shifts (we use the $U$ from 4Shooter).

CfA3 used only one reference image per SN. \citet{hicken12} note that this leads to an underestimate of the photometric uncertainties due to the imperfect host-galaxy subtraction process and use SNe with multiple reference images to provide an estimate of the additional uncertainty that should be added as a function of magnitude and filter to the CfA3 SNe. We follow the recommendation of \citet{hicken12} and add these uncertainties in quadrature with the light-curve uncertainties (although some amount of it must correlate between phases of the same SN in the same filter).

\subsubsection{CfA4} \label{subsec:CfA4}

\citet{hicken12} present natural-system light curves of 94 low-redshift SNe~Ia taken with Keplercam. The CfA4 data exhibit a change in $B$ and $r'$ color terms midway through the dataset (MJD $\sim 55058$), related to increasing deposits on the camera. These deposits are later baked out, essentially restoring the original color terms. \citet{hicken12} show $B$-band and $V$-band \bandpass scans before and after the bakeout; we make use of these \bandpasses as inital estimates before applying \bandpass shifts. For CfA4, we use the FLWO-provided $u$-band transmission (\url{http://www.sao.arizona.edu/FLWO/48/CCD.filters.html}), multiplied by the reflectivity of aluminum squared, a typical CCD response for the same type of CCD as Keplercam (Fairchild CCD 486, \url{http://linmax.sao.arizona.edu/FLWO/48/kepccd.html}), and the atmospheric transmission for KPNO from IRAF \citep{Tody1986}.

\subsubsection{CSP1 DR3} \label{subsec:CSP1}

\citet{krisciunas17} presented the complete sample of 123 Carnegie Supernova Project (CSP1) SNe (data release 3). We use only the $ugriBV$ optical data taken with Swope (not the NIR data), as only the optical is within the rest-frame wavelength range of SALT3. The \bandpasses of Swope have been measured with a monochromator-based flat-field illumination system \citep{Stritzinger2011}, and this delivers \bandpasses where synthetic color terms agree well with the observed color terms (Table~\ref{tab:FilterShifts} and \citealt{currie20} show only small \bandpass shifts).

\subsubsection{SCP Nearby SNe} \label{subsec:SCP99}
\citet{kowalski08} presented $BVRI$ photometry of nine low-$z$ SNe, taken with several different telescope/instrument combinations. The SN photometry is presented in the natural system, but the secondary standards are not, so there is no way to directly inter-compare field-star magnitudes in the natural systems. We calibrate this data on the \LandoltCALSPEC system and use the \bandpasses from \citet{kowalski08} without any additional \bandpass shifts for this small sample.

\subsubsection{LOSS} \label{subsec:LOSS}

\citet{ganeshalingam10} presented natural-system photometry of 165 SNe~Ia from KAIT (in four configurations) and the Nickel telescope (one configuration). These data are calibrated against standard stars from \citet{Landolt1992}. They present their estimates of the natural-system \bandpasses and also a set of \bandpass shifts to bring synthesized color terms into agreement with observed color terms. We take their \bandpasses, but derive our own shifts on top of theirs.

\citet{stahl19} presented a further 93 SNe~Ia observed with the last two configurations of KAIT, the same configuration of the Nickel telescope, and an additional configuration of the Nickel telescope. These data are calibrated against field stars that have been color-transformed to be on the \citet{Landolt1992} system. Again, they present estimates of the natural-system \bandpasses. However, our synthesized color terms only match the observed color terms in $V$ band. The problem is especially acute in $I$ band, where the \citet{stahl19} \bandpass is very different between the two Nickel configurations but the observed color term is identical. We use the given \bandpass for $B$ (after clipping some repeated transmission values from the \citealt{stahl19} table), but shift it about 70\ang to the red as shown in Table~\ref{tab:FilterShifts}. The $R$ and $I$ data have very similar color terms between the two Nickel configurations, so we just use the \citet{ganeshalingam10} \bandpasses for both (before applying \bandpass shifts).

\subsubsection{Foundation} \label{subsec:Foundation}

\citet{foley18} present 225 light curves of nearby SNe observed with Pan-STARRS1 as part of the Foundation survey. This survey has a number of unique advantages compared to many other low-z surveys: it is directly tied to CALSPEC, it is taken with an instrument that has had its \bandpasses measured (like CSP), it is (nearly) on the same magnitude system as an existing high-redshift SN survey (the PS1 medium-deep SNe, \citealt{Scolnic2018}), and the $3\pi$ survey provides SN-free reference (or ``template'') images immediately, without waiting $\sim 1$ year for the SN to fade.

As in \citet{Brout2022SuperCal}, we calibrate Foundation to PS1 aperture photometry, which \citet{currie20} find is more linear than the PS1 PSF photometry. We take their tertiary-star catalog (D. Brout, private communication) and perform a robust linear regression (Laplace uncertainties) between Foundation and PS1 aperture photometry as a function of $g-i$ color and {\it Gaia} $G$ magnitude (\citealt{GaiaCollaboration2021}). We use an independent magnitude from PS1 here to avoid Eddington bias \citep{Eddington1913} and remove the median $G$ magnitude for the sample (which has a similar magnitude to the mean SN) to decorrelate the magnitude slope and the AB offset. For the $g-i$ color, we do not remove the median, as we are trying to predict the offset near AB=0. We find magnitude slopes between PS1 aperture and Foundation of $\lesssim 2$~mmag/mag, and find small slopes in color equivalent to bandpass shifts $\lesssim 10$\ang. We thus take 2~mmag and 10\ang as $1\sigma$ systematic uncertainties and show our derived zeropoint offsets in Appendix~\ref{sec:MoreZeropoints}.

\subsubsection{CNIa0.02} \label{subsec:CNIa0.02}

\citet{Chen2020} presented the first 247 SNe~Ia of CNIa0.02, a program devoted to following up all SNe~Ia with $z<0.02$. The data are calibrated using Pan-STARRS1 or ATLAS-REFCAT2 \citep{Tonry2018} magnitudes of field stars, which are color transformed to $B$, $V$, $r_{\mathrm{SDSS}}$, and $i_{\mathrm{SDSS}}$. We thus use the \citet{Bessell2012} $B$ and $V$ \bandpasses and the SDSS $r$ and $i$ \bandpasses \citep{Doi2010} for this data. We do not use the {\it Swift} photometry (which has proven problematic in the past), leaving this for future work. Some upper limits are included in the light curves; by visual inspection, we implement these as $1 \sigma$ upper limits. It would be better to have these low S/N points as flux measurements in the future.

\subsubsection{LSQ + CSP} \label{subsec:LSQ}

\citet{Walker2015} presented 31 SNe found by the LaSilla/QUEST Variability Survey \citep{Baltay2013} and followed up with Swope. We assume that these SNe are on the same magnitude system as the CSP data (described in Section~\ref{subsec:CSP1}), as we see no evidence for a difference between the two in the tensions analysis in Section~\ref{sec:DeltaZPNearby}. We do see possible evidence for a higher unexplained dispersion in our cosmology analysis (Section~\ref{sec:unexplaineddispersion}), so it may be worth going back to the pixels and double-checking (perhaps a sample of) the photometry in the future.

\subsubsection{LSQ + LCO} \label{subsec:LSQLCOGT}

\citet{Baltay2021} presented 140 SNe~Ia discovered by LaSilla-QUEST, ASAS-SN, and others and followed by LCO. They present all light-curve data in the natural system (which is mildly heterogeneous, as the data might have been taken by any of the LCO 1m telescopes). The calibration is to the \citet{Landolt1992} and \citet{Smith2002} systems through field stars in common with APASS \citep{Henden2012}. We construct initial \bandpass estimates using the given natural-system filter + Sinistro CCD efficiencies, plus the KNPO atmospheric transmission from IRAF \citep{Tody1986} and two reflections off of aluminum. Even still, the synthesized color terms using the stellar libraries do not quite match the measured color terms, with the $V$ band in particular quite far off. The measured $V$ color term is small, yet the $V$ \bandpass is $\sim 120$~\ang bluer than the \citet{Bessell2012} $V$ \bandpass. Thus, our \bandpass shifts have a meaningful effect for this sample. Based on the Vega and BD+17$^\circ$~4708 magnitudes quoted, we assume the $BV$ data are on \citet{Landolt1992} and the $gri$ data are on \citet{Smith2002}.

\subsubsection{SCP WFPC2 + NICMOS} \label{subsec:SCP98}

\citet{knop03} presented observations of eleven supernovae discovered from the ground and observed with \HST WFPC2. Five of these supernovae were also observed with \HST NICMOS2 \citep[these observations were presented in][]{nobili09}, making these optical+NIR SNe an especially useful contribution to the high-redshift Hubble diagram (which is mostly from observer-frame optical observations). The optical data were presented transformed onto the \citet{Landolt1992} system, while the NICMOS2 data are on the Vega=0 system. We assume slightly smaller than usual color-transform uncertainties of \KnopCalUncM mag, as the original authors checked their ground-based photometry against the Vega=0 \HST system to this level and were careful in their treatment of the SNe in the WFPC2 filters.

\subsubsection{SDSS} \label{subsec:SDSS}

For the SDSS portion of the SDSS/SNLS Joint Light-Curve Analysis (JLA, \citealt{Betoule2013, betoule14}), we follow \citet{Sako2018} and calibrate using Photometric Telescope (PT) observations of three solar-analog stars (P330E, P177D, P041C) that are color-transformed to SDSS by \cite{Betoule2013}. We remove the offsets between AB and SDSS that were applied in \citet{Sako2018}: $u=-0.0679$, $g=0.0203$, $r=0.0049$, $i=0.0178$, $z=0.0102$, and replace them with updated values computed using 2023 CALSPEC: $u=-0.0844 \pm 0.0062$, $g=0.0173 \pm 0.0040$, $r=0.0069 \pm 0.0026$, $i=0.0265 \pm 0.0055$, $z=0.0196 \pm 0.0077$. The uncertainties on the mean are the (sample) standard deviation divided by $\sqrt{3}$. We use all SNe labeled `SNIa' or `SNIa?' and fit all light curves in flux ($\mu$Jy).

\subsubsection{SNLS JLA} \label{subsec:SNLS}

\citet{guy10} presented 252 SNe from the SuperNova Legacy Survey (SNLS) taken with MegaPrime/MegaCam on CFHT. These we recalibrated as part of the SDSS/SNLS Joint Light-Curve Analysis (JLA, \citealt{Betoule2013, betoule14}). We only are able to use supernovae that pass the JLA light-curve cuts, although based on our validation testing, we expect that this is a small effect. SNLS reached a survey depth of $i=24.3$ AB, although the spectroscopically confirmed sample we use here reaches a depth of $\sim i=24.0$ AB \citep{Perrett2010}.

The SNLS JLA photometric calibration is based on three paths: 1) three solar-analog stars observed directly with MegaPrime using short exposures. Unfortunately, the PSF from these short-exposure observations seems different on average than the longer-exposure PSF in the SN fields, and no same-exposure-time observations were taken to intercalibrate the solar-analog fields and the SN fields. We thus ignore this path for now. 2) Calibration through color-transforms to the \citet{Landolt1992} system. This path has large uncertainties for these dissimilar filter sets ($ugriz$ for MegaPrime, $UBVRI$ for Landolt), so we ignore it as well. 3) Intercalibration with SDSS to use the Photometric Telescope (PT) observations of three solar-analog stars (P330E, P177D, P041C). BD+17$^{\circ}$~4708 was originally included as well, but it appears to be mildly variable \citep{Bohlin2015, Marinoni2016, Rubin2022}. This only relies on three stars and relies on two transforms between photometric systems, so we use a newer option that was not available for JLA: 4) calibration through Pan-STARRS aperture photometry from the PS1 $3\pi$ survey.

To intercalibrate SNLS and PS1 aperture photometry, we perform an updated fit similar to those in \citet{currie20}. For each of $griz$, we collect CALSPEC observations in PS1 aperture photometry and match PS1 against the SNLS tertiary-star catalog in \cite{Betoule2013}. We use all tertiary stars bluer than $g-i = 2$. Then we compute synthetic PS photometry of those CALSPEC stars and also synthetic PS and MegaPrime photometry using the \citet{Pickles1998} stellar library. We use the stellar templates for dwarfs stars only, which is appropriate for most of the stars in these fields. For each filter, we then simultaneously infer:
\begin{itemize}
\item The AB offsets for both PS and MegaPrime.
\item An exponential warp in wavelength that can adjust the effective wavelength of the PS1 \bandpasses (Section~\ref{sec:MoreZeropoints} shows these values). We use this exponential rather than a simple filter shift, as this seems to work a bit better for PS1 $g$.
\item A 2D spline as a function of color and radius on the MegaPrime focal plane that describes the synthetic MegaPrime $-$ PS1 photometry of the spectral library. We use $g-i$ as the color for the $gri$ calibrations and $i-z$ for the $z$ calibration.
\item Gaussian mixture-model uncertainties for the MegaPrime $-$ PS1 observations. Thus our model naturally handles unrecognized variable stars and occasional problems in the photometry.
\item Separate per-star and per-radius uncertainty floors for the synthetic photometry of the stellar library.
\end{itemize}
We find AB offsets (in the sense of SNLS magnitudes minus AB magnitudes) of \MegaGMinusAB for $g$, \MegaRMinusAB for $r$, \MegaIMinusAB for $i$, and \MegaZMinusAB for $z$ and apply these to the SNLS SNe.

We combine this new calibration path with a new model for the MegaPrime \bandpasses. This model is constructed from a precise measurement of the transmission of its filters performed after their decommissioning on a dedicated measurement bench built by the Laboratoire des Mat\'eriaux Avanc\'es (LMA) in Lyon \citep{Sassolas2018}. The bench determines the transmissions at all positions with a 20~mm resolution and four different incidence angles. Averaging over all incidence angles in the occulted f/4 beam of the CFHT telescope and all positions gives the effective average transmission for all filters, which we multiply by the response curve for the telescope and camera optics and CCDs used in \citet{Betoule2013}. For our light-curve fitting and calibration, we average over the (small) azimuthal variation of each filter but take the radial dependence into account.

\subsubsection{ESSENCE} \label{subsec:ESSENCE}

\citet{narayan16} presented the photometry for the 213 spectroscopically confirmed SNe Ia from the Equation of State: Supernovae trace Cosmic Expansion (ESSENCE) survey. These data are calibrated to \citet{Landolt1992} standards and the light curves are kept in their natural system. The natural-system telescope--detector \bandpasses were established using a tunable-laser flat-field illumination system referenced to a calibrated photodiode \citep{Stubbs2007}. We place this data onto the \LandoltCALSPEC system using the provided linear color-color relations.

In addition, we include the \HST ACS and NICMOS observations of nine ESSENCE SNe from \citet{Krisciunas2005}. These magnitudes are on the CALSPEC system with Vega=0.

\subsubsection{Pan-STARRS} \label{subsec:PS1}

\citet{Jones2018} and \citet{Scolnic2018} presented 365 mid-redshift SNe observed with the Pan-STARRS1 telescope. The natural-system telescope--detector \bandpasses were established using a tunable-laser flat-field illumination system referenced to a calibrated photodiode \citep{Stubbs2010, Tonry2012}. \citet{Jones2022} presented data from RAISIN, some of which overlaps with the SALT3 rest-frame wavelength range; we include these light curves as well.

We calibrate to PS1 3$\pi$ aperture photometry, just as we do with Foundation (described in Section~\ref{subsec:Foundation}). We find magnitude slopes between PS1 aperture and the tertiary stars of $\lesssim 2$~mmag/mag, and find small slopes in color equivalent to bandpass shifts $\lesssim 15$\ang. We thus take 2~mmag and 10\ang as $1\sigma$ systematic uncertainties and show our derived zeropoint offsets in Appendix~\ref{sec:MoreZeropoints}.

\subsubsection{DES} \label{subsec:DES}

\citet{Brout2019Photometry} presented light curves of 251 spectroscopically confirmed SNe Ia from the Dark Energy Survey (DES). We handle DES as two separate surveys: shallow and deep (that share a calibration, but not the per-survey parameters unexplained dispersion or the selection probability with magnitude). The natural-system telescope--detector \bandpasses were established with a monochromator-based flat-field illumination system \citep{Marshall2013}. The data are placed on the AB system using observations of one CALSPEC star (C26202, specifically the spectrum c26202\_stisnic\_007), as described in \citet{Abbott2021}. As noted by the authors (also discussed in Section~\ref{sec:calspecuncertainties}), CALSPEC underwent a large change in both spectral shape and normalization in 2020 \citep{Bohlin2020} so we need to compute updated AB offsets for DES. Comparing synthetic photometry of c26202\_stisnic\_007 and the latest c26202\_stiswfcnic\_005, we find we need to add $-0.0127$ to $g$, $-0.0017$ to $r$, $0.0005$ to $i$, $-0.0012$ to $z$, and $-0.0192$ to $y$ to bring the DES photometry onto an up-to-date CALSPEC.

\subsubsection{HZT} \label{subsec:HZT03}

\citet{tonry03} presented eight high-redshift SNe observed with a mixture of ground-based optical and IR. We assume the magnitude upper limits are fluxes with measurements of 0 $\pm$ the magnitude limit to be conservative. The $VRI$ data are calibrated to Landolt, while the $Z$-band data are on a Vega=0 system. The $VRI$ data are transformed into the standard system, so we use \citet{Bessell2012} \bandpasses, while \citet{tonry03} provides the $Z$-band \bandpass. The ground-based IR data are calibrated to \citet{Persson1998} standards (which are linked to Vega through \citealt{Elias1982} standards) so these data are treated as being on the Vega=0 system.

\subsubsection{GOODS HZT} \label{subsec:HZT07}

\citet{riess07} presented new or updated \HST ACS photometry (many also with NICMOS2) for 40 SNe (including those from \citealt{riess04} and \citealt{Blakeslee2003}). We update the ACS zeropoints to those from \citet{Bohlin2016}, which we convert into Vega=0 zeropoints as described in Appendix~\ref{sec:MoreZeropoints}; this update only has a small effect. We do not update the NICMOS2 zeropoints to match \citet{rubin15a}, as when we spot check the photometry of some of these SNe, we see an overall offset in counts/second units, indicating that the normalization of the \citet{riess07} PSF was different from what was assumed by \citet{rubin15a}. This is likely related to the 0.057 magnitude offset in standard-star magnitudes discussed in \citet{conley11} and simply means that the encircled energy of the PSF was not defined to be 1 at a radius of infinity as it was in \citet{rubin15a}.

\subsubsection{WFPC2 SNe with Ground IR} \label{subsec:SCP01}

\citet{amanullah10} presented six high-redshift SNe discovered from the ground and observed in a mixture of ground-based optical and IR and space-based optical. For all but one (SN~2001cw has a poorly measured color), this gives better distance precision than the SNe observed in purely ground-based optical in the same redshift range from other datasets. The calibration path for the optical ground-based data is heterogeneous, spanning SDSS, Stetson \citep{Stetson2000}, and Landolt. We thus assume these data are on the Landolt system. The ground-based IR data are calibrated to \citet{Persson1998} standards (which are linked to Vega through \citealt{Elias1982} standards), so these data are treated as being on the Vega=0 system (as are the \HST data, which are calibrated through CALSPEC).

\subsubsection{CANDELS/CLASH SNe}  \label{subsec:MCT}

\citet{riess18} presented light curves for 15 SNe~Ia observed with \HST ACS and WFC3 IR as part of the Multi-Cycle Treasury (MCT) programs Cosmic Assembly Near-IR Deep Extragalactic Legacy Survey (CANDELS, \citealt{Grogin2011}) and Cluster Lensing And Supernova survey with Hubble (CLASH, \citealt{Postman2012}). The photometry was placed on the AB system using CALSPEC observations, so it must be updated to match modern CALSPEC. They do not specify which CALSPEC stars were used to compute AB magnitudes, so we use the average of the often-observed GD~71, GD~153, G~191B2B, P330-E, and KF06T2. Comparing late 2014 CALSPEC to 2023,\footnote{Specifically, our old references are gd71\_stisnic\_006, gd153\_stisnic\_006, g191b2b\_stisnic\_006, p330e\_stisnic\_006, and kf06t2\_stisnic\_003. Our new references are gd71\_stiswfcnic\_004, gd153\_stiswfcnic\_004, g191b2b\_stiswfcnic\_004, p330e\_stiswfcnic\_005, and kf06t2\_stiswfcnic\_005.}
 we find we must add the following magnitude offset to each filter: $-0.0048$ to $F125W$, $-0.0050$ to $F127M$, $-0.0048$ to $F139M$, $-0.0050$ to $F140W$, $-0.0048$ to $F153M$, $-0.0051$ to $F160W$, $-0.0093$ to $F606W$, $-0.0071$ to $F775W$, $-0.0066$ to $F814W$, and $-0.0051$ to $F850LP$.

\subsubsection{SCP \HST ACS/NICMOS SNe} \label{subsec:SCPHST}

\citet{suzuki12} presented 20 SNe observed with \HST ACS (many also with \HST NICMOS2 and one with ground-based $H$ from \citealt{Melbourne2007}). In addition, \citet{rubin13} presented ACS and NICMOS2 photometry of SN ``Mingus'', which at redshift 1.713 was the highest-redshift spectroscopically confirmed SN~Ia at the time. This SN had identically performed photometry to the \citet{suzuki12} sample, so we group it in.

We use the NICMOS Camera~2 zeropoints from \citet{rubin15a}: $25.296 \pm 0.022$ ST magnitudes for $F110W$ and $25.803 \pm 0.023$ for $F160W$. These zeropoints will be affected by \CALSPEC changes, so we compute the mean ST zeropoint move in $F110W$ and $F160W$ similarly as for the CANDELS/CLASH SNe. We find we need to add $-0.0071$ to $F110W$ and $-0.0035$ to $F160W$, but this is small compared to the uncertainties and would have to be propagated through updated WFC3 IR calibration for maximum accuracy, so we retain the old zeropoints. We also take the warped \bandpasses for $F110W$ accounting for a loss in sensitivity in the blue at low count rates ($\beta=2$ from \citealt{rubin15a}). 

For the ACS photometry, we use the ACS small-aperture \bandpasses, accounting for the preferential loss of flux at redder wavelengths in $F850LP$ as was done in both \citet{suzuki12} and \citet{rubin13}. We update the zeropoints to those from \citet{Bohlin2016}, which we convert into Vega=0 zeropoints in Appendix~\ref{sec:MoreZeropoints}; this update only has a small effect.

\subsection{Redshift and Host-Galaxy Information} \label{sec:redshiftsource}

To identify host galaxies for nearby ($z \lesssim 0.1$) SNe and retrieve heliocentric redshifts, we made use of SIMBAD \citep{Wenger2000}, HyperLeda \citep{HyperLEDA}, and the Transient Name Server (\url{https://www.wis-tns.org}).\footnote{We used HyperLEDA in particular to find original sources and cite the most common ones here: \citet{Saunders2000, daCosta1998, Doyle2005, Meyer2004, Campbell2014, Smith2000, Abazajian2004, Theureau1998, deVaucouleurs1991, Haynes2011, Aihara2011, Huchra1983, Humason1956, Wegner2003, Haynes2018, Jones2005, Makarov2011, 2007ApJS..172..634A, Jones2009, Springob2005, Alam2015, Falco1999, Courtois2009, Silverman2012, Huchra2012, Lavaux2011}.} We exclude all SNe with SN-derived redshifts (but no host-galaxy redshift) for $z < 0.1$, as SN-derived redshifts are lower precision and can be slightly biased by spectral-feature velocity mismatches compared to host-galaxy redshifts \citep{Zheng2008, Sako2018, Steinhardt2020}.

Observational systematic uncertainties in host-galaxy redshifts are in principle easier to control than systematic uncertainties in photometry, so our discussion is shorter than the rest of this section. See \citet{Davis2019} for a more detailed discussion. For optical/NIR redshifts, one needs a sufficiently high-resolution spectrograph, a light source (e.g., an arc lamp or sky lines) with a sufficient density of isolated emission lines of known wavelength to establish the wavelength calibration, an optical design that sends the calibration light through the spectrograph in a similar way to galaxy light, and an observing plan that interleaves calibration exposures with galaxy exposures (e.g., not taking all the calibrations at the end of the night, when the spectrograph may have cooled or flexed and changed its wavelength calibration). The median uncertainty on heliocentric host-galaxy redshifts is $3\times10^{-5}$ (8~km/s) and the SDSS in particular has validated its precision and repeatability to better than this level \citep{Bolton2012}. We take 8~km/s as a reasonable systematic uncertainty size.

In addition to observational systematic uncertainties, there can be astrophysical bias. \citet{Wojtak2015} find that Milky-Way-like observers typically have gravitational redshifts distributed around zero but with a spread of 6~km/s, biasing the redshift measurements of other galaxies.

Adding the 8~km/s observational systematic and the 6~km/s gravitational systematic in quadrature, we assume all $z < 0.1$ SNe share a correlated 10~km/s ($3.3\times10^{-5}$) redshift systematic uncertainty. We convert all redshift uncertainties to distance-modulus uncertainties with the empty-universe approximation used by \citet{Kessler2009}:

\begin{equation} \label{eq:dmudz}
    \Delta \mu = \Delta z \left( \frac{5}{\ln(10)} \right) \frac{1 + \zCMB}{\zCMB(1 + \zCMB/2)}\;.
\end{equation}
\noindent (Most $1+z$ terms in similar equations are really $1 + \zhelio$; this equation is an exception.) Section~\ref{sec:SmallPerturbations} shows how we include this uncertainty in our cosmology analyses.

\section{Union3 Compilation} \label{sec:UnionCompilation}

\subsection{Light Curve Fitting} \label{sec:lcfit}
We use the SALT3 light-curve model \citep{Kenworthy2021} implemented in \texttt{SNCosmo} \citep{sncosmo} for our results, as it covers a wider range of rest-frame wavelengths than SALT2. (We use the \citealt{Taylor2023} version, which has a bug fix in the calibration.) For each light-curve fit, we perform a nested iteration starting with 1) initializing an estimate for the date of maximum based on a combination of high signal-to-noise and high absolute flux. 2) cutting the data in a range $-15$ to $+45$ rest-frame days around this value. 3) Initializing the SALT model uncertainties to zero. 4) Optimizing the light-curve-fit parameters ${t_0,\ m_B,\ x_1,\ c}$. 5) Updating the SALT model uncertainties. 6) Going back to step 4) and repeating this loop 3--5 times until the fit converges. 7) Going back to step 2) with the new estimate for the date of maximum and looping until the date of maximum converges.

In general, we assume Gaussian uncertainties in flux for our light-curve fits. If a light-curve is given in magnitudes, we convert those magnitudes and their uncertainties to fluxes and flux uncertainties. In some cases (for roughly 2\% of the SNe in the sample), we noted single-date outliers from a light curve at the $\gtrsim 5~\sigma$ level when including SALT3 model uncertainties (these cases are distinct from many-date outliers which are frequently peculiar SNe that SALT3 does not model well). For this small fraction of SNe, we assumed the photometric uncertainties were a two-component Gaussian mixture (with the mixture uncorrelated point-to-point on the light curve). We fixed the mixture parameters to be a normal inlier distribution (normalized to 99\% and with the given uncertainties) and a $5\times$ broader distribution (normalized to 1\% and centered on the SALT3 model). Visual inspection indicates that these robust fits generally did a good job of deweighting suspect points.

As with the Union2 and Union2.1 compilations, we perturb each light-curve zeropoint by 0.01~magnitudes in turn and rerun the light-curve fit to compute \PDnofrac{m_B}{\mathrm{zp}}, \PDnofrac{x_1}{\mathrm{zp}}, and \PDnofrac{c}{\mathrm{zp}} for each zeropoint.
We verified the convergence of the light-curve fits by checking that $\Sigma_\mathrm{zp}$\PDnofrac{m_B}{\mathrm{zp}} is 1 and $\Sigma_\mathrm{zp}$\PDnofrac{x_1}{\mathrm{zp}} = $\Sigma_\mathrm{zp}$\PDnofrac{c}{\mathrm{zp}} = 0. If the derivatives did not sum to within 0.2 of these values, we expanded the derivative computation by using alternative steps of 0.0025 up to 1 magnitude. If these failed as well, we rejected the SN (\NSNefailingLCfits in all were rejected at this stage). Generally, failed convergence indicated poor light-curve coverage or SNe that did not match the SALT model well.

These derivatives are frequently used to propagate calibration uncertainties into the distance-modulus covariance matrix \citep{amanullah10, conley11}. However, our simulated-data testing (Section~\ref{sec:simulateddata}) showed biases with this approach. We determined that the source of this bias was Eddington bias \citep{Eddington1913} in the light-curve fits and their derivatives. As an example, suppose a SN is measured in rest-frame $U$, $B$, $V$, and $R$. Now suppose the SN fluctuates fainter in rest-frame $U$, either because of noise or SN-to-SN variation. This will move the color redder and thus push the SN brighter after standardization. It will also push the derivatives with respect to the rest-frame $U$ closer to zero, as the S/N will be lower. Thus there will be an artificial correlation between the derivative and the Hubble residual, which will bias the \DZP values (Section~\ref{sec:SmallPerturbations}) and thus the cosmological parameters.\footnote{A conventional covariance-matrix approach will not perform better as this approach is very similar to our \DZP approach here (\citealt{amanullah10} Appendix~C).}

The ultimate solution is to bring the light-curve fitting inside UNITY, but for now we must average out the per-SN derivatives to avoid this bias. One possibility is to average the derivatives between different SNe at similar redshifts.\footnote{\citet{betoule14} smoothed their derivatives in redshift to limit noise.} But this would be sub-optimal, especially at low redshift, where SNe are observed with different combinations of surveys. Much of the possibility for self calibration would thus be lost with such smoothing. So instead we smooth over per-SN variation. Before computing derivatives, we replace the fluxes for each SN with a model that has $M_B=-19.1$, $x_1=0$, $c=0$, then compute derivatives using these model fluxes. These are the derivatives used in Section~\ref{sec:SmallPerturbations} to propagate calibration uncertainties.

Finally, we also replace the light-curve fluxes for each SN (but as before, do not replace the light-curve flux uncertainties) with those for models with several values of $M_B$ and $c$. For each value, we refit the light curve and compute the light-curve-fit uncertainties. These light-curve-fit uncertainties are then interpolated in the predictive-posterior tests of UNITY (Section~\ref{sec:PPD}) to ensure that the uncertainties vary with SN magnitudes in a realistic way.

\subsection{SALT Validation and Quality Cuts}\label{sec:ValidQuality}

There are two redshift-dependent effects that affect light-curve fits in a single survey: 1) Higher-redshift SNe are fainter and thus tend to be observed at lower S/N. 2) Higher-redshift SNe are almost always observed at the same set of survey-specific observer-frame wavelengths and thus are observed bluer in the rest frame. Moreover, it remains the case that most of the lower-redshift SNe included here come from targeted observations and not rolling surveys. Thus, lower-redshift SNe tend to have light curves that start later. We thus need to make sure that our sample selection and light-curve fitting do not introduce biases as a function of all three of these parameters.

Our primary tool to ensure good performance as a function of redshift is to select well observed SNe and degrade their light curves to see when biases appear. These light-curve-fitter tests have a long history, e.g., \citet{Guy2005} compared the performance of the original SALT when measuring SNe in the rest-frame $U$ and $B$ compared to $B$ and $V$ to evaluate consistency as a function of redshift. \citet{kowalski08} took smoothly interpolated light curves (from \citealt{Strovink2007}) as the truth, creating simulated light curves with varied S/N and first phase of observations to determine what light-curve cuts to apply. We repeat similar tests here for SALT3.

\subsubsection{Phase Cuts}

For the phase coverage test, we select SNe with a light curve in at least one filter meeting the following criteria:
\begin{itemize}
    \item A point on the light curve at least ten rest-frame days before maximum with uncertainty $<~0.05$~mag
    \item A point on the light curve in the same filter at least 25 days after maximum with uncertainty $<~0.05$~mag
    \item At least ten points in the same filter with uncertainty $<~0.05$~mag
\end{itemize}
In addition, we require:
\begin{itemize}
    \item A light-curve shape parameter $|x_1|~<~3$
    \item A color parameter $|c|~<~0.3$.
\end{itemize}
For the 98 SNe meeting these criteria, we create copies with restrictions in phase coverage and see if we would recover the same light-curve parameters from these phase-restricted light curves. We restrict the copies' first phase and last phase in a grid from $-10$ to +10 days in first phase and 0 to +20 days last phase (with each grid point having all 98 SNe). After light-curve fitting, we consider distance-modulus changes from the original light-curve parameters to the new parameters after excluding light-curve fits failing the derivative test described in Section~\ref{sec:lcfit}. Encouragingly, many bad light-curve fits do fail this derivative test and need no other cuts to exclude them from the sample. Examining the clipped mean and NMAD (robust dispersion) as a function of grid position, and requiring the mean distance-modulus change to be $\lesssim 0.03$ magnitudes and the NMAD to be $\lesssim 0.1$ magnitudes, we settle on the following phase-coverage criteria:
\begin{itemize} 
    \item First phase $\leq +2$, last phase $\geq +8$, and last phase $-$ first phase $\geq 10$
    \item Or alternatively, first phase $\leq +6$, last phase $\geq +17$, and last phase $-$ first phase $\geq~15$.
\end{itemize}
The second criterion enables SNe with good phase coverage starting moderately after maximum to still be included.

\subsubsection{SALT3 Performance as a Function of Rest-Frame Wavelength} \label{sec:SALTRestFrame}

\begin{figure*}
    \centering
    \includegraphics[width=0.99 \textwidth]{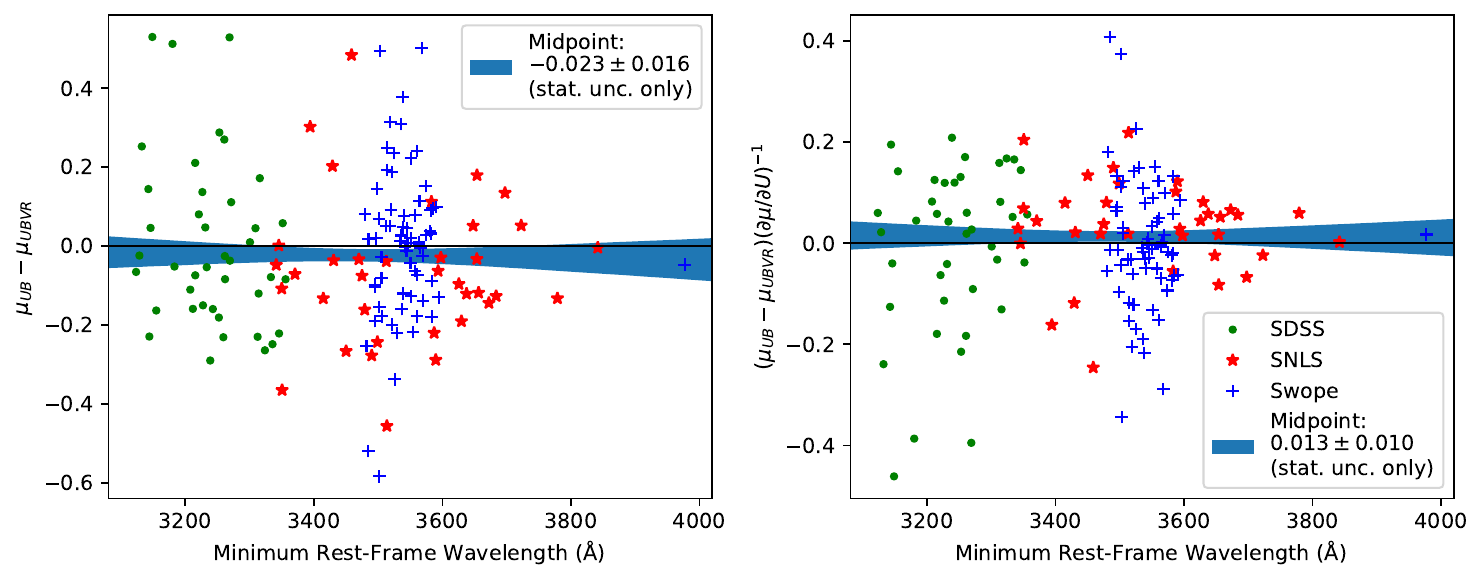}
    \caption{The results of a SALT3 validation test to see if different rest-frame wavelength ranges give consistent distance moduli. The {\bf left panel} shows the distance-modulus difference for each SDSS, SNLS, or Swope SN between rest-frame $UB$ and $UBVR$ as a function of the minimum rest-frame wavelength. The {\bf right panel} shows the same quantity divided by the sensitivity of the distance modulus to the rest-frame $U$. Neither case shows a significant trend. The midpoint values shown are the intercept values at the location where the slope with wavelength and the intercept are uncorrelated. These values should be similar to the robust mean over all data.\label{fig:UBUBVR}}
\end{figure*}

For the rest-frame wavelength test, we investigate the consistency of SALT3 distance moduli when using rest-frame $UB$ compared to $UBVR$ (we define rest-frame $U$ as 3000--4000\ang effective wavelength, $B$ as 4000--5000\ang, $V$ is 5000--6000\ang, and $R$ as 6000--7000\ang). We use three datasets which cover this wavelength range: SNLS (where $g$ provides rest-frame $U$ band coverage), SDSS, and CSP1. All three of these datasets have reasonably well-calibrated rest-frame $U$-band data (with CSP1 and SDSS agreeing to $\lesssim 0.02$ magnitudes, \citealt{Mosher2012}). We select SNe with aggregate S/N $>$ 15 in each rest-frame band and first phase $<$ 0, then simulate high redshift observations by including only the rest-frame $UB$. The left panel of Figure~\ref{fig:UBUBVR} shows the distance modulus difference between the $UB$ light-curve fits and the $UBVR$. We plot this difference against the minimum rest-frame wavelength covered in order to look for possible trends, but do not see either a slope or an offset (using a robust Student-t linear regression). The right panel of Figure~\ref{fig:UBUBVR} shows a similar quantity divided by the change in distance modulus with the bluest rest-frame band: $\partial \mu/\partial U$. This enables us to interpret the distance-modulus differences roughly in terms of the actual relative calibration of the rest-frame $U$ band with respect to $BVR$. Again, we see no evidence of slope or offset. We conclude that the \citet{Taylor2023} SALT3 rest-frame $U$-band is calibrated to $\sim 0.01$ magnitudes with respect to $BVR$, which we take as a systematic uncertainty (described in Section~\ref{sec:SALTCalibration}). Although we retrain SALT3 as a way to quantify training calibration uncertainties (also described in Section~\ref{sec:SALTCalibration}), we do not use our retrained SALT3 for cosmology fitting, as Figure~\ref{fig:UBUBVR} shows that our calibration has good consistency with the \citet{Taylor2023} calibration and this choice also maintains consistency with previous results.

\subsubsection{Final Quality Cuts for Union3}

Table~\ref{tab:NumberOfSNe} presents the number of SNe passing each selection cut. In addition to the cuts previously described, we require:
\begin{itemize}
    \item At least five light-curve points in at least two filters
    \item A cut on light-curve shape $|x_1| + \sigma_{x_1} < 5$, which helps remove peculiar SNe, non~Ia contamination, and failed light-curve fits
    \item A color cut $|c| < 0.3$, which helps remove peculiar SNe, non~Ia contamination, and highly extincted SNe where even small $R_V$ differences would add significant dispersion
    \item A color uncertainty cut $\sigma_c < 0.2$, which removes SNe which would have very little weight in the analysis
    \item Any highly peculiar SNe already noted by the original authors. As we want to apply cuts uniformly in redshift, we do not apply any cuts that would be difficult to match at high redshift, so we do not exclude 1991T-like SNe, but do exclude 1991bg-like, 2002cx-like, and ``super-Chandrasekhar'' SNe.
\end{itemize}
In the end, we have \totalNSNe SNe in the Union3 compilation.

\begin{deluxetable*}{llll}
 \tablehead{
 \colhead{Selection Cut} & \colhead{SNe Remaining} & \colhead{$z \leq 0.1$} & \colhead{$z > 0.1$}}
 \startdata
All SNe & 2711 (100.0 \%) & 1151 (100.0 \%) & 1560 (100.0 \%) \\ 
Enough Data & 2688 (99.2 \%) & 1132 (98.3 \%) & 1556 (99.7 \%) \\ 
Host-Galaxy Redshift (or $z>0.1$) & 2648 (97.7 \%) & 1092 (94.9 \%) & 1556 (99.7 \%) \\ 
CMB-Centric Redshift $> 0.01$ & 2537 (93.6 \%) & 981 (85.2 \%) & 1556 (99.7 \%) \\ 
MW $E(B-V) < 0.3$ & 2517 (92.8 \%) & 961 (83.5 \%) & 1556 (99.7 \%) \\ 
Converged LC Fits & 2416 (89.1 \%) & 896 (77.8 \%) & 1520 (97.4 \%) \\ 
Phase Range & 2206 (81.4 \%) & 785 (68.2 \%) & 1421 (91.1 \%) \\ 
$x_1$ Range & 2191 (80.8 \%) & 783 (68.0 \%) & 1408 (90.3 \%) \\ 
$c$ and $\sigma_c$ Limits & 2096 (77.3 \%) & 716 (62.2 \%) & 1380 (88.5 \%) \\ 
Remove Known Peculiar SNe & 2087 (77.0 \%) & 707 (61.4 \%) & 1380 (88.5 \%) \\ 
 \enddata
 \caption{Number of SNe passing cuts. \label{tab:NumberOfSNe}}
\end{deluxetable*}

\subsection{Inter-Survey Calibration Crosschecks} \label{sec:DeltaZPNearby}

Many nearby supernova datasets have incompletely quantified zeropoint uncertainties. Fortunately, there is now enough literature data to compare distances for the same objects using different instruments (an expanded version of the comparisons done by, e.g., \citealt{Stritzinger2002} or \citealt{hicken09}). For this purpose, we examine derived SALT3 distance-moduli differences based on groups/instruments that have observed at least one supernova in common (yielding \NCrossCheckSNe~SNe observed in at least two bands per group/instrument). To prevent data cuts from bringing any supernova into or out of the sample, we fix the $x_1$ and date of maximum in the individual-group light-curve fits from a joint fit of all of the data. We thus fit for magnitudes and colors using the data from each instrument. These give us a $\mu_{ij}$ value for each SN $i$, for each instrument $j$. We model these $\mu_{ij}$ values with a simple robust linear model:

\begin{equation}
        \mu_{ij} \sim \mathrm{Student\ t}(\mu_i + \Delta \mu_j,\ \sigma_j,\ 7)
\end{equation}
where $\mu_i$ is the inferred ``true'' distance modulus for SN $i$, $\Delta \mu_j$ is the calibration offset for instrument $j$ (we arbitrarily fix one instrument to have zero mean offset), $\sigma_j$ is the dispersion in calibration for instrument $j$, and we assume a Student t distribution with 7 degrees of freedom for robust inference. This model has hundreds of parameters ($2\, N_{\mathrm{instr}} - 1 + N_{\mathrm{SNe}}$), so we use the Hamiltonian Monte Carlo code Stan \citep{Carpenter2017} to efficiently sample the posterior (implemented with PyStan, \citealt{allen_riddell_2018_1456206}).

\newcommand{\ZeroDeltaMuSentence}{We arbitrarily pick CfA3/4 to have zero $\Delta \mu$ as its $\Delta \mu$ value has the smallest uncertainty; this choice does not affect the dispersions.\xspace}

\begin{figure*}
    \centering
    \includegraphics[width = 0.7 \textwidth]{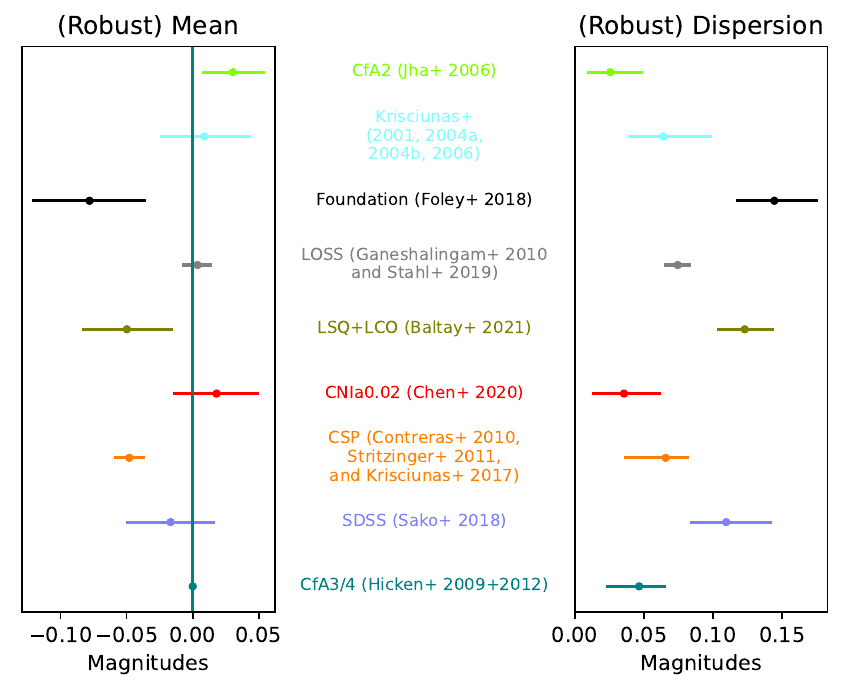}
    \caption{Robustly inferred instrumental distance-modulus offset ({\bf left panel}) and dispersion ({\bf right panel}) using nearby SNe observed by at least two instruments in at least two filters each. We show the posterior median and the 68\% credible interval for each instrument. \ZeroDeltaMuSentence Only statistical uncertainties are shown. In general, these offsets are consistent with our assigned zeropoint and \bandpass uncertainties. \label{fig:DeltaZP}}
\end{figure*}

Figure~\ref{fig:DeltaZP} shows the posteriors for each instrument: the $\Delta \mu_j$ in the left panel and the $\sigma_j$ in the right panel. We show the median of the posterior with a dot, and the 68.3\% credible interval with the error bar. \ZeroDeltaMuSentence We see that several of the nearby SNe datasets (CfA2, CfA3, CfA4, and LOSS) are self-consistent but offset from Swope (which is surprising, as Swope has measured \bandpasses and high-quality light curves). In general, these offsets are consistent with our assigned zeropoint and \bandpass uncertainties. We also see that Foundation, LCO, and SDSS have the largest dispersion (in other words, are the least consistent with the other observations for SNe in common). We leave further investigation to future work.

\section{UNITY} \label{sec:UNITY}

\begin{deluxetable*}{lrp{4.5cm}rp{4.5 cm}}
 \tablehead{
 \colhead{Version} & \colhead{Reference}  & \colhead{$x_1$, $c$ Population Distributions} & \colhead{Selection Effects} & \colhead{Purpose}  }
 \startdata
 UNITY1 & \citet{rubin15b} & Skew-Normal, Normal & Yes, assumed fixed & Unified Cosmological Analysis \\
 UNITY1.1 & \citet{hayden19} & Skew-Normal, Normal & No & Rise-Time Standardization \\
 UNITY1.2 & \citet{Rose2020} & Normal and Normal Mixtures & No & Standardization with Many Correlated Parameters \\
 UNITY1.5 & This Work & Exponential Convolved Normal (Approximated) & Yes, Marginalized & Improved Unified Cosmological Analysis \\
 \enddata
 \caption{Variants of UNITY analyses. In this context, ``selection effects'' refers to the cosmological bias remaining {\it after standardization} due to a survey selecting more luminous SNe conditional on light-curve parameters. This bias also modifies the $x_1$ and $c$ distributions, and UNITY1 and UNITY1.5 can infer estimates of the true distributions prior to selection.\label{tab:unityversions}}
\end{deluxetable*}

UNITY is a family of Bayesian Hierarchical Models used for SN standardization and cosmology. Table~\ref{tab:unityversions} presents analyses that have used a version of UNITY, with a summary of how each version has been different.

\begin{figure*}
    \centering
    \includegraphics[width=0.7 \textwidth]{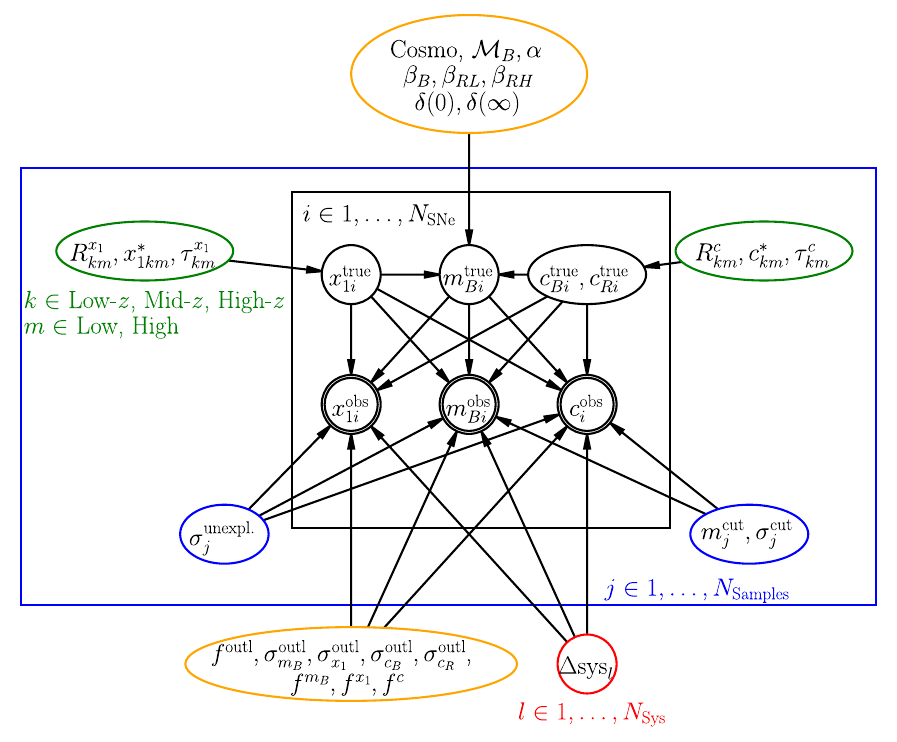}
    \caption{Probabilistic Graphical Model for UNITY1.5. We note that quantities assumed fixed (redshift and \Phigh) are not shown.
    \label{fig:PGM}}
\end{figure*}

The underlying philosophy of UNITY is to build a generative model of the \mBxc values from light-curve fits. If the model accurately includes all the effects in those data, then such a generative model optimally extracts the cosmological information. We have compiled from the literature a range of astrophysical and observational models for which there is some empirical evidence. Figure~\ref{fig:PGM} shows the UNITY1.5 probabilistic graphical model and Appendix~\ref{sec:Priors} lists its parameters with their priors, as motivated in the following subsections. For effects where we are not sure which model is correct (for example, whether the host-galaxy/luminosity correlation should change with redshift due to age effects as discussed by \citealt{Rigault2013, Childress2014}), we allow both models, and let the data choose which it prefers. This approach can give larger cosmological uncertainties than making an explicit choice, but we view it as less likely to be biased.

One example of this philosophy is that we do not make a hard distinction between statistical and systematic uncertainties. For example, a frequentist analysis has to make an explicit correction for selection effects (i.e., Malmquist bias, \citealt{Malmquist1922}), and the size of this correction for SNe~Ia depends on
\begin{equation}
    \mathrm{Malmquist\ bias} \propto \frac{(\mathrm{corrected\ dispersion})^2}{\mathrm{uncorrected\ dispersion}} \;.
\end{equation}
As a frequentist analysis has no reliable way to simultaneously estimate all the parameters that enter this equation, one has to estimate them separately, computing the bias correction and its uncertainties \citep[e.g.,][]{Perrett2010}, and then putting the uncertainties on the bias correction into the covariance matrix by hand (i.e., add explicit systematic-uncertainty covariance terms). UNITY is different, as it can simultaneously infer these parameters, so the size of the selection uncertainty will naturally decrease for larger samples of SNe. (Of course, the functional form of the posterior must be accurately specified to get the right answer, and we discuss UNITY1.5 validation in Section~\ref{sec:Validation}.)

UNITY1.5 has thousands of fit parameters (as did the original UNITY), so we again use the Hamiltonian Monte Carlo code Stan \citep{Carpenter2017, allen_riddell_2018_1456206} to efficiently sample the posterior. For most of the inference in this work, we use four hundred chains with 2500 samples per chain (the first half of which is discarded, so 500,000 samples). Convergence is usually excellent, with \citet{GelmanRubin92} $\hat{R} \lesssim 1.01$ for most parameters. In the event of bad sampling ($\hat{R} > 1.05$), we discard the affected chains and rerun.

\subsection{SN Standardization}

To summarize, UNITY1.5 uses a modified version of the \citet{Tripp1998} standardization equation:
\begin{eqnarray}
    m^{\mathrm{model}}_B & = & -\alpha\, x^{\mathrm{true}}_1 + \beta_B\, c^{\mathrm{true}}_B \nonumber \\
    & & + [\betaRL \, (1 - \Phigheff) +  \betaRH \, \Phigheff] c^{\mathrm{true}}_R  \nonumber \\
    &  & - \delta(0) \, \Phigheff \nonumber \\ 
                     & & + M_B + \mu(z,\ \mathrm{cosmology}) \;.  \label{eq:tripp}
\end{eqnarray}
$M_B$ is the rest-frame $B$-band absolute magnitude, which we combine with the $H_0$ in the distance modulus ($\mu$) following \citet{perlmutter97} as described in Appendix~\ref{sec:scriptM}. $\alpha$ is the $x_1$ standardization coefficient, $\beta_B$ is the color standardization coefficient associated with the Gaussian core of the color distribution, \betaRL and \betaRH are the color standardization coefficient of the red exponential tail in low-mass and high-mass hosts respectively, and $\delta(0)$ is the host-mass standardization coefficient at low redshift. \Phigheff is the probability of a host-galaxy behaving like a low-redshift host galaxy having high stellar mass (Equation~\ref{eq:Phigheff}). As described in the Introduction, the $\alpha$ and $\beta$ values are not multiplied against observed $x_1$ and $c$ values, but instead are multiplied against per-SN parameters \xonetrue, \cBtrue, and \cRtrue. These parameters are marginalized over for every SN in the analysis.

\subsubsection{Light-Curve Color Standardization}

As SN samples got larger, evidence built that a single linear color standardization was not adequate to describe the data, even for moderate values of reddening ($E(B-V) \sim 0.1$--0.2). A naive test is to separate colors into those bluer and redder than a given cut value and fit the color standardization coefficient on each half. Unfortunately, this test is foiled by Eddington bias \citep{Eddington1913}. A somewhat underluminous SN measured to be just bluer than the cut may actually be more likely to be a reddened SN whose color measurement just happened to have enough error to place it bluer than the cut. (Likewise for overluminous SNe just redder than the cut.) This Eddington bias causes large biases in both the slopes and intercepts on either side of the cut.

\citet{amanullah10} approached this problem with a frequentist model. They projected each SN onto its best-fit point along the global magnitude vs. color relation (taking into account uncertainties in both axes) and then divided the SNe on these projected values. They found dramatically different color standardization coefficients ($\beta$) among blue SNe ($1.1 \pm 0.3$) and red SNe ($3.0 \pm 0.1$).

In response, UNITY1 used two magnitude/color relations: a broken-linear relation for standardizing measured color, and an intrinsic covariance between magnitude and color in the unexplained dispersion. We simplify here, making the two color relations both explicit. UNITY1.5 now deconvolves the $c$ distribution into a Gaussian distribution and an exponential distribution \citep[e.g.,][]{jha06, Mandel2017}. This is very similar in detail to a broken-linear relation (\citealt{Mandel2017}, Figure~4) although the broken-linear relation is possibly less physically motivated. The deconvolution is modeled by having two latent colors for each SN: \cBtrue which is assumed to be drawn from a Gaussian distribution and \cRtrue which is assumed to be drawn from an exponential distribution. The sum of both latent colors predicts the observed color.

\subsubsection{Light-Curve Shape Standardization}

Similarly to color, UNITY1 used two magnitude/shape relations: a broken-linear, and intrinsic magnitude/shape correlation in the unexplained dispersion. Again, we simplify and use a simple linear relation for now similarly to other cosmology analyses. However, we note the relations between, e.g., $x_1$ and $\Delta m_{15}$ \citep{Phillips1993} or stretch \citep{perlmutter97} are nonlinear (\citealt{Guy2007} presents the nonlinear transformation equations between these quantities), a linear standardization relation in $x_1$ cannot be linear in the others.

\subsubsection{Host-Galaxy Correlations} \label{sec:hostgalaxycorrelations}

There is strong evidence that SALT SN~Ia distances are affected by host-galaxy properties, implying that SALT does not standardize all the astrophysical behavior of SNe~Ia (e.g., progenitor metallicity or carbon/oxygen ratio,  \citealt{Hoflich2010}) and/or the host-galaxy environment (e.g., host-galaxy extinction). Most of the first studies correlating SN standardized absolute magnitudes with their hosts focused on host-galaxy morphology \citep{Sullivan2003, hicken09} and host-galaxy stellar mass \citep{kelly10, sullivan10, Lampeitl2010}, as these were the most accurate quantities to measure for high-redshift galaxies (though this does not necessarily indicate accuracy in capturing the relevant astrophysics, \citealt{Boone2021B, Briday2022}). The \citet{conley11}, \citet{Sullivan2011}, and \citet{suzuki12} cosmology analyses used a SN luminosity ``step'' in stellar mass, allowing different standardized absolute magnitudes above and below a cutoff stellar mass.

However, \citet{Rigault2013} presented evidence that the local specific star-formation rate drives the host-mass correlation, and that the underlying variable is more strongly correlated with age (\citealt{Neill2009} also suggested using age for standardization based on a set of SN hosts observed with GALEX and \citealt{Childress2013} suggested age based on the cutoff stellar mass). If this is the case, then the host-mass step should decrease with increasing redshift, as an increasing number of systems should be young at high redshift, irrespective of host-galaxy stellar mass. \citet{Rigault2013} presented a fitting formula for this decrease (\citealt{Childress2014} presented a similar formula). Those authors' proposed host-mass-standardization evolution predicts that the mass-standardization coefficient will approach zero at high redshift; UNITY1 instead assumed the coefficient smoothly approaches a possibly non-zero quantity, $\delta(\infty)$ starting from the low-redshift value $\delta(0)$. UNITY1 took a flat prior on $\delta(\infty)/\delta(0)$ from 0 to 1, allowing the mass standardization to be constant or declining with redshift, spanning all of the claims in the literature \citep{rubin15b}.

\newcommand{\mstar}[1]{\ensuremath{m^{\mathrm{#1}}_{\star}}\xspace}

Following \citet{suzuki12} Equation~2, we write the probability of a host galaxy being high stellar mass as
\begin{eqnarray} \label{eq:Phigh}
    \Phigh \equiv P(\mstar{true} > \mstar{threshold}|\mstar{obs}) = \\
\int_{\mstar{threshold}}^{\infty} P(\mstar{obs}|\mstar{true})P(\mstar{true}) d\mstar{true}\nonumber \;,
\end{eqnarray}
where the prior $P(\mstar{true})$ is a fixed, sample-dependent distribution based on whether the bulk of the sample was massive, targeted galaxies or untargeted field galaxies. As in \citet{suzuki12} and \citet{rubin15b}, we assume that \Phigh is a fixed quantity and do not marginalize over \mstar{true} in UNITY, which we essentially get away with because host stellar mass has a much smaller impact on luminosity than $x_1$ or $c$. The UNITY1 treatment of $\delta(z)$ is equivalent to replacing \Phigh with \Phigheff, a redshift- and host-mass-dependent probability of a host galaxy behaving like a low-redshift host galaxy having high stellar mass:

\begin{equation}
    \Phigheff \equiv \Phigh \left[ \frac{1.9}{0.9 \cdot 10^{0.95 \,z} + 1} \left[ 1 - \frac{\delta(\infty)}{\delta(0)}\right]  + \frac{\delta(\infty)}{\delta(0)} \right] \;. \label{eq:Phigheff}
\end{equation}

In addition, increasing evidence has pointed to different color-magnitude relations \citep{sullivan10, Sullivan2011, Childress2013, Mandel2017, Brout2021} as a function of host-galaxy mass, possibly related to different dust extinction $R_V$. So we again update our model to accommodate this possibility as well. UNITY1.5 now allows for a different color standardization slope $\beta_R$ (for the red tail described by $c^{\mathrm{true}}_R$) according to \Phigheff:

\begin{equation}
    \beta_R \equiv \betaRL \, (1 - \Phigheff) +  \betaRH \, \Phigheff \;.
\end{equation}
We do see evidence of two different $\beta_R$ values, discussed more in Section~\ref{sec:uncertaintyanalysis}.

\subsection{Model for Light-Curve Shape and Color ($x_1$ and $c$) Populations} \label{sec:ShapeAndColorPopulation}

As discussed in Section~\ref{sec:Introduction}, informative priors are necessary for the true light-curve shape ($x_1$) and color ($c$) parameters for each SN. \citet{Gull1989} suggests Gaussian priors are reasonable choices for linear regression (even if the underlying distributions are not Gaussian). However, UNITY uses nonlinear standardization, and accurately modeling the populations of $x_1$ and $c$ is necessary for modeling selection effects (discussed in Section~\ref{sec:MalmquistBias}), so more accurate distributions may be helpful (e.g., by getting higher moments of the distributions correct). UNITY1 used normal distributions for $x_1$ and skew-normal distributions for $c$ (as did the related \Steve model, \citealt{Hinton2019}). UNITY1.5 uses the convolution of an exponential distribution and a normal distribution for both $x_1$ and $c$. Integrals of this new model can be computed accurately in the presence of selection effects (see Section~\ref{sec:MalmquistBias} and Appendix~\ref{sec:colorpop} for details). These integrals were only approximated in UNITY1 \citep{rubin15b} and \Steve \citep{Hinton2019}.

UNITY1 and \Steve both used splines as a function of redshift and SN sample to describe how the $x_1$ and $c$ populations vary with redshift. UNITY1 assumed the selection efficiencies of each survey were fixed, so any redshift variation in the observed $x_1$ and $c$ distributions of selected SNe incompatible with the assumed selection efficiency could be described by the splines. Here, we are marginalizing the survey selection depths, so we make two simplifying assumptions to give reliable inference. 1) the $x_1$ and $c$ populations within each survey do not vary with redshift. Thus, any redshift variation in the observed $x_1$ and $c$ distributions inside a survey can be directly interpreted as the result of selection. 2) We sort each survey into low-, mid-, and high-redshift and assume each survey in each redshift range has the same population parameters (before selection). For correctness, we also allow the $x_1$ and $c$ population distributions to be different according to \Phigh in addition to redshift category, as light-curve shape in particular is sensitive to host-galaxy properties \citep{Pskovskii1967, Hamuy1995, Branch1996, Hamuy1996}, see Appendix~\ref{sec:Priors}. In principle, we could allow the $x_1$ and $c$ population distributions to vary as \Phigheff rather than \Phigh, but the low-, mid-, and high-redshift categories already capture any variation with redshift.

\begin{figure*}[h!tbp]
    \centering
    \includegraphics[width=0.75\textwidth]{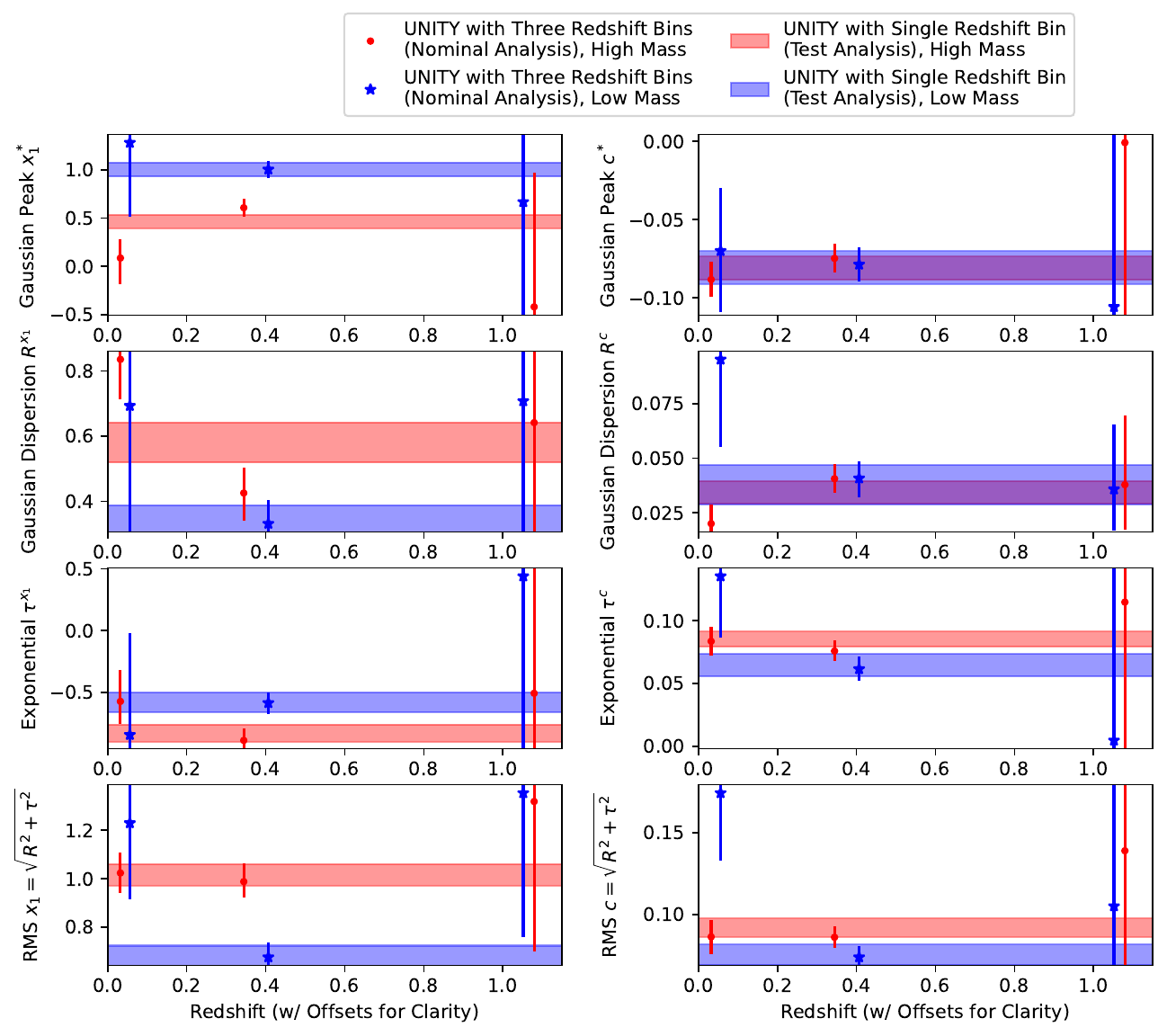}
    \caption{68\% credible intervals for the parameters of two $x_1$ and $c$ population models: our nominal three-redshift-bin analysis and a redshift-independent comparison. The {\bf left panels} show $x_1$ and the {\bf right panels} show $c$. {\bf Top to bottom}, the rows are the peak of the Gaussian component, the width of the Gaussian component, the length scale of the exponential component, and the RMS of the population. The red dots or shaded regions show the model for high-stellar-mass hosts and the blue stars or shaded regions show the model for low-stellar-mass hosts. Only moderate evidence for redshift evolution is seen and the cosmological results between the two models are essentially the same.
    \label{fig:PopVsZ}}
\end{figure*}

As a crosscheck, we also try running UNITY1.5 with redshift-independent population distributions. For cosmological models, we consider both flat \LCDM (so a single cosmological parameter) and flat $w$CDM (so two cosmological parameters). We find the same posterior medians as the low/mid/high-$z$ population model for both cosmological models. Figure~\ref{fig:PopVsZ} shows both the nominal three-redshift-bin population model and this constant-in-redshift test. This crosscheck provides the complete answer to the population question investigated in \citet{rubin16}: it is adequate to assume population parameters that are redshift-independent (before selection effects), as long as one also models and marginalizes over the selection effects of each sample. It is not clear that this crosscheck will continue to pass with larger, next-generation SN surveys, so we do recommend continuing to assume redshift-dependent populations in the future.

\subsection{Unexplained Dispersion} \label{sec:unexplaineddispersion}

Unfortunately, even after standardization, SNe~Ia are not perfect standard candles. The unexplained dispersion must be included in the SN-distance uncertainties to give accurate cosmological-parameter uncertainties from the final set.\footnote{This is among many other problems that inaccurate uncertainties cause, like biased standardization coefficients and an inaccurate Malmquist-bias correction.} In a Bayesian Hierarchical Model like UNITY1.5, including part of the uncertainties as parameters is straightforward.

We follow Union \citep{kowalski08} and allow each dataset to have its own unexplained dispersion. We also marginalize over the fraction of unexplained variance in $m_B$, $x_1$, and $c$ (these are global parameters), using a unit simplex $\{f^{m_B},\ f^{x_1},\ f^c\}$ for these fractions to force them to sum to 1. If the unexplained dispersion has a similar size in magnitudes in $m_B$, $x_1$, and $c$, then the proper procedure is to scale these values by $\{1,\ \alpha^{-2},\ \beta^{-2}\}$ to obtain the unexplained variance in \mBxc. But if we did that, then the uncertainty propagation into distance-modulus uncertainty (multiplying the variance in \mBxc by $\{1,\ \alpha^{2},\ \beta^{2}\}$) would cancel $\alpha$ and $\beta$, leaving all the unexplained dispersion in $m_B$. So we follow UNITY1 and scale these values by nominal coefficients $\{1,\ 0.14^{-2},\ 3^{-2}\}$ instead.\footnote{We simplify from UNITY1 by only allowing unexplained dispersion on the diagonal of the covariance matrix. We choose to include off-diagonal terms with per-SN nuisance parameters that are marginalized over. \citet{amanullah10} Appendix C discusses the correspondence between covariances and nuisance parameters that enter the model linearly and have Gaussian priors. For example, intrinsic scatter between $m_B$ and $c$ can be represented with the \cBtrue parameters. This choice makes the results easier to interpret.} In short, the \mBxc unexplained dispersion for dataset $j$ is

\newcommand{\sigunexpl}{\ensuremath{\sigma^{\mathrm{unexpl.}}}\xspace}

\begin{equation}
    \left\{ \sigunexpl_j \sqrt{f^{m_B}},\ \sigunexpl_j \frac{\sqrt{f^{x_1}}}{0.14},\ \sigunexpl_j \frac{\sqrt{f^{c}}}{3} \right\} \;.
\end{equation}

\newcommand{\diagnosticssentence}{In general, all the samples are consistent with the same absolute magnitude (any trend in redshift would be nearly degenerate with \Om). However, some datasets have noticeably larger unexplained dispersions than others.\xspace}

\begin{figure*}[h!tbp]
    \centering
    \includegraphics[width=0.95\textwidth]{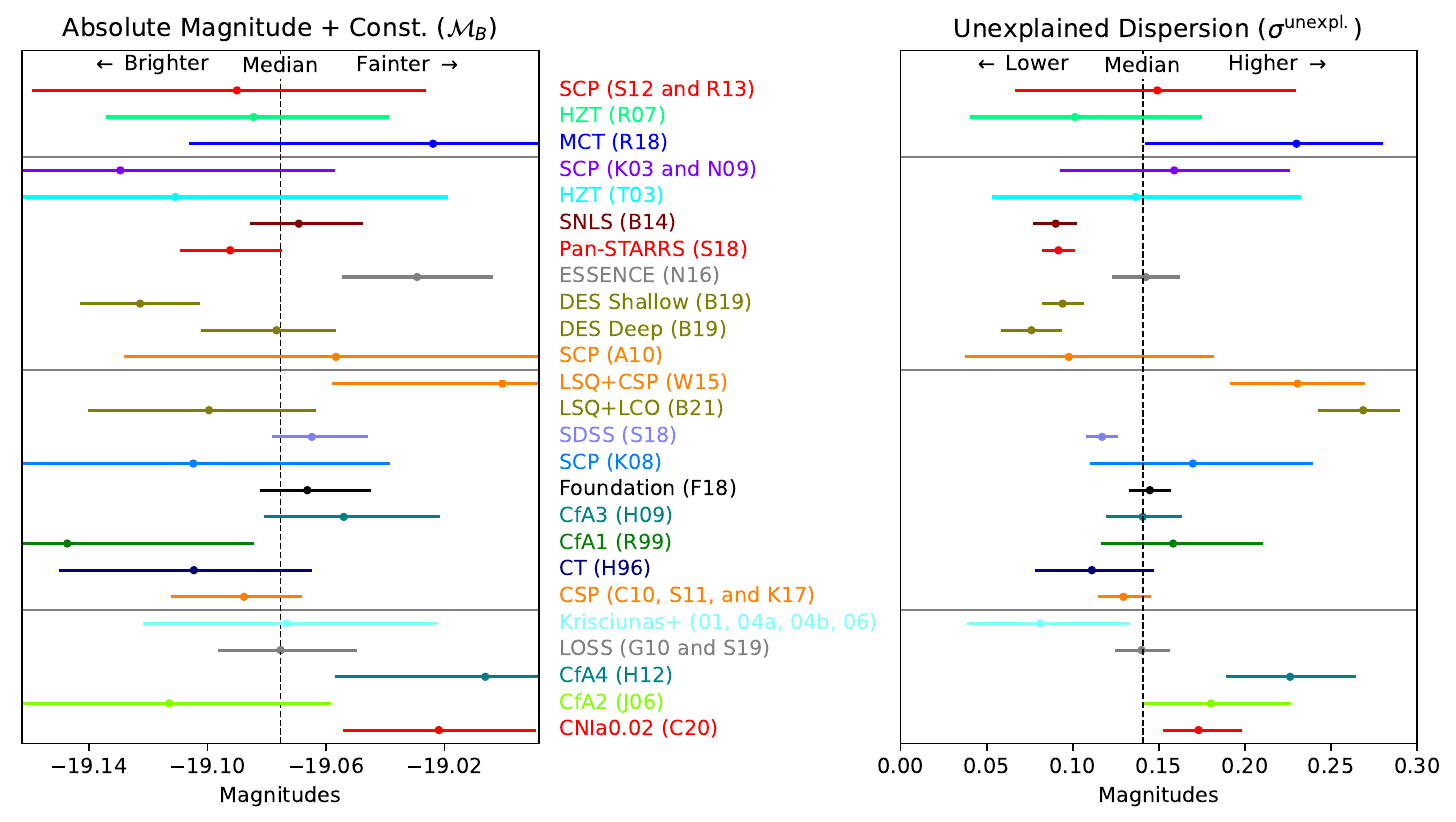}
    \caption{Diagnostics plot showing the 68\% credible intervals for \scriptM ({\bf left panel}) and unexplained dispersion ({\bf right panel}) for a UNITY model with one absolute magnitude per sample (and fixed \Om). The unexplained dispersion computed by UNITY does not include the model uncertainties added by SALT3, which are treated as measurement uncertainties (these amount to several hundredths of a magnitude when converted into distance-modulus uncertainties). We show the median value for each with a dashed vertical line. \diagnosticssentence}
    \label{fig:Diagnostics}
\end{figure*}

As a crosscheck, Figure~\ref{fig:Diagnostics} shows 68\% credible intervals on \scriptM (defined in Equation~\ref{eq:scriptM}) and unexplained dispersion for a UNITY variant where \scriptM is allowed to vary dataset-to-dataset (and \Om is fixed to the best-fit value of 0.36). As there is covariance between the \scriptM values, we force the median \scriptM for each posterior sample to be equal to the median posterior \scriptM over all samples. This removes the correlated additive uncertainty, shrinking the uncertainties and making for a more sensitive test. \diagnosticssentence This figure suggests that it is worth reexamining the photometry of at least some SNe with large unexplained dispersions (going back to the images of SNe and standard stars) to see if there are any problems, but we leave this for future work as it is only a small fraction of the data and the images are not public in general.

\begin{figure*}[h!tbp]
    \centering
    \includegraphics[width = 0.95\textwidth]{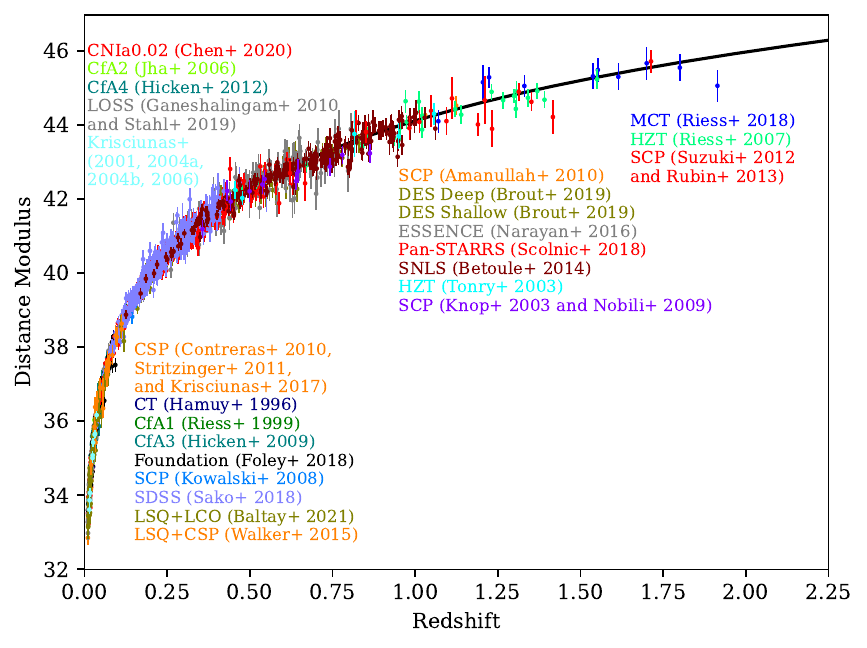}
    \caption{A simple frequentist Hubble diagram for Union3 as a sanity check. \label{fig:HubbleDiagram}}
\end{figure*}

Figure~\ref{fig:HubbleDiagram} shows a frequentist Hubble diagram as a sanity check of our results. For this, we simply use $\beta_B$ for SNe with observed color $< 0$ and $\beta_R$ for SNe with observed color $> 0$ and construct distance-modulus estimates for each SN.

\subsection{Small SN-Parameter Perturbations} \label{sec:SmallPerturbations}

\newcommand{\selfcalibration}{The reduction in scale is due to UNITY inferring the values of the systematics nuisance parameters using the data (partial ``self calibration'') and is especially visible in the peculiar-velocity model.\xspace}

There is a set of uncertainties (e.g., calibration uncertainties) that effectively perturb the light-curve fits from the true values by small amounts. Following \citet{rubin15b}, we parameterize each of these with a nuisance parameter (\DZP), such that the \mBxc for each supernova is given by:
\begin{equation} \label{eq:derivatives}
\left(\begin{array}{c}
 m_B \\
 x_1 \\
 c 
 \end{array} \right) \rightarrow \left(\begin{array}{c}
 m_B + \sum_l \PD{m_B}{\DZP_l} \DZP_l \\
 x_1 + \sum_l \PD{x_1}{\DZP_l} \DZP_l \\
 c + \sum_l \PD{c}{\DZP_l} \DZP_l
 \end{array} \right)\;.
\end{equation}
\noindent The \DZP values have a prior around zero with an appropriate size and correlation structure, as discussed in the following subsection. \citet{amanullah10} Appendix C shows that our approach is equivalent to the standard covariance matrix approach
\begin{equation} \label{eq:partialderiv}
    C_{\mu_i \mu_j} = \sum_l \frac{\partial \mu_i}{\partial \DZP_l}\frac{\partial \mu_j}{\partial \DZP_l} \sigma_{\DZP_l}^2
\end{equation}
in the limit of perfectly linear systematics (derivatives constant for each SN) and Gaussian priors and uncertainties. For computational ease, we internally multiply the derivatives by $\sigma_{\DZP_l}$ to normalize them so that the priors around zero are all unit normal.

\begin{figure*}[h!tbp]
    \centering
    \includegraphics[width = 0.99 \textwidth]{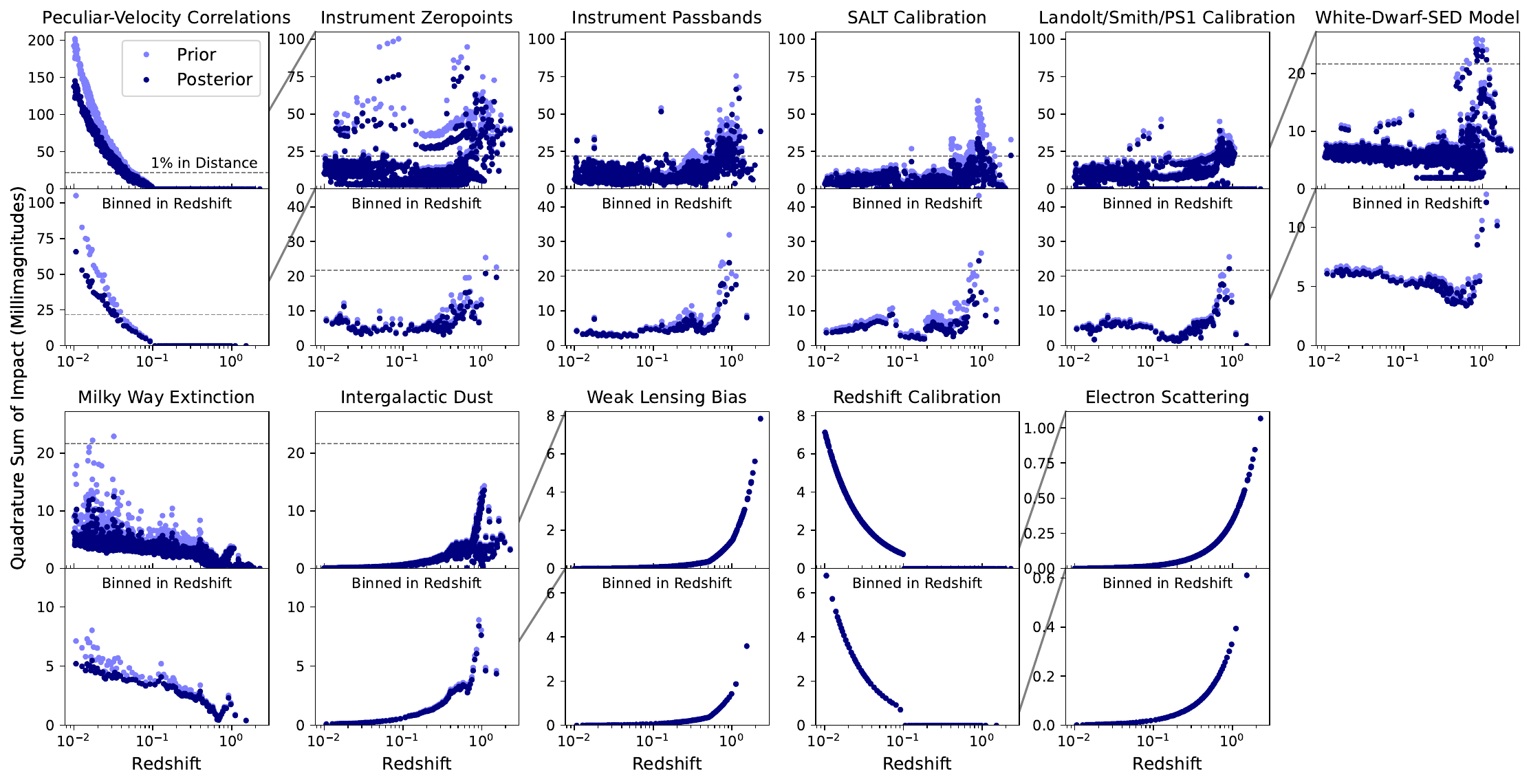}
    \caption{Quadrature sum of $[\partial(m_B + 0.15\, x_1 - 3.1\, c)/\partial \DZP] \sigma_{\DZP}$ (c.f. Equation~\ref{eq:partialderiv}) for each SN plotted against redshift, where we take the sum over each $\DZP$ in each of 11 categories. The {\bf panels} order the impacts from largest to smallest in terms of their effect on magnitude. This is not the same as their impact on cosmological parameters, which is also sensitive to their redshift dependence and is discussed in Section~\ref{sec:uncertaintyanalysis}. The light points show the sum computed with the prior $\sigma_{\DZP}$ while the dark points use the width of the posterior. \selfcalibration Some systematics (e.g., instrument zeropoints) show SNe from individual surveys clustering. In the {\bf lower part of each panel} we show the same quantity averaged in redshift in 100 roughly equal-SN-number bins. These panels show that many systematics do not strongly correlate between SNe at the same redshift and average down. Note the zoom lines which are show when the panels change scale. \label{fig:DeltaSysZ}}
\end{figure*}

The following subsections provide descriptions of each perturbative uncertainty and Figure~\ref{fig:DeltaSysZ} shows how SNe at different redshifts respond to these uncertainties by showing quadrature sums of the on-diagonal uncertainty terms in Equation~\ref{eq:partialderiv} as a function of redshift. Figure~\ref{fig:DeltaSysZ} shows both the standard deviation taken from the prior and taken from the posterior. \selfcalibration The panels order the \DZP impacts from largest to smallest in terms of their effect on magnitude. This is not the same as their impact on cosmological parameters, which is also sensitive to their redshift dependence and is discussed in Section~\ref{sec:uncertaintyanalysis}. For example, the redshift dependence of the intergalactic-extinction uncertainty is very similar to that of a change in $\Omega_m$ for flat \LCDM and therefore this becomes a large systematic on $\Omega_m$.

\subsubsection{Zeropoint Uncertainties} \label{sec:zeropointuncertainties}

We incorporate all zeropoint uncertainties from Section~\ref{sec:PhotometryCompilation} using \DZP terms (Equation~\ref{eq:derivatives}) for each zeropoint. The derivative computation is described in Section~\ref{sec:lcfit}.

\subsubsection{Bandpass Uncertainties}

We represent each \bandpass uncertainty as an uncertainty on the effective wavelength only. Higher-order \bandpass uncertainties (\bandpass width, etc.) generally only have a fraction of the impact of effective wavelength (depending on filter and redshift). We evaluate the impact of a \bandpass uncertainty similarly to the zeropoint uncertainties. For each \bandpass in each light-curve fit, we warp the \bandpass by $\exp({\lambda/(1 \mu\mathrm{m}}))$ (which modifies the \bandpass by a typical uncertainty size of tens of \ang), compute the change in effective wavelength for the \bandpass, and re-fit the light curve. The change in \mBxc divided by the change in effective wavelength gives $\partial/\partial \Delta \mathrm{\ang}$ for Equation~\ref{eq:derivatives}. Our \bandpass uncertainties are discussed in Section~\ref{sec:PhotometryCompilation}.

\subsubsection{CALSPEC Uncertainties} \label{sec:calspecuncertainties}

\newcommand{\CALSPECSentence}{The left panel shows the absolute comparison, while the right panel shows the comparison normalized to the same wavelength that defines the CALSPEC flux scale (5557.5~\ang).\xspace}
\newcommand{\CALSPECSentenceTwo}{Encouragingly, the changes with time look similar to the quoted uncertainty.\xspace}

\begin{figure*}[h!tbp]
    \centering
    \includegraphics[width = \textwidth]{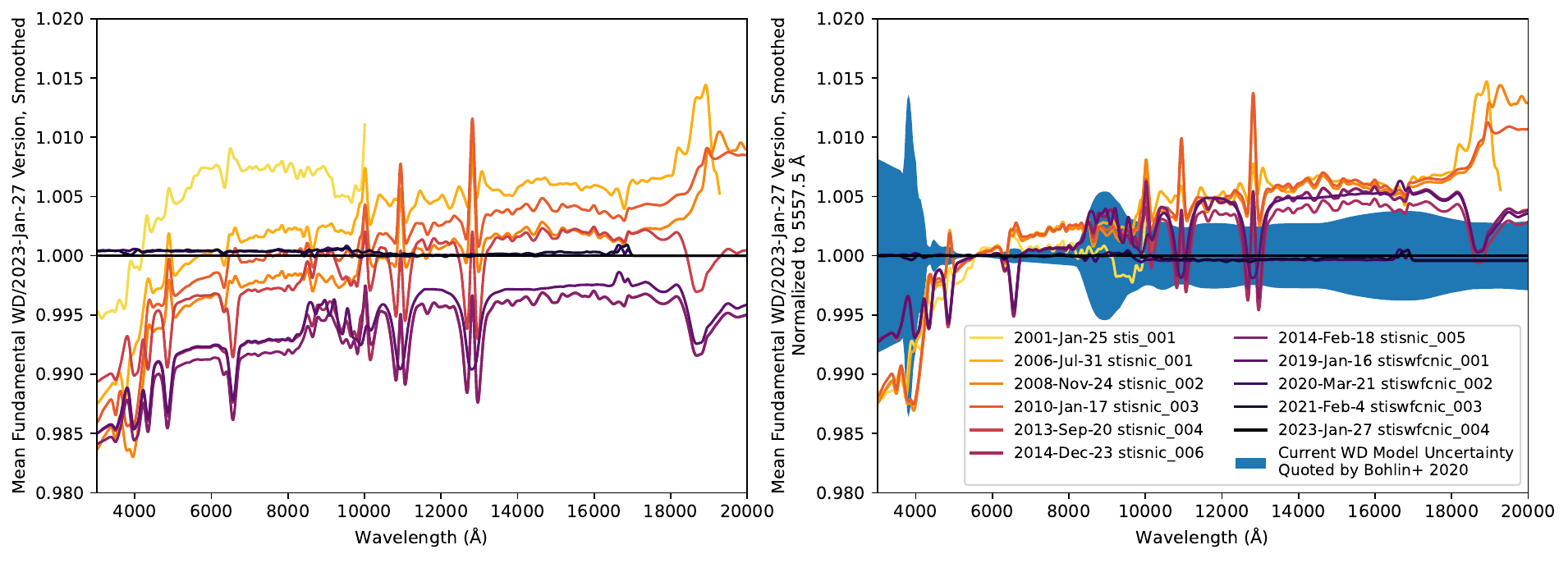}
    \caption{Changes in the CALSPEC calibration of the mean of the three fundamental white dwarfs over time. \CALSPECSentence \CALSPECSentenceTwo \label{fig:CALSPECchanges}}
\end{figure*}

As noted in Section~\ref{sec:FundamentalCalibration}, NLTE atmosphere models for three fundamental white dwarfs define the wavelength-to-wavelength flux calibration of the CALSPEC system. For the first time, \citet{Bohlin2020} provide an estimate of the uncertainty covariance matrix from these WD models. It is worth reviewing previous changes to the CALSPEC system to compare the scale of this uncertainty. Figure~\ref{fig:CALSPECchanges} shows the full history of the mean of the fundamental WDs on CALSPEC, with each WD referenced to its latest version. \CALSPECSentence It is this right panel that is most relevant for SN cosmology, as the overall wavelength-independent scale of the fluxes is degenerate with \scriptM. One can see how poorly the CALSPEC uncertainties match a simple slope in wavelength (as assumed in e.g., \citealt{Betoule2013}). A better approach is to have more flexible uncertainties as a function of wavelength \citep[e.g.,][]{amanullah10}. \CALSPECSentenceTwo

In addition, \citet{Rubin2022} shows that the white-dwarf models seem to be inconsistent in $U-V$ and $B-V$ by 3--5~mmags with other CALSPEC stars as judged by the SuperNova Integral-Field Spectrograph (SNIFS, \citealt{Lantz2004}), although whether this is due to scattered light in SNIFS, scattered light in STIS, or some other effect is not clear. (We note that \HST GO~17207 is observing eleven new nearby hot white dwarfs with STIS to improve the statistics, so more light should be shed on this soon.)

In the end, we assume CALSPEC uncertainties relative to 5000\ang--6000\ang of 10~mmag from 3000\ang--4000\ang, 5~mmag from 4000\ang--5000\ang, 2~mmag from 6000\ang--8000\ang, 5~mmag above 8000\ang, and an additional 5~mmag above 10000\ang (the switch from \HST STIS to \HST WFC3 IR and NICMOS). These uncertainties are assumed uncorrelated although Figure~\ref{fig:CALSPECchanges} shows that previous CALSPEC updates do have correlations in wavelength (e.g., correlated Balmer and Paschen breaks). We propagate the CALSPEC uncertainties using the same derivatives from Section~\ref{sec:zeropointuncertainties}.

\subsubsection{SALT Calibration Uncertainties} \label{sec:SALTCalibration}

As noted in Section~\ref{sec:SALTRestFrame}, the relative rest-frame photometric calibration of the SALT model matters in terms of measuring consistent distances as a function of rest-frame wavelength (and thus redshift). One could consider two approaches for estimating the calibration uncertainties of SALT: validation testing (Section~\ref{sec:ValidQuality}) and propagating calibration uncertainties from the SALT training. It is not a given that these would give the same answer. In fact, running the validation test of Section~\ref{sec:ValidQuality} with SALT2-4 and the JLA data \citep{betoule14} or with the original version of SALT3 \citep{Kenworthy2021} shows SALT biases in excess of those expected from propagating training calibration uncertainties. We therefore chose to go with the results from the validation testing since it is a direct test of performance with the data, and then use the training uncertainties as a comparison and to sharpen the wavelength dependence, as discussed below.

From our validation testing, the mean uncertainty on rest-frame $U-B$ for a given $B-V$ color is about 0.01~magnitudes, so we take this as a correlated systematic for bands bluer than 4000\ang in the rest-frame. There is limited data for verifying the calibration of SALT redder than the $R$ band, so we also take a correlated 0.01~magnitude uncertainty for bands redder than 7000\ang in the rest-frame.

\begin{figure}[h!tbp]
    \centering
    \includegraphics[width = 0.5 \textwidth]{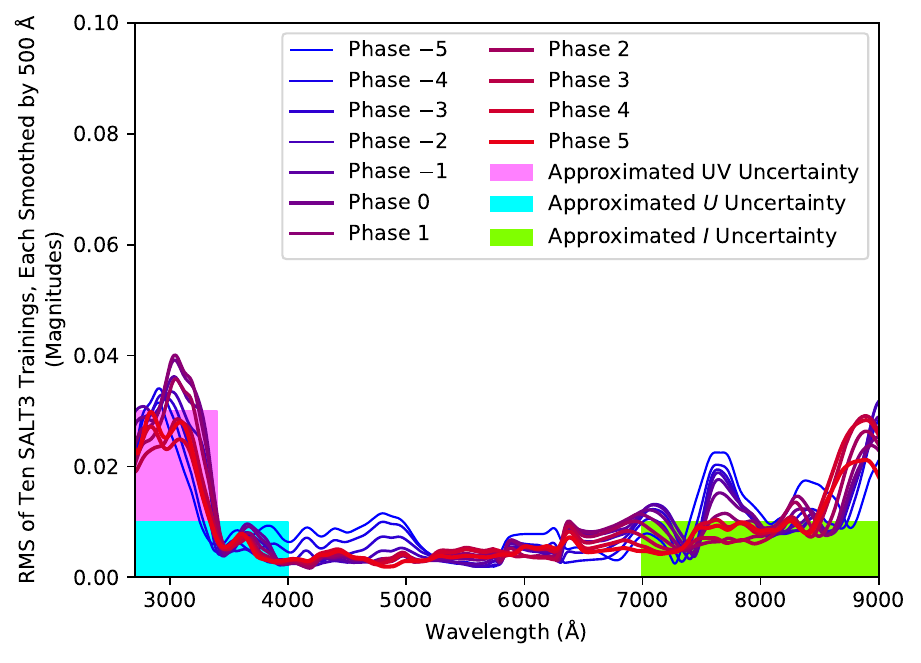}
\caption{RMS as a function of phase and wavelength for ten trainings of SALT3 with randomly perturbed zeropoints after normalization in magnitude and SALT color. We show our assumed binned calibration uncertainties as shaded rectangles. \label{fig:SALT3training}}
\end{figure}

Propagating SALT training uncertainties sharpens these values, but gives consistent results. We train SALT3 ten times, each with every zeropoint in the training data perturbed by a Gaussian draw with our modal uncertainty size (for the most important datasets) of 0.01~magnitudes. We normalize the mean SN template (\texttt{salt3\_template\_0.dat}) between each run by fitting for a relative scaling and relative color (which multiplies the SALT color law) to remove the changes in normalization and color, both of which have arbitrary zeropoints. After normalization, the standard deviation of the mean SN of these ten SALT3 training runs (shown in Figure~\ref{fig:SALT3training}) essentially depends only on wavelength and matches the values from our validation testing well over most wavelengths. (The consistency of our results here shows that, at this point, one could consider simply using different SALT training runs as a measure of uncertainty as was done by \citealt{Brout2022SuperCal} rather than our binned-in-wavelength approach.) The dispersion increases for the bluest wavelengths, so we add another 0.03 magnitudes of correlated uncertainty below 3400\ang in the rest frame to capture this. We propagate the SALT calibration uncertainties using the same derivatives from Section~\ref{sec:zeropointuncertainties}. 

\subsubsection{(Line-of-Sight) Peculiar Velocities}

Through Equation~\ref{eq:dmudz}, peculiar-velocities along the line of sight affect the observed redshifts and thus impact the inferred distances.\footnote{We note that peculiar-velocity model does not affect the observed heliocentric redshifts used in the light-curve fitting.} If this effect were fully independent SN-to-SN, it would be a statistical uncertainty on redshifts. However, over and underdensities in the universe produce correlated motions in the surrounding galaxies. Many analyses use galaxy-density maps to predict and remove this correlated signal \citep{Hudson2004, Neill2007}. It is difficult to capture the full covariance matrix of the remaining velocities after applying this method (e.g., the popular \citealt{Carrick2015} model does not have such uncertainties), so we take a different approach: building a SN-only peculiar-velocity model simultaneously with all other parameters. The top left panel of Figure~\ref{fig:DeltaSysZ} shows that much of our assumed peculiar-velocity uncertainty is actually constrained by the SN distances without the need for galaxy-density information.\footnote{The optimal approach is a simultaneous density/velocity analysis, which we are pursuing \citep{Kim2024}.}

We include the linear-theory expectation of the peculiar-velocity covariances for the first time in a Union/UNITY analysis, as computed with the PairV code \citep{Hui2006, Davis2011}, which produces a distance-modulus covariance matrix from the SN coordinates and redshifts. We limit this covariance matrix to $\zCMB<0.1$, both because the effect of peculiar velocities is small beyond this redshift, and because the correlation scale of peculiar velocities is much smaller than the $\zCMB=0.1$ universe, so they cease to be an important correlated uncertainty between SNe. In addition, we include 10~km/s of correlated redshift uncertainty as discussed in Section~\ref{sec:redshiftsource}.

\begin{figure*}[h!tbp]
    \includegraphics[width = \textwidth]{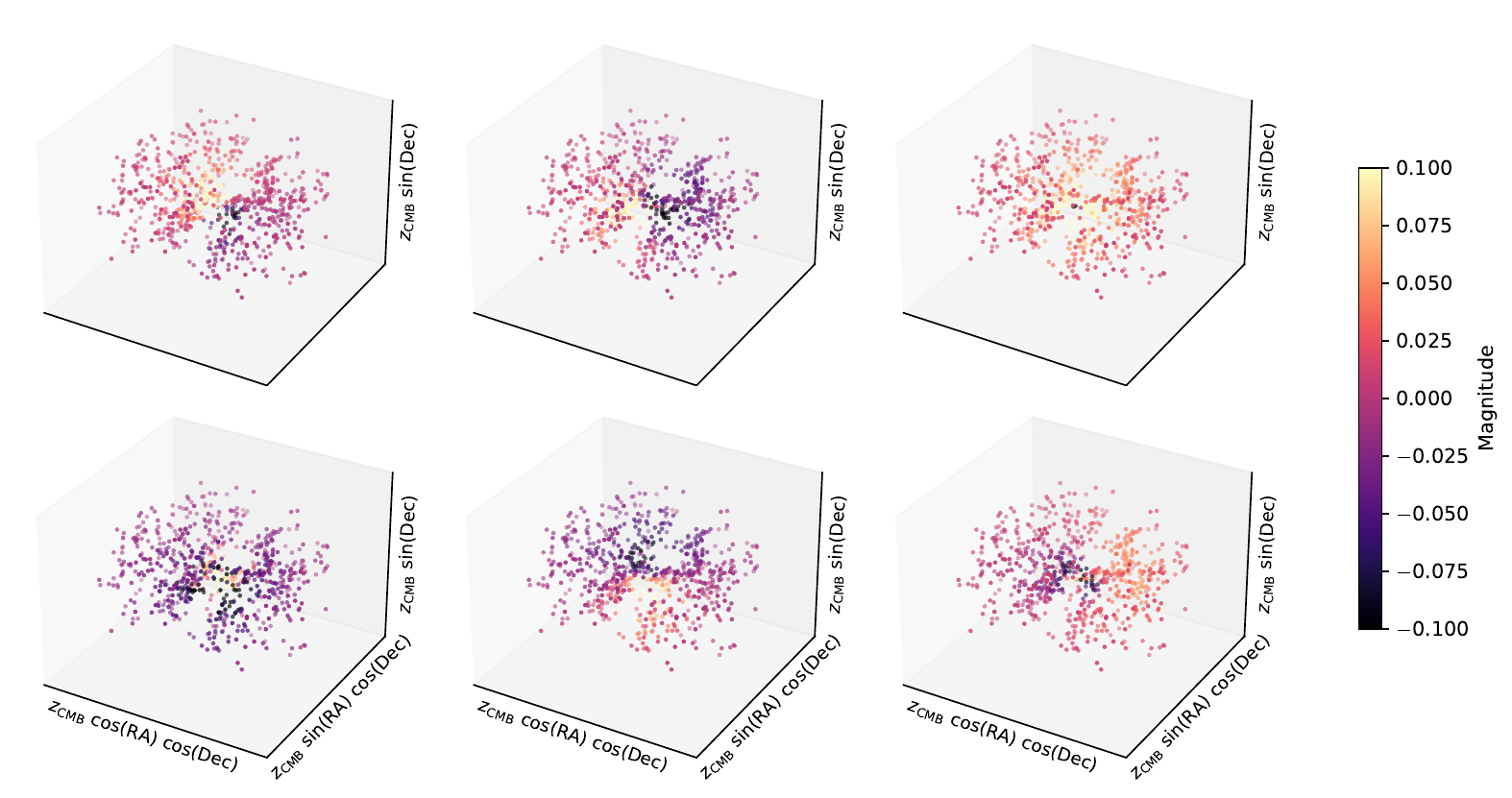}
    \caption{The first six peculiar-velocity eigenvectors shown with a scale from $-0.1$ to +0.1 magnitudes (the sign is arbitrary) and a range $\zCMB < 0.03$ for clarity (the eigenvectors are computed out to $\zCMB=0.1$). Note the correlated motions.}
    \label{fig:PeculiarEigen}
\end{figure*}

\begin{figure}[h!tbp]
    \centering
    \includegraphics[width = 0.47 \textwidth]{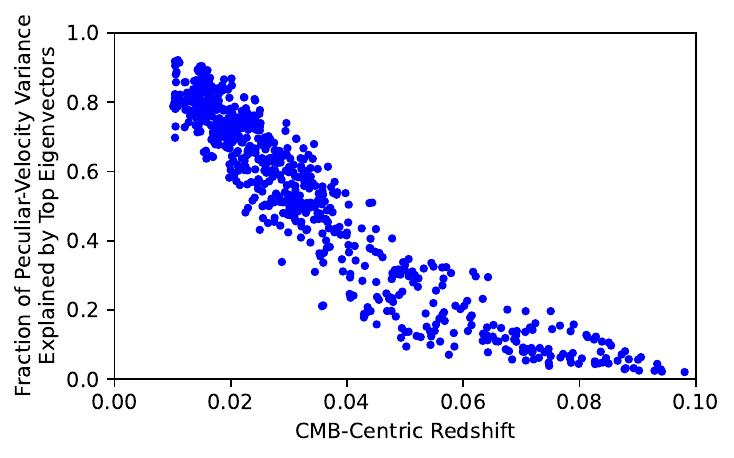}
    \caption{The fraction of line-of-sight peculiar-velocity variance explained by the top 100 eigenvectors computed for this work. The remaining variance is added to the $m_B$-$m_B$ term in the covariance matrix for each SN.}
    \label{fig:EigenExplain}
\end{figure}

As discussed in Section~\ref{sec:SmallPerturbations}, Hamiltonian Monte Carlo generally handles more parameters with uncorrelated data faster than fewer parameters with correlated data. Thus, similarly to \citet{rubin16}, we decompose PairV's distance-modulus covariance matrix into eigenvectors, keep the first \nPecVelEigen (capturing more than 90\% of the inverse variance of the full matrix), and marginalize over the projection onto each one with a set of \DZP parameters.\footnote{More accurately, we decompose the off-diagonal portion of the covariance matrix. We add the remaining diagonal portion in quadrature to the $m_B$ uncertainties.} Figure~\ref{fig:PeculiarEigen} shows the first six eigenvectors in 3D coordinates (converted from coordinates on the sky and \zCMB); these show smoothly varying correlated bulk motions. Figure~\ref{fig:EigenExplain} shows the fraction of the peculiar-velocity variance explained by the \nPecVelEigen peculiar-velocity eigenvectors considered in this work (showing that by $\zCMB=0.1$, peculiar velocities are essentially uncorrelated SN-to-SN).\footnote{There has been some discussion in the literature (e.g., \citealt{RubinHeitlauf2020}) of the CMB-centric correction for SN cosmology. This correction is based on assuming that essentially all of the observed CMB dipole is due to the motion of the solar system (rather than intrinsic) and that no large-scale bulk flows shift the mean reference frame of the nearby SNe away from the CMB frame. As a test, we introduce 300 km/s shifts along each axis on the sky and verify that the eigenvector projections can absorb them. Thus, marginalizing over our eigenvector projections is equivalent to solving for the reference frame of the solar system which has been done explicitly in some other analyses \citep{Gordon2008, Horstmann2022}.}

\newcommand{\PecVelSentence}{2D slice through the recovered peculiar-velocity field along the celestial equator, interpolated with nearest-neighbor interpolation (in 3D). The left panel shows the posterior median; the other four panels show draws from the posterior to give a sense of the size of the uncertainties.\xspace}

\begin{figure*}[h!tbp]
    \centering
    \includegraphics[width = \textwidth]{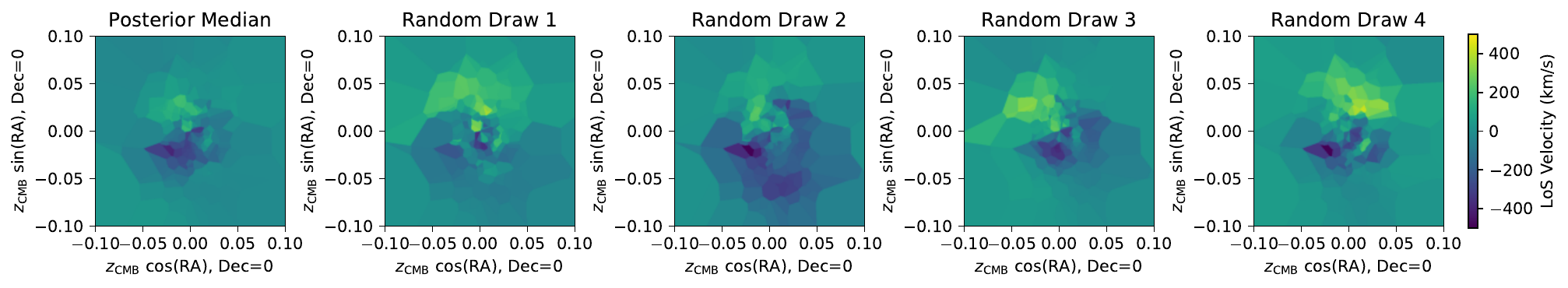}
    \caption{A \PecVelSentence}
    \label{fig:VelocityField}
\end{figure*}

In addition to accounting for peculiar-velocity uncertainty, UNITY1.5 thus produces the full posterior of the peculiar-velocity field with all SN uncertainties modeled. Figure~\ref{fig:VelocityField} shows a \PecVelSentence

Another source of systematic uncertainty due to peculiar velocities at low redshift are a set of Eddington-like biases:
\begin{itemize}
    \item For a homogeneous universe and a (assumed to be complete) very nearby sample of SNe, the number of SNe per small redshift step should scale as $z^2$. Thus, for any given redshift, more SNe scatter from higher to lower redshift due to peculiar velocities than lower to higher.
    \item SNe scattering lower in redshift are given less weight than SNe scattering higher in redshift, as the observed redshifts are used to determine the peculiar-velocity uncertainties.
    \item Finally, distance modulus is a concave-down function of redshift. Thus SNe scattering lower in redshift have larger distance-modulus residuals than those scattering higher in redshift.
\end{itemize}
Simulations show that these biases (taken together) are subdominant to our assumed peculiar-velocity uncertainties, so we ignore them for now. However, future cosmology analyses should explicitly marginalize over the true redshift of each SN to properly treat these effects.\footnote{\citet{Roberts2017} and \citet{Hayden2023} consider photometric-redshift uncertainties and show how this marginalization can be done.}

\subsubsection{Milky Way Extinction}
We use the \citet{Schlafly2011} recalibration of the \citet{Schlegel1998} Milky-Way extinction map for our MW extinctions (that is, we scale the SFD98 map by 0.86). We use the \citet{Fitzpatrick1999} extinction curve with $R_V = 3.1$. We assume each MW extinction has 16\% statistical uncertainty and a correlated 10\% systematic uncertainty in the normalization of the full map. Furthermore, we assume an additive zeropoint uncertainty of the map of 5~mmags in $E(B-V)$ motivated by a comparison between HI and reddening at low column density \citep{Lenz2017}. We include these uncertainties by computing \PDnofrac{\mBxci}{E(B-V)} for each SN $i$ and constructing two eigenvectors, one from

\begin{equation}
    \PD{\mBxci}{E(B-V)} \, 0.10 \, E(B-V)_i \;,
\end{equation}
and one from
\begin{equation}
    \PD{\mBxci}{E(B-V)} 0.005 
\end{equation}
and marginalizing over the projections (\DZP) onto these two eigenvectors with a unit normal prior on each. We also compute
\begin{equation}
    \PD{\mBxci}{E(B-V)} \, 0.16 \, E(B-V)_i \;,
\end{equation}
and add the outer product of this vector with itself to the covariance matrix for each SN $i$ to propagate the statistical uncertainty.

\subsubsection{Intergalactic Dust Extinction} \label{sec:intergalacticextinction}

\citet{Menard2010} made a detection of extinction of background quasars due to the outer halos of foreground galaxies at $z \sim 0.3$. This result is generally consistent with the results based on background elliptical galaxies of \citet{Peek2015} and the estimate of \citet{Zhang2007}. Similarly to \citet{amanullah10} and \citet{suzuki12}, we include this effect as a systematic uncertainty, but (due to its somewhat uncertain size) do not attempt to correct for it.

We take the \citet{Menard2010} estimate of the average comoving number density of dusty galaxies $n=0.037\, h^3/$Mpc$^3$. Assuming an inner radius of $20\, h^{-1}$kpc, and an outer radius of $110\, h^{-1}$, \citet{amanullah10} found that the light from an average $z=1$ SN intercepts $\sim 7$ halos on its way to us and that the average rest-frame $V$-band extinction per halo at $z \sim 0.3$ is about 4~mmag. As in \citet{amanullah10}, we scale this value by $[(1 + z)/(1 + 0.3)]^{-1.1}$ \citep{Menard2008}. We also assume the extinction has an $R_V$ of 3.1 in the rest-frame of each halo. We compute the average extinction as a function of wavelength and SN redshift and apply it as a systematic term.

\newcommand{\IGExtinctSentence}{We plot a solid line for each redshift which is the best-fit SALT3 color law scaled by $\Delta c$ plus a constant $\Delta m$ (values given in the legend). Subtracting $(\beta = 3.1) \Delta c$ from the $\Delta m$ values (to mimic SALT standardization) gives the quoted $\Delta \mu$ and does not generally remove the full $\Delta m$; the remaining uncorrected extinction contributes as a systematic.\xspace}

\begin{figure}[h!tbp]
    \centering
    \includegraphics[width = 0.49 \textwidth]{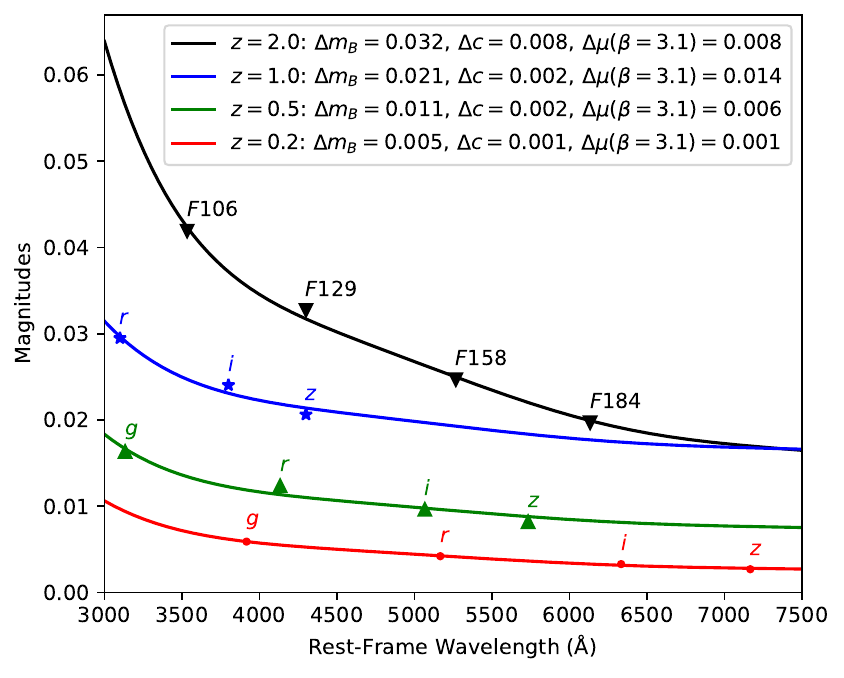}
    \caption{An illustration of the impact of our assumed intergalactic dust extinction model for four different redshifts: 0.2, 0.5, 1.0, and 2.0. The total intergalactic dust extinction along the line of sight is shown on the y-axis; the rest-frame wavelength is shown on the x-axis. For redshift 0.2/0.5/1.0, we show $griz$ filters; for redshift 2.0, we show \RomanSpelled $F106$/$F129$/$F158$/$F184$ filters. \IGExtinctSentence}
    \label{fig:IGExtinct}
\end{figure}

Figure~\ref{fig:IGExtinct} shows the average extinction as a function of redshift and rest-frame wavelength. \IGExtinctSentence This turns out to be a large source of uncertainty in our analysis, as discussed in Section~\ref{sec:uncertaintyanalysis}. Future analyses should thus explicitly consider the impact parameters for each galaxy along the line of sight and compute a per-SN extinction estimate to enable self-calibration of this uncertainty.

\subsubsection{Thomson Scattering}

There will also be an extinction component due to Thomson scattering \citep{Griffiths1999, Zhang2008}. This scattering is gray (wavelength-independent) and would not be corrected for with the color standardization. Fortunately, it is quite small, $\sim 3$~mmags to redshift 1, and its uncertainty is much smaller. Assuming flat $\Lambda$CDM, the average optical depth in the late universe will scale as
\begin{equation}
    \tau(z) = \frac{c \, \sigma_T}{H_0} \int_0^z \frac{n_{e}(z^{\prime}) \, d z^{\prime}}{(1 + z^{\prime}) \sqrt{\Omega_m (1 + z^{\prime})^3 + (1 - \Omega_m)}} \;,
\end{equation}
and assuming the ionization state of the universe does not change over the redshift range of interest
\begin{equation}
    n_e(z^{\prime}) = n_{e}(0) (1 + z^{\prime})^3 \;,
\end{equation} then 
\begin{equation}
    \tau(z) = \frac{2 c \, \sigma_T \, n_{e}(0) [\sqrt{\Omega_m (1 + z)^3 + (1 - \Omega_m)} - 1]}{3 H_0 \, \Omega_m} \;.
\end{equation}
We set the normalization by requiring that the optical depth to $z=7.82$ is 0.0561 \citep{PlanckCollaboration2020} but find virtually the same normalization with WMAP9 \citep{Hinshaw2013}, which has a higher $\tau$ with a higher reionization redshift. We apply this correction with 10\% uncertainty (and for the first time in a Union/UNITY analysis).

\subsubsection{Gravitational Lensing}

Gravitational lensing by halos near the line of sight can amplify or deamplify a SN with respect to a homogeneous universe. Some analyses attempt to estimate the masses along the line of sight and remove the effect of lensing \citep{Jonsson2007, Kronborg2010, Jonsson2010}, although we avoid this here as it offers limited statistical gain and could cause a bias. We include a redshift-dependent $0.055\, z$ magnitudes of weak gravitational lensing dispersion \citep{Jonsson2010}. We assume this is a Gaussian distribution, and add it in quadrature to the $m_B$ uncertainties for each SN. This functional form is inappropriate for redshifts $\gg 1$, where the dispersion should asymptote (e.g., Equation~1 of \citealt{Aldering2007}), but most of our SNe are below redshift 1, so we use it for now.

In addition to the statistical dispersion, gravitational lensing will cause a redshift-dependent bias on the Hubble diagram. Although lensing conserves flux when averaging across all lines of sight, the lensing PDF skews positive (lensing can amplify the flux by a factor $\gg 1$, but cannot scale flux down by a factor $\ll 1$), so the PDF will not average out when magnitudes are used \citep{Holz2005}. The proper treatment will be to include the lensing PDF inside UNITY, but for now, we evaluate the size of the bias and find it to be small.

We use the functional form of the lensing PDF from Equation~12 of \citet{Linder2008} assuming $s=1$. We shift the mean to require that the average lensing is exactly 1. After analytically converting the PDF to magnitudes (instead of lensing amplification), we find that the mean bias in magnitudes is described by $0.5\; \sigma_{\mathrm{lensing}}^2$ and is thus less than 0.01~magnitudes at all redshifts considered in this work. We do not make this correction, and simply include it as a correlated uncertainty on magnitude for all SNe. We caution that this is probably something of an underestimate, as the \citet{Linder2008} parameterization does not reproduce the high-amplification tail accurately, but we leave further refinement of the lensing PDF (or object-by-object lensing correction) to future work.

\subsection{Selection-Effect Model} \label{sec:MalmquistBias}

Practical considerations imply that both imaging and spectroscopic followup do not adequately measure every SN~Ia in the areal footprint of the survey. If the missing SNe were missing at random, this would affect the rate of useful SN production, but cause no cosmological bias. However, the missing SNe are preferentially those that appeared fainter than a given observer-frame magnitude range. Such ``selection effects'' are optimally treated with a model of both measured SNe and missing SNe \citep[e.g.,][]{Gelman2004}. By marginalizing over the number of missing SNe and their properties, inference can proceed even with biased SN samples. \citet{rubin15b} presented a first such model for SN cosmology (see also \citealt{March2018} and \citealt{Hinton2019}), which we improve on here.

UNITY1.5 uses the same basic assumptions as UNITY1 (an error-function probability of a SN~Ia being selected as a function of magnitude), but follows \citet{Hinton2019} in marginalizing the depth of each survey instead of fixing these values as UNITY1 did. Although the UNITY1 model was applicable to observer-frame magnitudes, the Union2.1 reanalysis in \citet{rubin15b} used rest-frame $B$ magnitudes instead (a mostly adequate approximation). UNITY1.5 now uses the observer-frame magnitudes given in Table~\ref{tab:DatasetSummary} by converting those magnitudes as a function of redshift for each sample to an affine combination of $m_B$ and $c$ and using that affine combination in Equation~B2 of \citet{rubin15b}.

For example, the SNLS selection is essentially a function of observer-frame $i$-band magnitude. At around $z=0.8$, the affine relation is $m_i = m_B -0.8$, i.e., $i$-band matches well to rest-frame $B$ up to a $K$-correction. But at higher redshift, the $i$-band shifts bluer in the rest frame, such that color plays a role in the transformation. At $z=1.1$, the affine relation is $m_i = m_B -0.5 + 1.0\,c$ (at lower redshift, the sign on $c$ is flipped). These relations allow UNITY1.5 to combine each survey's observer-frame magnitude limits with rest-frame populations of SNe.

\begin{figure*}
    \centering
    \includegraphics[width=0.7\textwidth]{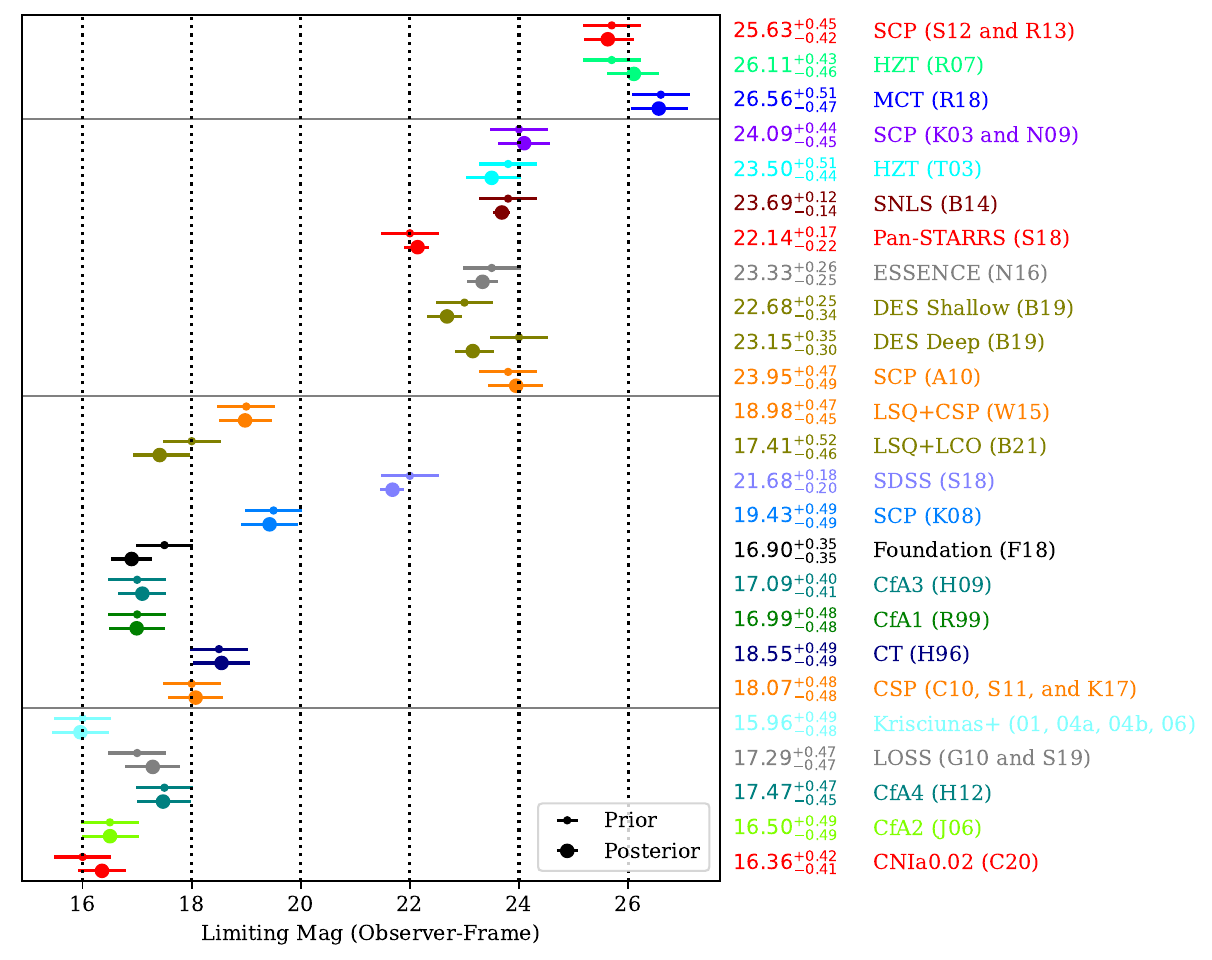}
    \caption{Limiting observer-frame magnitudes for each sample, showing both our priors and the UNITY1.5 posteriors. We show the median with the plot point, and the error bar spans the percentiles 15.9 to 84.1 ($\pm 1 \sigma$ for a Gaussian distribution). These values are also shown numerically to the right of the plot. In general, we see good agreement between the priors and posteriors. One sees that for larger surveys the constraints tend to be tighter than their priors, while smaller surveys rely more on their priors. As noted in Section~\ref{sec:datasets}, the exact width chosen for the priors does not drive the cosmology results or their uncertainties.}
    \label{fig:MagntiudeLimits}
\end{figure*}

Figure~\ref{fig:MagntiudeLimits} shows the input priors and the recovered limiting magnitudes of each sample. In general, our priors are reasonably consistent with the data (see also the validation with the predictive posterior distribution in Section~\ref{sec:PPD}), although the two DES samples (from the shallow fields and the deep fields) can be seen to have similar depths that are shallower than the priors because both are limited by the depth of spectroscopic selection in this analysis.

\subsection{Non-Ia Contamination}
As in the original UNITY, we use an outlier model to limit the impact of non-Ia SNe (and peculiar SNe Ia) in the sample \citep[e.g.,][]{Kunz2007}. As Union3 is a collection of spectroscopically confirmed SNe (which are generally $\gtrsim 98\%$ pure), we do not have to build a perfect model of the outlier distribution. The outlier model simply has to be significantly broader than the inlier (SN~Ia) model so that SNe that are outliers have a high likelihood of being from that distribution. We use an uncorrelated Gaussian distribution in \mBxcc for the outlier distribution. The widths of this Gaussian are fit parameters: $\sigma_{m_B}^{\mathrm{outl}}$, $\sigma_{x_1}^{\mathrm{outl}}$, and $\sigma_{c_B}^{\mathrm{outl}}$, and $\sigma_{c_R}^{\mathrm{outl}}$.

We assume each SN~Ia has an identical prior probability of being an outlier: $f^{\mathrm{outl}}$, which is also a fit parameter (the posterior is $f^{\mathrm{outl}} = \foutlconstraint$). This is an assumption that is easy to change if UNITY is run with photometrically classified SNe which have an associated probability of being a SN~Ia based on fits to the light-curve.

\subsection{UNITY1.5 Validation} \label{sec:Validation}

\subsubsection{Simulated-Data Testing of UNITY1.5} \label{sec:simulateddata}

Simulated-data testing is key for any analysis to validate that the analysis framework produces reasonable results given known inputs. For SN cosmology analyses, these simulations can take the form of simulated SNe at the pixel level (e.g., \citealt{Astier2013, Brout2019Photometry, Rubin2021}), simulated light-curve or spectral measurements (e.g., \citealt{Kessler2009SNANA, hayden19}), or simulated light-curve-fit results (e.g., \citealt{March2011,rubin15b, March2018}). We have no ability to repeat pixel-level checks of the photometry, so we opt to test with simulated light curves.

The basic UNITY framework was investigated with simulated light-curve-fit results by \citet{rubin15b}, who simulated a realistic compilation of four datasets (similar to low-redshift SNe, SDSS SNe, SNLS SNe, and \HST SNe). \Steve, a similar model, was tested on simulations of two datasets (low-redshift and DES3-like to test the DES3 analysis, \citealt{Hinton2019}).

As the selection-effect model is new, we opt to test the new UNITY1.5 population and selection-effect model by simulating datasets with selection effects large enough to move the cosmological parameters $\sim 2\times$ the entire rest of the uncertainty budget if they were not modeled (see Tables~\ref{tab:SimulatedSummaryLHV} and \ref{tab:SimulatedSummary}). For each of 100 independent realizations, we simulate three magnitude-limited datasets which we call low-$z$, mid-$z$, and high-$z$. We use \texttt{SNCosmo} \citep{sncosmo} to simulate simple rolling surveys for each dataset. For each of the 100 realizations, we perform two simulated-data cosmology analyses with UNITY1.5. The first (and more similar to Union3) test is to recover $w_0$ and $w_a$ from the compilation of all three datasets (here, we fix $\Omega_m$ to avoid having to consider external cosmological constraints). The second (more challenging) test is to recover $\Omega_m$ from the mid-$z$ dataset alone, essentially differencing the lower-redshift, more complete half of the sample against the highly biased, higher-redshift half of the sample. As UNITY1.5 advances selection-effect and population modeling, this test is of interest as a ``stress test'' of the model.

\begin{itemize}
\item For the low-$z$ simulated SNe, we simulate $griz$ light curves with a $5 \sigma$ depth of 20.0 AB and a cadence of four days (this is roughly similar to the Foundation survey, although we simulate a faster cadence but lower depth per point). We nominally simulate 8,000 square degrees and 200 visits, although the goal here is only to overproduce SNe for spectroscopic selection (described below). We draw the dates of maximum for the SNe uniformly between two cadence steps (eight days) after the start of the survey and two cadence steps before the end of the survey.
\item For the mid-$z$ simulated SNe, we simulate $griz$ light curves with $5 \sigma$ depths of 24.0 AB (similar to the deep tier of the spectroscopically confirmed DES3 SNe). Again, we simulate a cadence of four days and 200 visits, but only simulate 10 square degrees. (Again, the purpose here is just to overproduce SNe for the spectroscopic selection.). We draw the dates of maximum for the SNe uniformly between two cadence steps after the start of the survey and two cadence steps before the end of the survey.
\item For the high-$z$ simulated SNe, we simulate ACS~$F775W$, ACS~$F850LP$, WFC3~$F125W$, and WFC3~$F160W$. We simulate a cadence of 17 days, six visits, and a $5 \sigma$ depth of 26.14 AB (all similar to MCT). We draw the dates of maximum uniformly between one cadence step after the start of the survey and three cadence steps before the end. We simulate four square degrees, which is equivalent to 3,000 \HST WFC3 IR pointings in an untargeted search. This is larger than the actual MCT survey (although there are other \HST datasets) but this gives us sufficient statistics for a sensitive test of breaking the $w_0$/$w_a$ degeneracy. After selection (described below), this gives an average of 87 high-$z$ SNe per realization.
\end{itemize}
We draw the $x_1$ population from an exponentially modified normal ($\mu = 0.8$, $\sigma = 0.5$, $\tau = -0.8$) and the color population from an exponentially modified normal ($\mu = -0.07$~mag, $\sigma = 0.05$~mag, $\tau = 0.07$~mag). While we draw these as the same population for all three samples, we do not assume that the three samples have the same population in the UNITY analysis. For the light-curve uncertainties, we also include the SALT3 model uncertainties (using \texttt{SNCosmo}'s \texttt{source.bandflux\_rcov}) but do not assume any correlated scatter between bands. We simulate outliers with an average rate of 2\% (prior to discovery and selection cuts) and assume the outliers have a broad Gaussian population distribution with widths 2 in $x_1$, $0.2$ in color, and $0.5$ in absolute magnitude. We center the outliers on zero color, zero $x_1$, and $-19.1$ AB $B$-band absolute magnitude (the same absolute magnitude as the SNe~Ia). We use flat \LCDM as our input cosmology model with $\Omega_m = 0.3$. We simulate independent zeropoint uncertainties of 5~mmag for each filter. UNITY1/1.5 models selection effects as an error function in magnitude; for increased realism, we do not match this assumption for the larger low-$z$ and mid-$z$ datasets. Instead, for each cadence step, the three brightest SNe in observer-frame $r$ band (low-$z$) or $i$ band (mid-$z$) are selected that have not already been observed, giving a simulated dataset of 600 SNe for each of low- and mid-$z$. These choices replicate finite uniform spectroscopic follow-up over the survey duration. For the high-$z$ dataset, we simulate a magnitude limit of 26.0 with a width of 0.25 magnitudes in $F125W$. I.e., a SN is found if it gets brighter than 26.0 + a random Gaussian of width 0.25 magnitudes in any visit. (This still does not quite match UNITY1.5's selection model, which is based on modeled magnitude at maximum light, not at the phases that are actually observed, so it is still an interesting test.) We simulate 0.12 magnitudes of total unexplained dispersion, allocated between $m_B$, $x_1$, and $c$ as described below. We add $0.055\, z$ lensing dispersion in quadrature to the unexplained dispersion in magnitude. We simulate using $\alpha = 0.15$, $\beta = 3.1$, and $\delta(0) = 0.08$~mag. (Here, our main concern with respect to $\beta$ is that biases in UNITY1.5 might make a linear/uniform color standardization look nonlinear, so we chose to make all $\beta$ values the same in the simulated data to see if that is what is recovered.)

In general, Bayesian Hierarchical Models like UNITY give accurate uncertainties \citep{Kelly2007, March2011, hayden19}, but we also test for accurate uncertainties here. To be able to use our 100 realizations to examine whether the reported uncertainties are accurate, we must vary some of the parameters realization to realization.\footnote{In general, the parameters we must vary are parameters where the priors matter. Any parameter where the data are insufficiently constraining will be correlated realization-to-realization by the prior. For example, a parameter that has no constraint from the data will have exactly the same posterior (equal to the prior) in every realization.} Between the 100 realizations, we randomly scatter the zeropoint values by their uncertainties, scatter $\delta(\infty)/\delta(0)$ uniformly between 0 and 1, and randomly draw the fraction of unexplained variance in $m_B$, $x_1$, and $c$ from a simplex (rather than picking discrete models, e.g., \citealt{guy10, Chotard2011}).

\newcommand{\simdatasetence}{Each column shows a different analysis variant and each row shows a parameter. For each entry, four values are shown. The first value is the mean (of the posterior medians) of all simulated realizations, the first uncertainty is the uncertainty on that mean using 100 realizations, the second uncertainty is the mean uncertainty for each of the 100 realizations, and parenthetical values are the RMS pull of each parameter around the true value\xspace}

\newcommand{\simdatasetencetwo}{This quantity is expected to be 1 if the pulls are unit normal and has a $1 \sigma$ uncertainty of $1/\sqrt{200} = 0.07$.\xspace}
\newcommand{\pullsaregood}{In general, the RMS is within a few percent of 1 (horizontal line), the value expected for a unit normal.\xspace}

\begin{figure*}[h!tbp]
    \centering
    \includegraphics[width=0.98\textwidth]{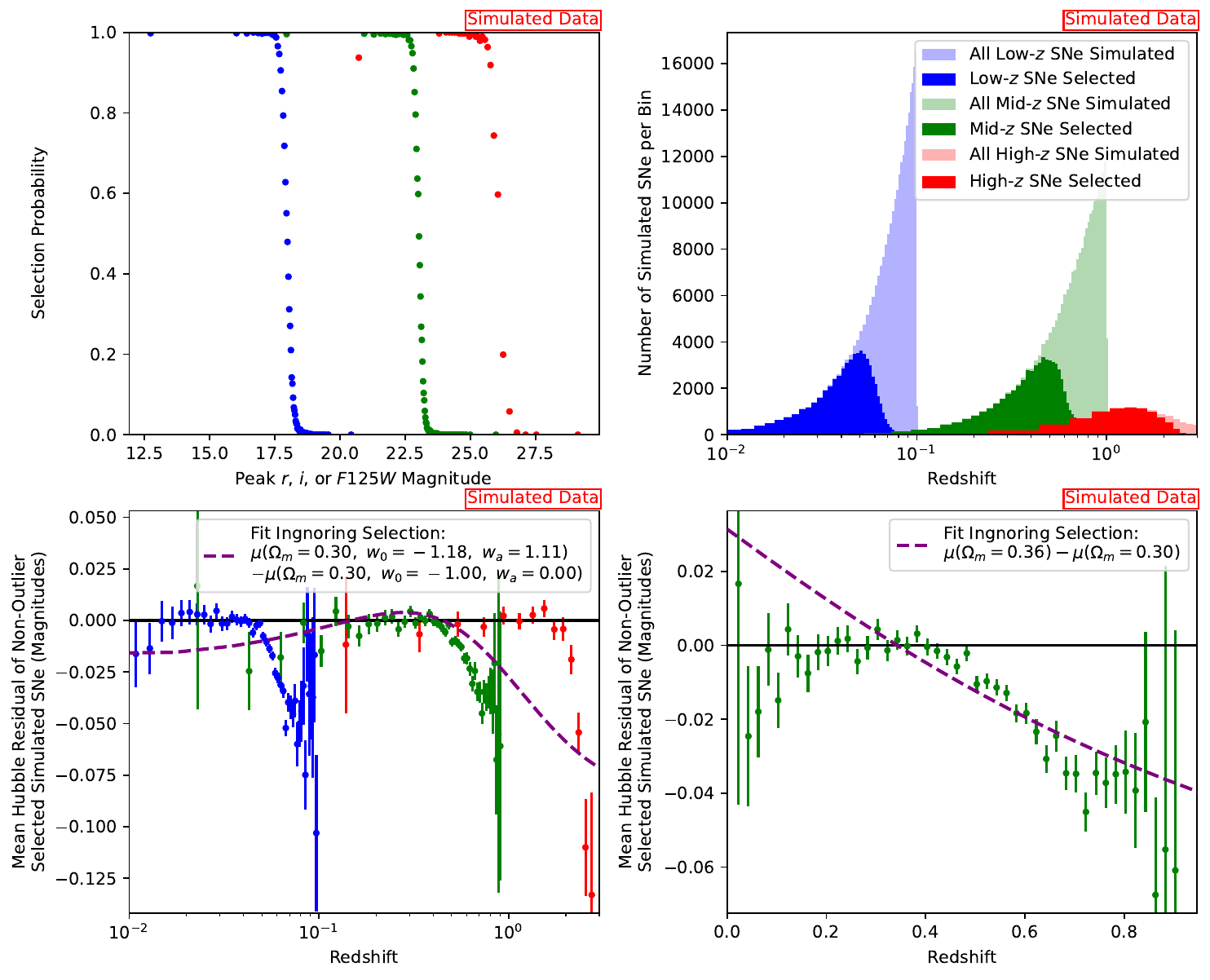}
    \caption{Summary statistics for the combined 100 realizations of three simulated datasets (low-, mid-, and high-$z$) used for testing UNITY1.5. {\bf Top-left panel:} probability of selecting a SN as a function of peak magnitude. The three magnitude limits for each dataset are visible. {\bf Top-right panel:} redshift histograms of SNe generated (lightly shaded bars) and selected (filled bars). {\bf Bottom-left panel:} Binned mean Hubble-diagram residual vs. redshift for SNe from all three simulated datasets (low-, mid-, and high-$z$). Also shown is the approximate recovered (highly biased from the input \Om = 0.3) cosmological model when fitting the simulated SNe with a version of UNITY with no selection-effect model (purple dashed curve). {\bf Bottom-right panel:} Binned mean Hubble-diagram residual vs. redshift for SNe selected in the analysis from just the realizations of the mid-$z$ dataset. Again, we show the bias from ignoring selection effects in the model when they are present in the data (purple dashed curve). As expected, UNITY1.5 outperforms this assumption (Tables~\ref{tab:SimulatedSummaryLHV} and \ref{tab:SimulatedSummary}). 
    \label{fig:simdata}}
\end{figure*}

Figure~\ref{fig:simdata} shows summary statistics for the simulated datasets, including selection efficiency as a function of magnitude, the redshift distribution, and the mean Hubble-diagram residual as a function of redshift of SNe making it into this analysis. As expected, our procedure generates simulated datasets with severe Malmquist bias of up to $\sim 0.1$ magnitudes. Appendix~\ref{sec:validLCfit} uses the light-curve fits to investigate any biases or deviations from Gaussian uncertainties and generally finds good agreement between the scatter around the true \mBxc and the quoted uncertainties.

Tables~\ref{tab:SimulatedSummaryLHV} and \ref{tab:SimulatedSummary} summarize the results of our testing. \simdatasetence:
\begin{equation}
\sqrt{\frac{1}{100} \sum_{i = 1}^{100} \frac{[(\mathrm{posterior\ median})_i - (\mathrm{true\ value})_i]^2}{(\mathrm{posterior\ uncertainty})_i^2}}\;.
\end{equation}
\simdatasetencetwo This quantity can exceed 1 either because of underestimated uncertainties or because of a bias. The first column shows our nominal UNITY1.5 model, the second column shows UNITY1.5 with the improved parameter limits on the red-color outlier model discussed in Appendix~\ref{sec:Priors}, the third column shows a model that does not include selection effects and assumes constant-in-redshift $x_1$ and $c$ populations (c.f., \citealt{March2011}), and the last column shows a model that does not include selection effects but assumes redshift-dependent populations (c.f., \citealt{rubin16}).

We begin by examining the three-simulated-dataset cosmology results in Table~\ref{tab:SimulatedSummaryLHV}. The first three rows show the recovery of cosmological parameters: $w_0$, the uncorrelated-with-$w_a$ $w_0 + 0.15 w_a$, and $w_a$. For the nominal UNITY1.5 model (first column of values), we see a bias on $w_0 + 0.15 w_a$ of $+0.012 \pm 0.005$ or $0.24 \pm 0.10 \sigma$, i.e., statistically significant only after averaging over 100 simulated realizations. Appendix~\ref{sec:validLCfit} shows that a good amount of this is due to subtle biases on the light-curve fitting. However, this light-curve fitting bias is much smaller than the calibration uncertainties and the cosmological impact of this bias is much smaller than the cosmological uncertainties on any one realization. We thus take this as strong validation of our overall light-curve-fitting + UNITY approach for the current samples. $w_a$ shows no evidence of biases and both $w_0 + 0.15 w_a$ and $w_a$ show accurate uncertainties. Both of the no-selection-effect models show large cosmological biases, as expected.

\begin{figure*}[h!tbp]
    \centering
    \includegraphics[width=0.49\textwidth]{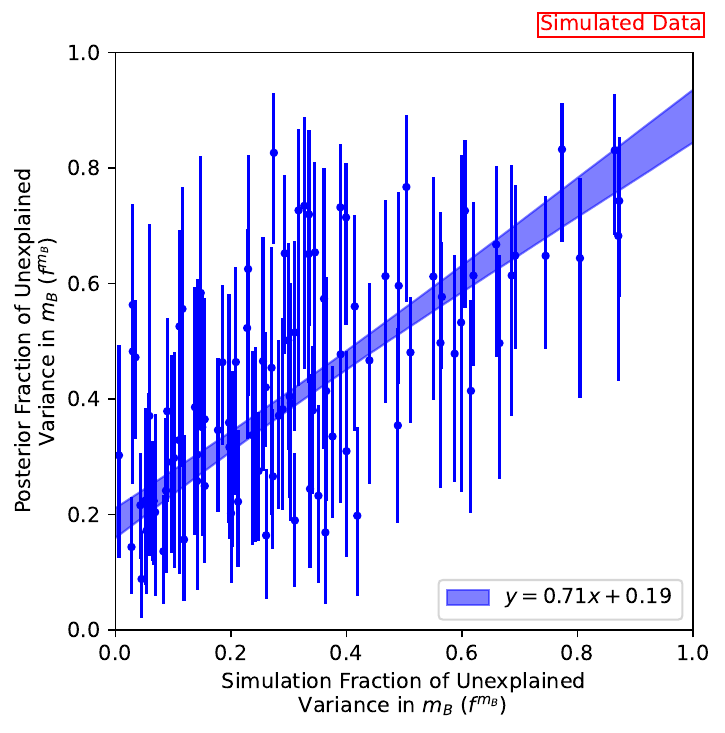}
    \includegraphics[width=0.49\textwidth]{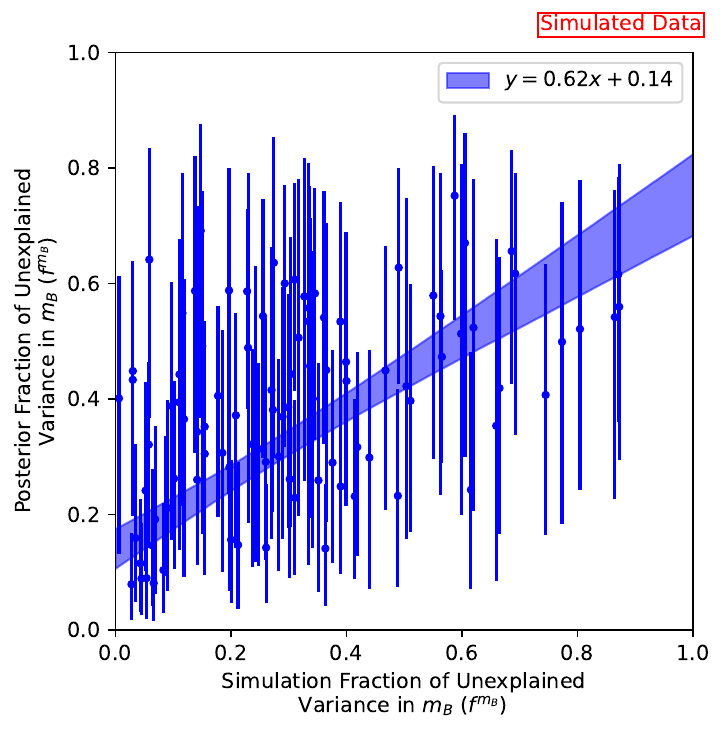}
    \caption{Recovery of the fraction of unexplained dispersion in $m_B$ (as opposed to $x_1$ or $c$) from simulated-data testing. The {\bf left panel} shows the results from 100 realizations that include low-$z$, mid-$z$, and high-$z$ datasets, while the {\bf right panel} shows results from the 100 realizations with just the mid-$z$ dataset. For each UNITY result, we summarize the posterior as the median (blue dots) and the 15.9th--84.1st percentile range (blue lines). We also show the best-fit linear relation and its uncertainties (this is fit assuming the posteriors are skew-normal distributions) as a blue shaded region. With simulated datasets of this size, the impact of the 3D simplex prior is clearly visible (i.e., the posteriors cannot go below 0 and are unlikely to approach 1), but the frequentist coverage is pretty good, especially for the larger, more realistic three-dataset simulations ({\bf left panel}). This in turn yields good but not perfect frequentist coverage for $\beta_B$, which correlates with the unexplained dispersion in color.
    \label{fig:simfmB}}
\end{figure*}

Next, we examine the three-simulated-dataset results on other parameters in Table~\ref{tab:SimulatedSummaryLHV}. In general, these are unbiased and show reasonable uncertainties, with two exceptions. 1) We find mild biases (mild in the sense of being much smaller than the uncertainties from any one realization) in the direction of UNITY1.5 putting too much unexplained dispersion in magnitude compared to color. This is further explored in Figure~\ref{fig:simfmB}. The bias for any one realization is modest, the uncertainties are fairly accurate, and it is encouraging that UNITY1.5 can recover the unexplained dispersion $m_B$/$x_1$/$c$ distribution at all. However, putting too much uncertainty into magnitude and not enough into color does bias $\beta_B$ downward, especially for the nominal UNITY1.5 model (UNITY1.5 with improved parameter limits on the red-color outlier model performs better). 2) The more statistically significant (but less practically significant) bias is that the outlier fraction is generally biased low. As we know which SNe were simulated as outliers, we know the true outlier fraction for each realization after selection cuts (the range of this true outlier fraction is quoted in the table). In general, this biased outlier fraction is of little practical significance in the following sense. For a 1D Gaussian mixture model (e.g., the original BEAMS, \citealt{Kunz2007}) with an inlier population width of 0.15 magnitudes and a centered, 2\% outlier population with width of 0.5 magnitudes, a SN is more likely to be an outlier than an inlier if it is more than 0.50 magnitudes from the central value. If a 1\% outlier fraction is assumed instead (a large factor of two change), the crossover value changes only mildly to 0.54 magnitudes. This would only result in large changes to the posterior probabilities of a few percent of the outliers (which would only be $\sim0.1\%$ of SNe overall). However, in the future, the outlier model may need refinement in order to include photometrically classified SNe in UNITY.

We next consider the cosmological results from the mid-$z$ simulated dataset, shown in Table~\ref{tab:SimulatedSummary}. $\Omega_m$ (the first row of values) looks excellent in both central values and uncertainties for UNITY1.5.\footnote{The comparison between UNITY and traditional simulation-based methods of removing bias will vary with the simulated dataset and the assumptions, but UNITY1.5 shows generally good performance. For example, the JLA analysis (which quotes each component of their covariance matrix) has an estimated uncertainty on their bias correction of about 50\% (although it is not clear how to assess statistical vs. systematic uncertainty on this 50\%).} When fit with a model assuming no selection effects, \Om is biased severely high (by $+0.2$) compared to the input value of 0.3. The selection-effect bias is actually enhanced by the BHM nature of UNITY; the assumption of a constant-in-redshift population distribution pushes the inferred colors (and to a smaller extent, the inferred $x_1$ values) of bluer SNe selected at high redshift towards the (lower-redshift, redder) mean. This results in under-standardization and thus an enhancement of the selection effects. Using a no-selection-effect model with redshift-dependent $x_1$ and $c$ populations avoids much of this bias (e.g., \citealt{Wood-Vasey2007, rubin16, RubinHeitlauf2020}) although arguably not enough. Such a model yields an average recovered \Om of 0.36 (last column).

Finally, we say a few words about the other parameters in Table~\ref{tab:SimulatedSummary}. Between the smaller sample size (600~SNe) and large selection effects, the $\beta$ values are not well measured and the $\beta_R$ values are biased by about $1\sigma$ (the posteriors are also not very Gaussian); this bias is highly significant with 100 realizations. If this were not an auxiliary ``stress-test'' of the model, it might be worth running simulated mid-$z$ datasets with more than 600 SNe to shrink (and presumably make more Gaussian) the posteriors to investigate how much of this is finite-sample effects and how much is UNITY1.5 biases in the presence of strong selection effects. However, we do not concern ourselves with this here, as the cosmological inference looks reasonable since the inferred \Om matches its input. Finally, we note the outlier fraction is also biased low, as it was for the three-simulated-dataset test previously described.

\movetabledown=5cm
\begin{rotatetable*}
\begin{deluxetable*}{lr|rr|rr}
\tablecaption{Simulated-data testing for the combined low-$z$, mid-$z$ and high-$z$ simulations}
\tablehead{\colhead{Parameter} & \colhead{Input} & \colhead{\fbox{Nominal UNITY1.5 Model}} & \colhead{Improved Outlier Limits} & \colhead{No Selection Effects} & \colhead{No Sel. Eff., $z$-Dep. Pop.} }
\startdata
\hline 
\multicolumn{6}{c}{Cosmology Parameters}\\ 
\hline 
$w_0$ & $-1.000$ & $-0.982 \pm 0.010 \pm 0.118 \ (0.90)$ & $-0.981 \pm 0.011 \pm 0.119 \ (0.91)$ & $-1.059 \pm 0.008 \pm 0.086 \ (1.13)$ & $-1.184 \pm 0.010 \pm 0.108 \ (1.90)$\\
$w_0 + 0.15\;w_a$ & $-1.000$ & $-0.988 \pm 0.005 \pm 0.051 \ (0.96)$ & $-0.987 \pm 0.005 \pm 0.051 \ (0.97)$ & $-0.952 \pm 0.003 \pm 0.037 \ (1.60)$ & $-1.019 \pm 0.005 \pm 0.050 \ (0.98)$\\
$w_a$ & $0.000$ & $-0.021 \pm 0.062 \pm 0.709 \ (0.88)$ & $-0.026 \pm 0.062 \pm 0.711 \ (0.88)$ & $0.721 \pm 0.042 \pm 0.479 \ (1.82)$ & $1.109 \pm 0.050 \pm 0.584 \ (2.14)$\\
\hline 
\multicolumn{6}{c}{Other Parameters}\\ 
\hline 
$\alpha$ & $0.150$ & $0.147 \pm 0.001 \pm 0.010 \ (1.04)$ & $0.147 \pm 0.001 \pm 0.011 \ (1.00)$ & $0.147 \pm 0.001 \pm 0.011 \ (1.03)$ & $0.145 \pm 0.001 \pm 0.011 \ (1.09)$\\
$\beta_B$ & $3.100$ & $2.899 \pm 0.027 \pm 0.246 \ (1.39)$ & $2.972 \pm 0.026 \pm 0.255 \ (1.20)$ & $2.747 \pm 0.037 \pm 0.257 \ (2.40)$ & $2.815 \pm 0.034 \pm 0.243 \ (2.03)$\\
$\beta_{RL}$ & $3.100$ & $3.122 \pm 0.021 \pm 0.196 \ (1.06)$ & $3.100 \pm 0.021 \pm 0.198 \ (1.07)$ & $3.125 \pm 0.039 \pm 0.217 \ (1.13)$ & $3.067 \pm 0.046 \pm 0.214 \ (1.20)$\\
$\beta_{RH}$ & $3.100$ & $3.109 \pm 0.020 \pm 0.224 \ (0.87)$ & $3.089 \pm 0.020 \pm 0.230 \ (0.87)$ & $3.083 \pm 0.048 \pm 0.241 \ (0.93)$ & $3.107 \pm 0.020 \pm 0.224 \ (0.88)$\\
$\delta(0)$ & $0.080$ & $0.080 \pm 0.002 \pm 0.017 \ (0.91)$ & $0.081 \pm 0.002 \pm 0.017 \ (0.92)$ & $0.079 \pm 0.002 \pm 0.017 \ (0.93)$ & $0.079 \pm 0.002 \pm 0.017 \ (0.94)$\\
$\delta(\infty)/\delta(0)$ & $\mathcal{U}(0,\ 1)$ & $0.484 \pm 0.017 \pm 0.267 \ (0.96)$ & $0.484 \pm 0.017 \pm 0.269 \ (0.97)$ & $0.499 \pm 0.020 \pm 0.244 \ (0.98)$ & $0.495 \pm 0.017 \pm 0.268 \ (0.99)$\\
$m_{50}$ Low-$z$ & \nodata & $17.885 \pm 0.009 \pm 0.072 \ ($\nodata$)$ & $17.883 \pm 0.009 \pm 0.072 \ ($\nodata$)$ & \nodata & \nodata\\
$\sigma_m$ Low-$z$ & \nodata & $0.233 \pm 0.004 \pm 0.029 \ ($\nodata$)$ & $0.234 \pm 0.004 \pm 0.029 \ ($\nodata$)$ & \nodata & \nodata\\
$\sigma^{\mathrm{unexpl}}$ Low-$z$ & $0.120$ & $0.119 \pm 0.000 \pm 0.006 \ (0.81)$ & $0.120 \pm 0.000 \pm 0.006 \ (0.83)$ & $0.117 \pm 0.001 \pm 0.006 \ (0.98)$ & $0.117 \pm 0.000 \pm 0.006 \ (0.91)$\\
$m_{50}$ Mid-$z$ & \nodata & $22.998 \pm 0.006 \pm 0.046 \ ($\nodata$)$ & $22.997 \pm 0.006 \pm 0.047 \ ($\nodata$)$ & \nodata & \nodata\\
$\sigma_m$ Mid-$z$ & \nodata & $0.178 \pm 0.003 \pm 0.023 \ ($\nodata$)$ & $0.179 \pm 0.003 \pm 0.023 \ ($\nodata$)$ & \nodata & \nodata\\
$\sigma^{\mathrm{unexpl}}$ Mid-$z$ & $0.120$ & $0.119 \pm 0.001 \pm 0.006 \ (1.13)$ & $0.120 \pm 0.001 \pm 0.006 \ (1.15)$ & $0.116 \pm 0.001 \pm 0.006 \ (1.28)$ & $0.117 \pm 0.001 \pm 0.006 \ (1.22)$\\
$m_{50}$ High-$z$ & \nodata & $26.040 \pm 0.016 \pm 0.248 \ ($\nodata$)$ & $26.041 \pm 0.016 \pm 0.248 \ ($\nodata$)$ & \nodata & \nodata\\
$\sigma_m$ High-$z$ & \nodata & $0.330 \pm 0.007 \pm 0.137 \ ($\nodata$)$ & $0.329 \pm 0.007 \pm 0.137 \ ($\nodata$)$ & \nodata & \nodata\\
$\sigma^{\mathrm{unexpl}}$ High-$z$ & $0.120$ & $0.118 \pm 0.002 \pm 0.022 \ (0.95)$ & $0.119 \pm 0.002 \pm 0.022 \ (0.96)$ & $0.117 \pm 0.002 \pm 0.022 \ (1.06)$ & $0.116 \pm 0.002 \pm 0.022 \ (1.06)$\\
$f^{m_B}$ & Simplex & $0.438 \pm 0.019 \pm 0.164 \ (1.31)$ & $0.404 \pm 0.018 \pm 0.172 \ (1.13)$ & $0.504 \pm 0.021 \pm 0.157 \ (1.88)$ & $0.487 \pm 0.020 \pm 0.156 \ (1.73)$\\
$f^{x_1}$ & Simplex & $0.314 \pm 0.022 \pm 0.066 \ (0.95)$ & $0.315 \pm 0.022 \pm 0.066 \ (0.94)$ & $0.324 \pm 0.023 \pm 0.071 \ (0.88)$ & $0.314 \pm 0.022 \pm 0.070 \ (0.92)$\\
$f^{c}$ & Simplex & $0.242 \pm 0.014 \pm 0.154 \ (1.27)$ & $0.276 \pm 0.015 \pm 0.161 \ (1.10)$ & $0.160 \pm 0.008 \pm 0.143 \ (2.05)$ & $0.191 \pm 0.011 \pm 0.143 \ (1.77)$\\
$f^{\mathrm{outl}}$ & 0.010--0.028 & $0.014 \pm 0.000 \pm 0.004 \ (1.57)$ & $0.013 \pm 0.000 \pm 0.003 \ (1.97)$ & $0.016 \pm 0.000 \pm 0.004 \ (1.28)$ & $0.015 \pm 0.000 \pm 0.004 \ (1.30)$\\
\enddata
\tablecomments{
\begin{minipage}[t]{1.22 \textwidth}
Summary of simulated-data testing using 100 realizations of low-, mid-, and high-$z$ simulated datasets. \simdatasetence. \simdatasetencetwo The first line shows the recovery of $w_0$, the second line shows an uncorrelated-with-$w_a$ $w_0 + 0.15\, w_a$ (often called $w_{\mathrm{pivot}}$), and the third line shows $w_a$; UNITY1.5 (shown in the first two columns) removes essentially all the bias seen in the models without treatment of selection effects (shown in the last two ``No Selection Effects'' columns). Figure~\ref{fig:simmeanresid} shows that much of the possible bias on $w_0 + 0.15\, w_a$ is due to small ($\lesssim$~1~mmag) biases in the light-curve fitting, rather than from UNITY.
\end{minipage}
\label{tab:SimulatedSummaryLHV}}
\end{deluxetable*}
\end{rotatetable*}

\movetabledown=5cm
\begin{rotatetable*}
\begin{deluxetable*}{lr|rr|rr}
\tablecaption{Simulated-data testing for the mid-$z$-only simulations (``stress-test'')}
\tablehead{\colhead{Parameter} & \colhead{Input} & \colhead{\fbox{Nominal UNITY1.5 Model}} & \colhead{Improved Outlier Limits} & \colhead{No Selection Effects} & \colhead{No Sel. Eff., $z$-Dep. Pop.} }
\startdata
\hline 
\multicolumn{5}{c}{Cosmology Parameters}\\ 
\hline 
$\Omega_m$ & $0.300$ & $0.303 \pm 0.004 \pm 0.042 \ (0.92)$ & $0.303 \pm 0.004 \pm 0.042 \ (0.92)$ & $0.516 \pm 0.004 \pm 0.047 \ (4.57)$ & $0.356 \pm 0.005 \pm 0.055 \ (1.25)$\\
\hline 
\multicolumn{5}{c}{Other Parameters}\\ 
\hline 
$\alpha$ & $0.150$ & $0.150 \pm 0.002 \pm 0.015 \ (1.01)$ & $0.150 \pm 0.002 \pm 0.016 \ (1.04)$ & $0.143 \pm 0.002 \pm 0.016 \ (1.13)$ & $0.145 \pm 0.002 \pm 0.016 \ (1.18)$\\
$\beta_B$ & $3.100$ & $3.004 \pm 0.037 \pm 0.302 \ (1.23)$ & $3.011 \pm 0.036 \pm 0.297 \ (1.23)$ & $2.842 \pm 0.036 \pm 0.329 \ (1.46)$ & $2.814 \pm 0.040 \pm 0.287 \ (1.88)$\\
$\beta_{RL}$ & $3.100$ & $2.440 \pm 0.117 \pm 0.750 \ (1.32)$ & $2.381 \pm 0.121 \pm 0.728 \ (1.35)$ & $2.553 \pm 0.103 \pm 0.590 \ (1.35)$ & $2.965 \pm 0.091 \pm 0.563 \ (1.16)$\\
$\beta_{RH}$ & $3.100$ & $2.257 \pm 0.137 \pm 0.809 \ (1.30)$ & $2.198 \pm 0.138 \pm 0.869 \ (1.28)$ & $2.578 \pm 0.122 \pm 0.682 \ (1.16)$ & $2.595 \pm 0.111 \pm 0.725 \ (1.15)$\\
$\delta(0)$ & $0.080$ & $0.078 \pm 0.003 \pm 0.037 \ (0.92)$ & $0.077 \pm 0.004 \pm 0.037 \ (0.96)$ & $0.077 \pm 0.003 \pm 0.035 \ (0.97)$ & $0.079 \pm 0.003 \pm 0.034 \ (0.93)$\\
$\delta(\infty)/\delta(0)$ & $\mathcal{U}(0,\ 1)$ & $0.437 \pm 0.012 \pm 0.311 \ (0.99)$ & $0.440 \pm 0.012 \pm 0.313 \ (0.99)$ & $0.433 \pm 0.012 \pm 0.309 \ (0.97)$ & $0.387 \pm 0.012 \pm 0.312 \ (1.19)$\\
$m_{50}$ & \nodata & $22.998 \pm 0.006 \pm 0.048 \ ($\nodata$)$ & $22.997 \pm 0.006 \pm 0.048 \ ($\nodata$)$ & \nodata & \nodata\\
$\sigma_m$ & \nodata & $0.179 \pm 0.003 \pm 0.023 \ ($\nodata$)$ & $0.180 \pm 0.003 \pm 0.023 \ ($\nodata$)$ & \nodata & \nodata\\
$\sigma^{\mathrm{unexpl}}$ & $0.120$ & $0.117 \pm 0.001 \pm 0.007 \ (1.17)$ & $0.117 \pm 0.001 \pm 0.007 \ (1.14)$ & $0.117 \pm 0.001 \pm 0.007 \ (1.10)$ & $0.116 \pm 0.001 \pm 0.006 \ (1.21)$\\
$f^{m_B}$ & Simplex & $0.397 \pm 0.016 \pm 0.207 \ (1.03)$ & $0.394 \pm 0.016 \pm 0.209 \ (1.04)$ & $0.407 \pm 0.016 \pm 0.203 \ (1.08)$ & $0.351 \pm 0.018 \pm 0.182 \ (1.22)$\\
$f^{x_1}$ & Simplex & $0.339 \pm 0.022 \pm 0.103 \ (0.89)$ & $0.332 \pm 0.022 \pm 0.104 \ (0.90)$ & $0.328 \pm 0.021 \pm 0.102 \ (0.94)$ & $0.351 \pm 0.021 \pm 0.101 \ (1.05)$\\
$f^{c}$ & Simplex & $0.243 \pm 0.010 \pm 0.192 \ (1.17)$ & $0.253 \pm 0.010 \pm 0.191 \ (1.19)$ & $0.246 \pm 0.010 \pm 0.190 \ (1.18)$ & $0.280 \pm 0.013 \pm 0.178 \ (1.34)$\\
$f^{\mathrm{outl}}$ & 0.003--0.031 & $0.013 \pm 0.000 \pm 0.004 \ (1.75)$ & $0.013 \pm 0.000 \pm 0.004 \ (1.83)$ & $0.014 \pm 0.000 \pm 0.004 \ (1.60)$ & $0.013 \pm 0.000 \pm 0.004 \ (1.73)$\\
\enddata
\tablecomments{
\begin{minipage}[t]{1.14 \textwidth}
Summary of the simulated-data testing using just the mid-$z$ simulated dataset. \simdatasetence. \simdatasetencetwo The first line shows the recovery of \Om (simulated with a value of 0.3); UNITY1.5 (shown in the first two columns of values) removes essentially all the bias seen in the models without treatment of selection effects (shown in the last two ``No Selection Effects'' columns).
\end{minipage}
\label{tab:SimulatedSummary}
}
\end{deluxetable*}
\end{rotatetable*}

\subsubsection{Predictive Posterior Distribution Checks} \label{sec:PPD}

\newcommand{\PPDConditional}{Specifically, we consider the PPD for each SN conditioned on the redshift and host-galaxy \Phigh (Equation~\ref{eq:Phigh}) for that SN.\xspace}

\newcommand{\PPDinterestingfinding}{One interesting finding hinted by the data is that the lowest-$x_1$ SNe are undercorrected by a single linear $\alpha$. This was also seen in \citet{rubin15b} but only affects a few percent of SNe, so it seems likely that a nonlinear $x_1$ standardization will only have a small effect and leave it out for computational simplicity.\xspace}

\newcommand{\PPDresultsentence}{In general, we see good agreement between the simulated distributions and observed distributions in all figures. This shows that the models of populations, standardization, and selection are working well and that UNITY1.5 generally has the fidelity to match the data.\xspace}

As a validation check to see if UNITY1.5 matches the real data well, we examine the predictive posterior distributions (PPDs). These are essentially simulated datasets conditioned on the data and the posterior of the parameters. \PPDConditional For each MCMC posterior sample, for each SN, we draw $x^{\mathrm{true}}_1$, $c^{\mathrm{true}}_B$, $c^{\mathrm{true}}_R$ from the population distributions for that SN's dataset, redshift, and \Phigh. Then we compute $m_B^{\mathrm{true}}$ using Equation~\ref{eq:tripp}. We add the $c_B^{\mathrm{true}}$ and $c_R^{\mathrm{true}}$ values to obtain $c$, and convolve the true \mBxc by appropriate measurement uncertainties (including the unexplained dispersion) using the interpolated light-curve-fit uncertainties as a function of magnitude and color described in Section~\ref{sec:lcfit}. We also include the self-calibration from the actual data; we add the sum in Equation~\ref{eq:derivatives} to the $m_B$, $x_1$, and $c$. We use the SALT model to synthesize peak observer-frame magnitudes and simulate whether or not the SN was found (brighter than the selection threshold). We record only $m_B$, $x_1$, $c$ when the simulated SN is selected, and we also record the selection probability over all trials. We note that the selection thresholds are both uncertain (Figure~\ref{fig:MagntiudeLimits}) and stochastic (as described in Section~\ref{sec:MalmquistBias}). We only simulate SNe with inlier posteriors greater than outlier posteriors, and do not draw from the outlier distributions in the PPDs.

\begin{figure*}[h!tbp]
\centering
    \includegraphics[width = 0.8 \textwidth]{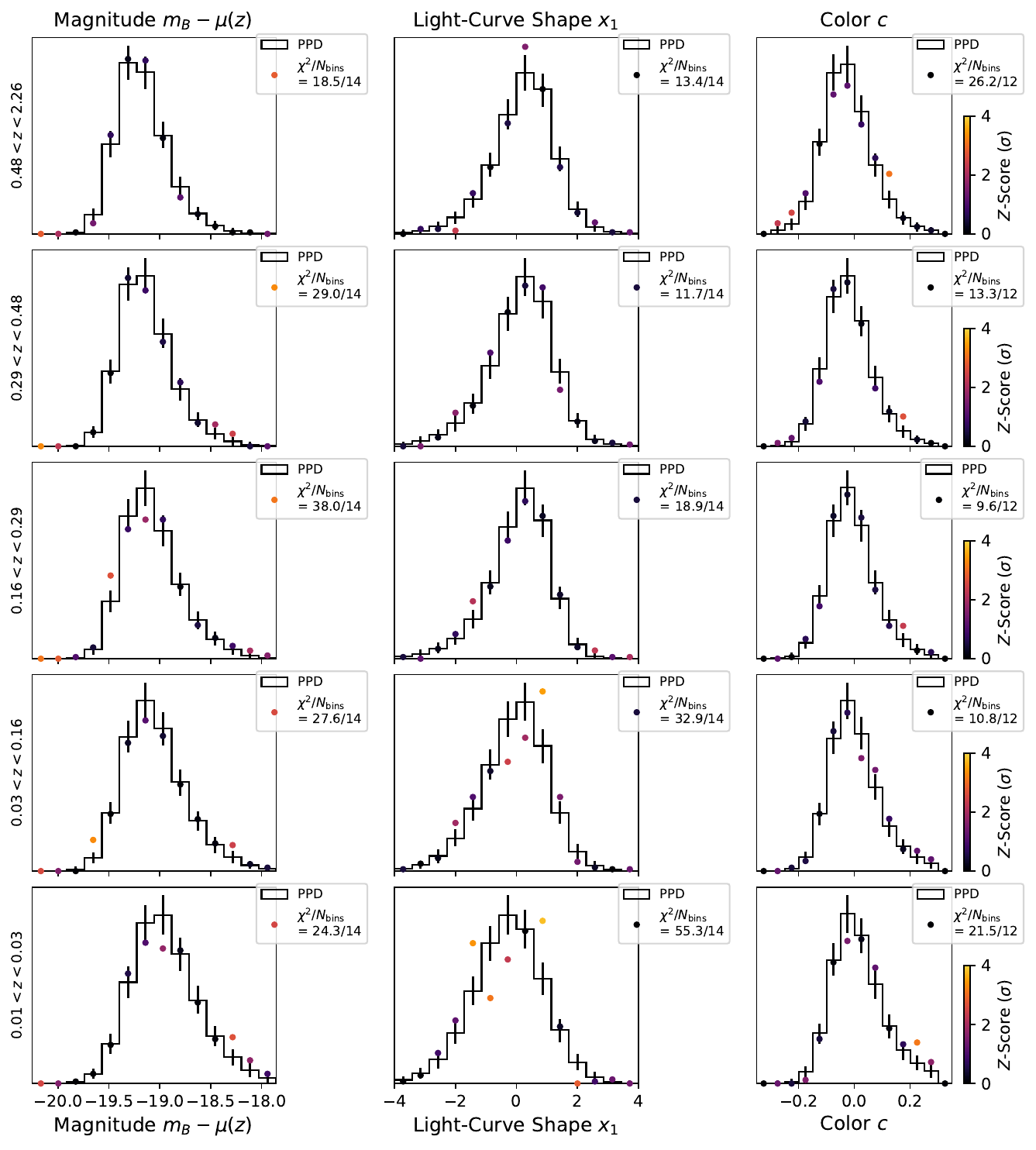}
    
\caption{This figure compares the predictive-posterior-distribution of Union3 + UNITY1.5 against the real data. Each column shows a light-curve-fit parameter: $m_B - \mu(z)$ ({\bf left panels}), $x_1$ ({\bf middle panels}) or $c$ ({\bf right panels}) compared to simulations based on the UNITY1.5 posterior. {\bf Each row} shows a different range in redshift. The histograms (with Poisson uncertainties) are the predictive-posterior distributions and the points are the data. Each panel shows the Poisson-equivalent $\chi^2$ and the number of bins, with the equivalent $Z$ score for each bin indicated by the plot point color. Broadly speaking, these results show that UNITY1.5's population, standardization, and selection-effects models are working well, with some of the larger $\chi^2$ values due to bins with very few events or the mild bimodality in $x_1$ at low redshift noted in the text.\label{fig:PPDz}}
\end{figure*}

\begin{figure*}[h!tbp]
\centering
    \includegraphics[width = 0.8 \textwidth]{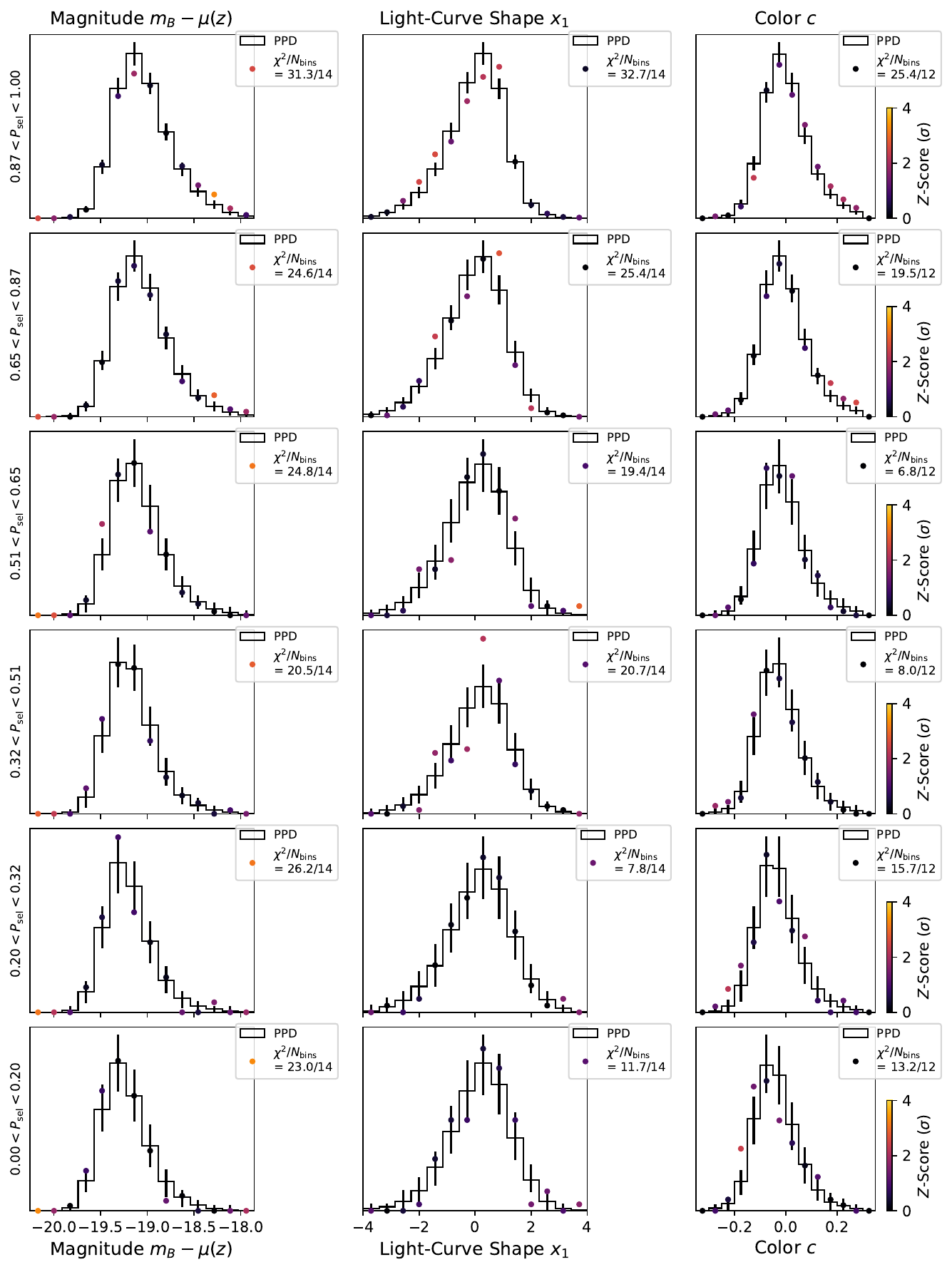}
    
\caption{This figure compares the predictive-posterior-distribution of Union3 + UNITY1.5 against the real data. Each column shows a light-curve-fit parameter: $m_B - \mu(z)$ ({\bf left panels}), $x_1$ ({\bf middle panels}) or $c$ ({\bf right panels}) compared to simulations based on the UNITY1.5 posterior. {\bf Each row} shows a different range in selection probability. The histograms (with Poisson uncertainties) are the predictive-posterior distributions and the points are the data. Each panel shows the Poisson-equivalent $\chi^2$ and the number of bins, with the equivalent $Z$ score for each bin indicated by the plot point color. Again, these results show that UNITY1.5's population, standardization, and selection-effects models are working well, with some of the larger $\chi^2$ values due to bins with very few events.\label{fig:PPDsel}}
\end{figure*}

\begin{figure*}[h!tbp]
\centering
    \includegraphics[width = 0.8 \textwidth]{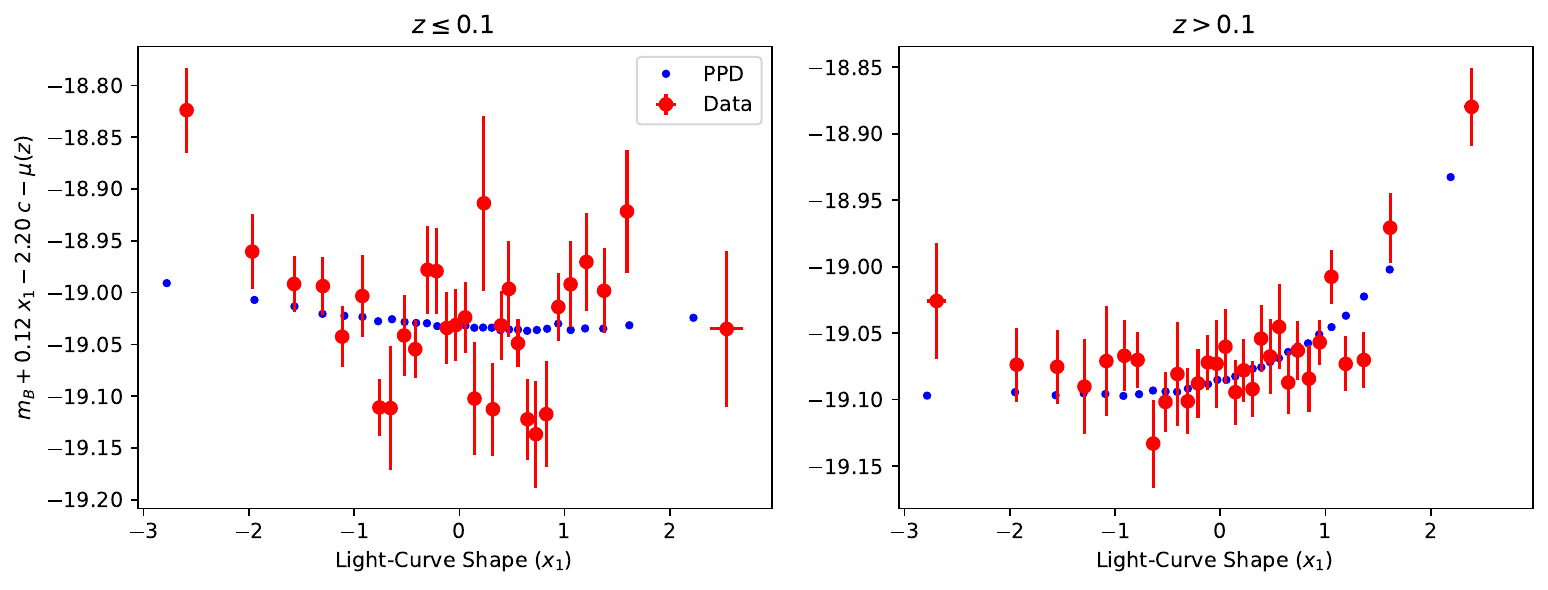}
    
        \includegraphics[width = 0.8 \textwidth]{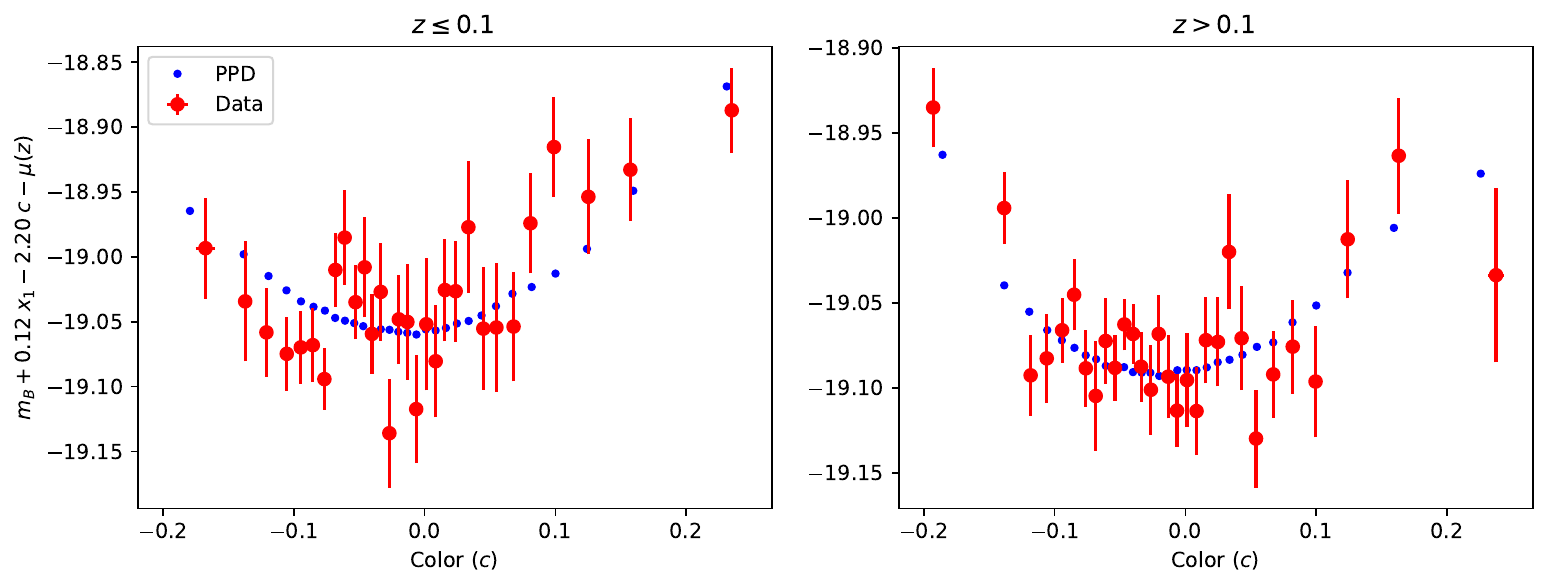}

\caption{This figure compares the predictive-posterior-distribution of Union3 + UNITY1.5 against the real data. Each panel shows a \citet{Tripp1998}-like Hubble residual binned in $x_1$ ({\bf top panels}) or $c$ ({\bf bottom panels}), with the data in larger red points (with uncertainties) and the predictive-posterior-distribution in smaller blue points. We divide the data into 30 roughly equally sized bins. We show separate results for $z \leq 0.1$ ({\bf left panels}) and $z > 0.1$ ({\bf left panels}) to help show any mismatches that might bias the cosmology. These binned plots are affected by a type of Eddington bias (called regression dilution) due to uncertainties in $x_1$ and $c$, so the indicated relations are biased towards zero compared to $\alpha$ and the $\beta$'s. For the same reason, the relations can appear nonlinear, even for $x_1$ where UNITY1.5 models the relation linearly. But the relations should be biased similarly for real data and the predictive-posteriors and generally speaking, they are. Impressively, one can see that UNITY1.5 correctly models different observed relations as a function of redshift due to the different population distributions and signal to noise. Again, these results show that UNITY1.5's population, standardization, and selection-effects models are working well. \PPDinterestingfinding \label{fig:PPDHR}}
\end{figure*}

\begin{figure*}[h!tbp]
\centering
    \includegraphics[height = 0.9 \textheight]{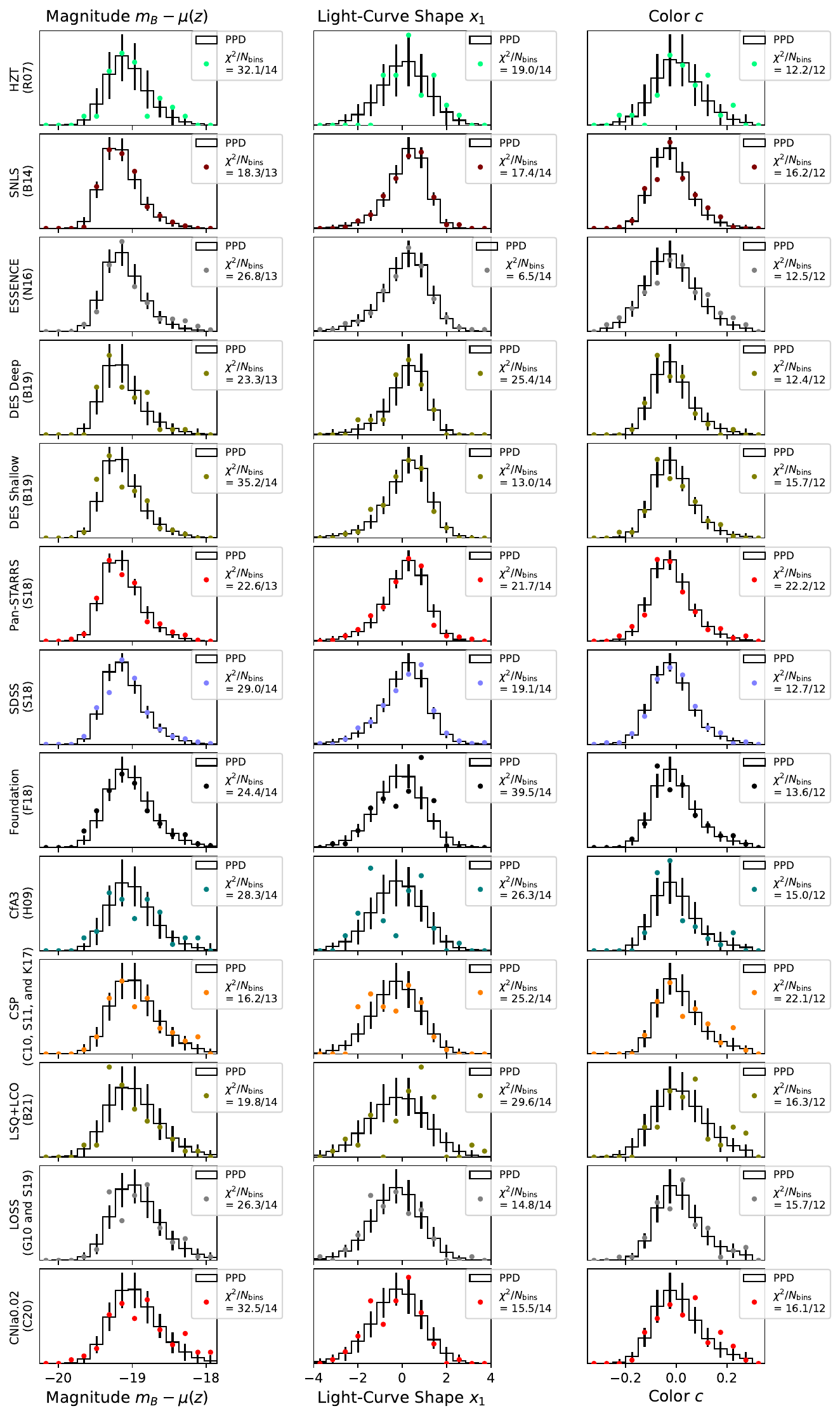}
\caption{This figure compares the predictive-posterior-distribution of Union3 + UNITY1.5 against the real data. The {\bf left panels} show the comparisons for magnitude with the modeled flat-$\Lambda$CDM distance modulus subtracted, the {\bf center panels} show the comparisons for $x_1$, and the {\bf right panels} show the comparisons for $c$. Each row shows a different SN sample (only samples with at least 30 SNe are shown). The histograms (with Poisson uncertainties) are the predictive-posterior distributions and the points are the data. Each panel shows the Poisson-equivalent $\chi^2$ and the number of bins. Again, these results show that UNITY1.5's population, standardization, and selection-effects models are working well. \label{fig:PPDvssamp}}
\end{figure*}

There are several comparisons of interest and Figures~\ref{fig:PPDz} through \ref{fig:PPDvssamp} go through these comparisons in detail for $m_B - \mu(z)$, $x_1$, and $c$. Figure~\ref{fig:PPDz} shows the comparison as a function of redshift, finding good consistency albeit with minor discrepancies at low redshift for $x_1$ discussed below. Figure~\ref{fig:PPDsel} shows the comparison as a function of selection probability, again showing good consistency. Figure~\ref{fig:PPDHR} shows binned \citet{Tripp1998}-like Hubble residuals as a function of $x_1$ and $c$ for both $z\leq0.1$ and $z > 0.1$ to investigate standardization in more detail. \PPDinterestingfinding Finally, Figure~\ref{fig:PPDvssamp} shows the PPD by SN sample. \PPDresultsentence\footnote{We note that the $\chi^2$ values and binned Hubble residual uncertainties shown on these predictive-posterior figures are quoted assuming no correlations bin-to-bin, even though such correlations will exist due to uncertainty on the parameters and systematics that correlate SNe. It is not trivial to put  uncertainties on the predictive-posterior distributions, as different SNe will be discovered/pass cuts in different MCMC samples. So the covariance is not straightforward to evaluate because there are likely zero samples with all the SNe present. Of course, this challenge only affects the predictive posteriors; UNITY1.5 does propagate correlated uncertainties into the cosmology fits.}

One moderately significant result for a few of the low-$z$ samples dominated by old galaxies is the bimodality visible in $x_1$ that is not captured by UNITY1.5 \citep{Nicolas2021}. However, this has little impact on UNITY, as the first few moments of the distribution are approximated well (which are the most important, \citealt{rubin15b}). In fact, in the limit of linear standardization, Gaussian uncertainties, and a complete sample, only inferring the mean and variance of the Gaussian population distribution correctly is important. This holds true even for very non-Gaussian independent-variable (e.g., $x_1$) distributions.

\clearpage

\section{Cosmological Results} \label{sec:CosmologyConstraints}

With Union3 and UNITY1.5 in hand, we unblinded the SN cosmological constraints. We also combine with other cosmological probes as described here. Section~\ref{sec:uncertaintyanalysis} examines the Union3+UNITY1.5 uncertainty decomposition, showing the correlations between cosmological parameters and other SN parameters in the analysis. Section~\ref{sec:CosmoProbeLike} describes our treatment of external data and its combination with SNe. Finally, Section~\ref{sec:cosmoexternalconstraints} shows our parameter constraints for each combination of cosmological model and data.

\subsection{Uncertainty Analysis} \label{sec:uncertaintyanalysis}

\newcommand{\cosmoparam}{\mathrm{cosmo\ parameter}\xspace}
\newcommand{\partialcosmoparam}{\frac{\partial \, \cosmoparam}{\partial \, \mathrm{parameter}}}

Unlike most other cosmology analyses, we look at the full impact of all parameters (including those traditionally classified as statistical or systematic) on cosmology parameters. We define this impact as:
\begin{equation}
     \partialcosmoparam \sigma_{\mathrm{parameter}} \;,
\end{equation}
where $\partialcosmoparam$ is computed assuming Gaussian uncertainties:
\begin{equation}
    \partialcosmoparam = \frac{\mathrm{Cov}(\cosmoparam,\ \mathrm{parameter})}{ \sigma_{\mathrm{parameter}}^2  }
\end{equation}
so the impact simplifies to:
\begin{equation} \label{eq:impact}
    \frac{\mathrm{Cov}(\cosmoparam,\ \mathrm{parameter})}{\sigma_{\mathrm{parameter}}} \;.
\end{equation}
This equation ignores correlations between parameters. For example, uncertainty on the fundamental calibration impacts SN distances and therefore cosmology, but it also impacts the absolute magnitude \scriptM (defined in Equation~\ref{eq:scriptM}), which impacts cosmology. Thus we do not expect the quadrature sum of all impacts to be exactly equal to the uncertainty on a cosmological parameter (it will, in general, be a bit larger).

Table~\ref{tab:UncertainyAnalysis} presents each impact on flat-universe $\Omega_m$ (or equivalently, $\OL$), ordered from largest to smallest, the fraction of total variance for each impact, and the cumulative fraction of variance for each. A variety of sources dominate the uncertainty budget. First is the absolute magnitude (\scriptM), which is related to the number of available nearby SNe, how well calibrated they are, and how well their selection functions are understood. Photometric calibration directly makes up two out of the top four slots. But astrophysical effects matter as well: intergalactic dust, peculiar velocities, and host-galaxy correlations all have an impact.

\begin{deluxetable*}{cccr}[h!tbp]
\tablehead{\colhead{Impact (Units of $\Omega_m$)} & \colhead{Fraction of Variance} & \colhead{Cumulative}  & \colhead{Term}} 
\startdata
0.0114	&	0.166	&	0.166	&	Absolute Magnitude (for $h = 0.7$) $\scriptM$ \\
0.0095	&	0.115	&	0.281	&	Intergalactic Dust \\
0.0093	&	0.111	&	0.392	&	SALT Calibration \\
0.0079	&	0.081	&	0.473	&	Selection Effects \\
0.0072	&	0.066	&	0.540	&	Instrument Zeropoints \\
0.0063	&	0.051	&	0.591	&	Peculiar-Velocity Correlations \\
0.0063	&	0.051	&	0.642	&	$\delta(z = \infty)$ \\
0.0058	&	0.043	&	0.685	&	Color Population \\
0.0058	&	0.043	&	0.728	&	White-Dwarf-SED Model \\
0.0055	&	0.039	&	0.767	&	Outlier Model \\
0.0054	&	0.038	&	0.805	&	Instrument Passbands \\
0.0054	&	0.038	&	0.843	&	$\beta_R \equiv 0.5(\betaRH + \betaRL)$ \\
0.0050	&	0.032	&	0.875	&	Unexplained Scatter \\
0.0047	&	0.028	&	0.903	&	$\delta(z = 0)$ \\
0.0046	&	0.028	&	0.931	&	Milky Way Extinction \\
0.0046	&	0.027	&	0.958	&	Landolt/Smith/PS1 Calibration \\
0.0041	&	0.021	&	0.980	&	$\beta_B$ \\
0.0027	&	0.009	&	0.989	&	$x_1$ Population \\
0.0020	&	0.005	&	0.994	&	$\alpha$ \\
0.0016	&	0.003	&	0.997	&	$\Delta \beta_R \equiv \betaRH - \betaRL$ \\
0.0013	&	0.002	&	0.999	&	Weak Lensing Bias \\
0.0006	&	$ < 0.001$	&	1.000	&	Redshift Calibration \\
0.0004	&	$ < 0.001$	&	1.000	&	Electron Scattering \\
\enddata
\caption{ Uncertainty decomposition on flat $\Lambda$CDM \Om (or equivalently, \OL) following Equation~\ref{eq:impact} for SNe alone. We order each impact from largest to smallest.
\label{tab:UncertainyAnalysis}
}
\end{deluxetable*}

To investigate how the uncertainty decomposition changes for different cosmological parameters, Table~\ref{tab:UncertainyAnalysiswa} shows the uncertainty decomposition for $w_a$ when also including BAO and CMB data (Section~\ref{sec:cosmoexternalconstraints}). The importance of terms relevant for low-redshift SNe actually rises compared to the decomposition of \Om. Evidently, some of the constraint on flat-universe \Om comes from mid-redshift SNe that are used more to help measure $w_a$ now that additional cosmological parameters are included.

\begin{deluxetable*}{cccr}[h!tbp]
\tablehead{\colhead{Impact (Units of $w_a$)} & \colhead{Fraction of Variance} & \colhead{Cumulative}  & \colhead{Term}} 
\startdata
0.1321	&	0.274	&	0.274	&	Absolute Magnitude (for $h = 0.7$) $\scriptM$ \\
0.1271	&	0.253	&	0.527	&	Peculiar-Velocity Correlations \\
0.0673	&	0.071	&	0.598	&	SALT Calibration \\
0.0587	&	0.054	&	0.652	&	$\delta(z = \infty)$ \\
0.0572	&	0.051	&	0.704	&	Instrument Passbands \\
0.0567	&	0.050	&	0.754	&	Selection Effects \\
0.0502	&	0.039	&	0.793	&	Landolt/Smith/PS1 Calibration \\
0.0467	&	0.034	&	0.828	&	Instrument Zeropoints \\
0.0441	&	0.031	&	0.858	&	Color Population \\
0.0393	&	0.024	&	0.882	&	$\delta(z = 0)$ \\
0.0383	&	0.023	&	0.905	&	$\beta_R \equiv 0.5(\betaRH + \betaRL)$ \\
0.0366	&	0.021	&	0.926	&	Unexplained Scatter \\
0.0354	&	0.020	&	0.946	&	Intergalactic Dust \\
0.0274	&	0.012	&	0.958	&	Milky Way Extinction \\
0.0259	&	0.011	&	0.968	&	White-Dwarf-SED Model \\
0.0246	&	0.009	&	0.978	&	Outlier Model \\
0.0216	&	0.007	&	0.985	&	$x_1$ Population \\
0.0192	&	0.006	&	0.991	&	$\alpha$ \\
0.0180	&	0.005	&	0.996	&	$\beta_B$ \\
0.0118	&	0.002	&	0.998	&	Redshift Calibration \\
0.0110	&	0.002	&	1.000	&	$\Delta \beta_R \equiv \betaRH - \betaRL$ \\
0.0004	&	$ < 0.001$	&	1.000	&	Electron Scattering \\
0.0001	&	$ < 0.001$	&	1.000	&	Weak Lensing Bias \\
\enddata
\caption{ Uncertainty decomposition on $w_a$ following Equation~\ref{eq:impact} including BAO and CMB constraints. We order each impact from largest to smallest.
\label{tab:UncertainyAnalysiswa}
}
\end{deluxetable*}

\clearpage

\newcommand{\coefficentcorrelationsentence}{These parameters are all somewhat correlated with $\Omega_m$. $\Delta \beta_R$ is more strongly non-zero than $\delta(0)$, indicating stronger evidence for a correlation between $\beta_R$ and host-galaxy mass than between standardized luminosity and host-galaxy mass. We also see evidence of nonlinear $\beta$ with color, i.e., $\beta_B < \beta_R$. \xspace}

\newcommand{\uncertaintycorrelationsentence}{None of these parameters show a strong correlation with $\Omega_m$.\xspace}

\begin{figure*}[h!tbp]
\centering
\includegraphics[width=\textwidth]{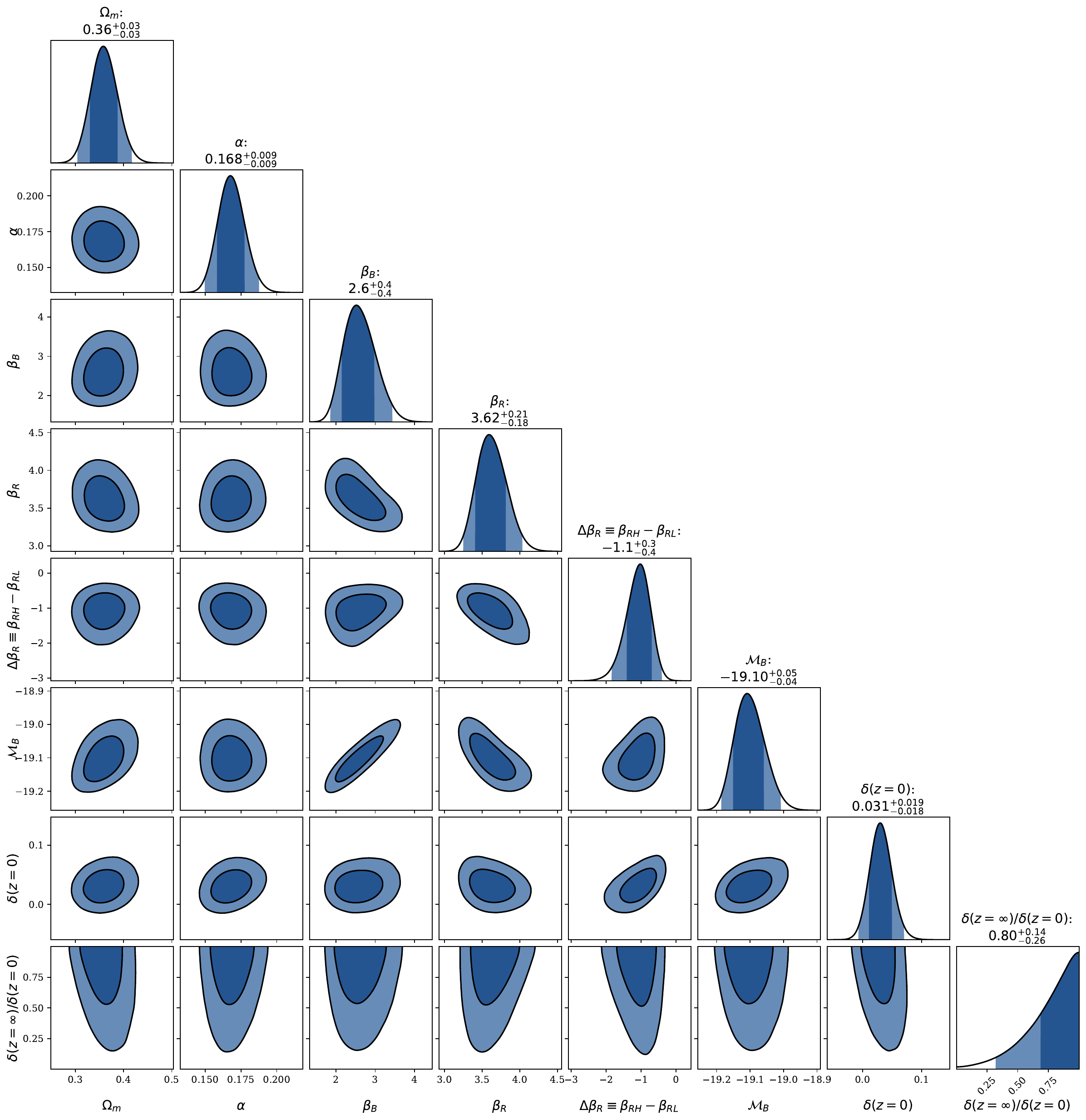}
\caption{Correlations between $\Omega_m$ and the standardization coefficients. We show the $x_1$ standardization parameter ($\alpha$), the blue-color standardization parameter ($\beta_B$), the mean (across high and low host mass) red-color standardization parameter ($\beta_R$), the difference in $\beta_R$ between high and low host mass ($\Delta \beta_R$), the Hubble-constant-free absolute magnitude ($\mathcal{M}_B$), the host-mass standardization parameter at low redshift ($\delta(0)$), and the relative host-mass standardization parameter at high redshift ($\delta(\infty)/\delta(0)$). \coefficentcorrelationsentence \label{fig:cornercoeffs}}
\end{figure*}

Figure~\ref{fig:cornercoeffs} shows the correlations between $\Omega_m$ and the standardization coefficients assuming SN data alone and flat \LCDM. As in \cite{rubin15b}, we make this figure by running a KDE of the MCMC samples and solving for the least-area contours that enclose 68.3\% or 95.4\% of the posterior. For each parameter, we also show the 1D posterior with the least-length shading that encloses 68.3\% or 95.4\%. We also list 1D numerical credible intervals based on the 50th percentile, 15.9th, and 84.1st (for a Gaussian posterior, this is the same as the least-length shading). 

First, it is worth noting that we find reasonably strong evidence that all three $\beta$ values are different from each other, with $\beta_R$ larger in low-stellar-mass host galaxies and $\beta_B$ much lower than the mass-averaged $\beta_R$. Our $\beta_B$ and $\beta_R$ values are similar to those of \citet{Mandel2017} and higher than those found using the first version of SALT2-1 by \citet{amanullah10}, \citet{suzuki12}, and \citet{rubin15b}. The difference in $\beta_R$ with mass has been previously seen in other models \citep{sullivan10, Sullivan2011, Brout2021} at about the same size.

It is also worth noting that we find greater evidence for a non-zero host-mass/$\beta_R$ correlation ($\Delta \beta_R$ non-zero) than evidence for a host-mass luminosity correlation ($\delta(0)$ non-zero). We also find no evidence of redshift evolution of the host-galaxy correlations ($\delta(\infty)/\delta(0)$ consistent with 1) although our constraints are weak.

\newcommand{\unexplainedallocation}{We find that about half the unexplained dispersion should be in color and significantly less in magnitude, roughly in between the \citet{guy10} and \citet{Chotard2011} models as converted to SALT magnitude and color by \citet{Kessler2013}.\xspace}

\begin{figure*}[h!tbp]
\centering
\includegraphics[width=\textwidth]{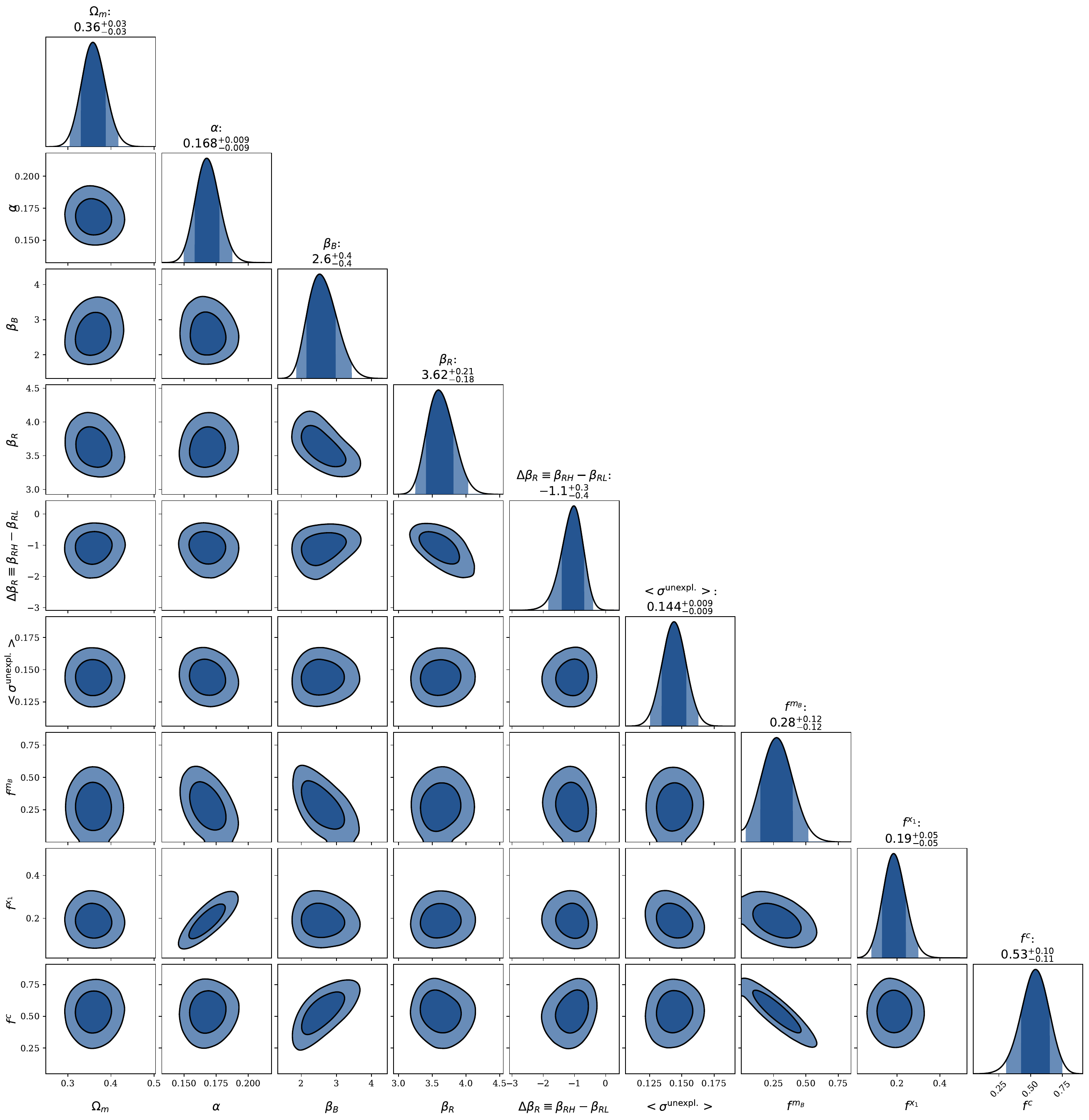}
\caption{Correlations between some of the standardization coefficients and the fraction of the unexplained dispersion in $m_B$, $x_1$, and $c$. As one would expect even in a frequentist analysis, assigning more uncertainty to $x_1$ increases $\alpha$ and assigning more uncertainty to $c$ increases $\beta_B$. The $\beta_R$ values (for redder SNe) are less affected. We also show the overall unexplained dispersion ($\sigma^{\text{unexplained}}$). Each SN sample has its own value, so this is the average over all samples. \label{fig:cornerstandardization}}
\end{figure*}

Figure~\ref{fig:cornerstandardization} shows the correlations between standardization coefficients and the fraction of unexplained dispersion in $m_B$, $x_1$, and $c$. As expected, it shows strong correlations between $\alpha$ and the fraction in $x_1$ and $\beta_B$ and the fraction in $c$. These correlations support UNITY's approach of marginalizing over the unexplained dispersion distribution as opposed to picking discrete possibilities and only considering those (e.g., picking the \citealt{guy10} or the \citealt{Chotard2011} models). \unexplainedallocation

\begin{figure*}[h!tbp]
\centering
\includegraphics[width=\textwidth]{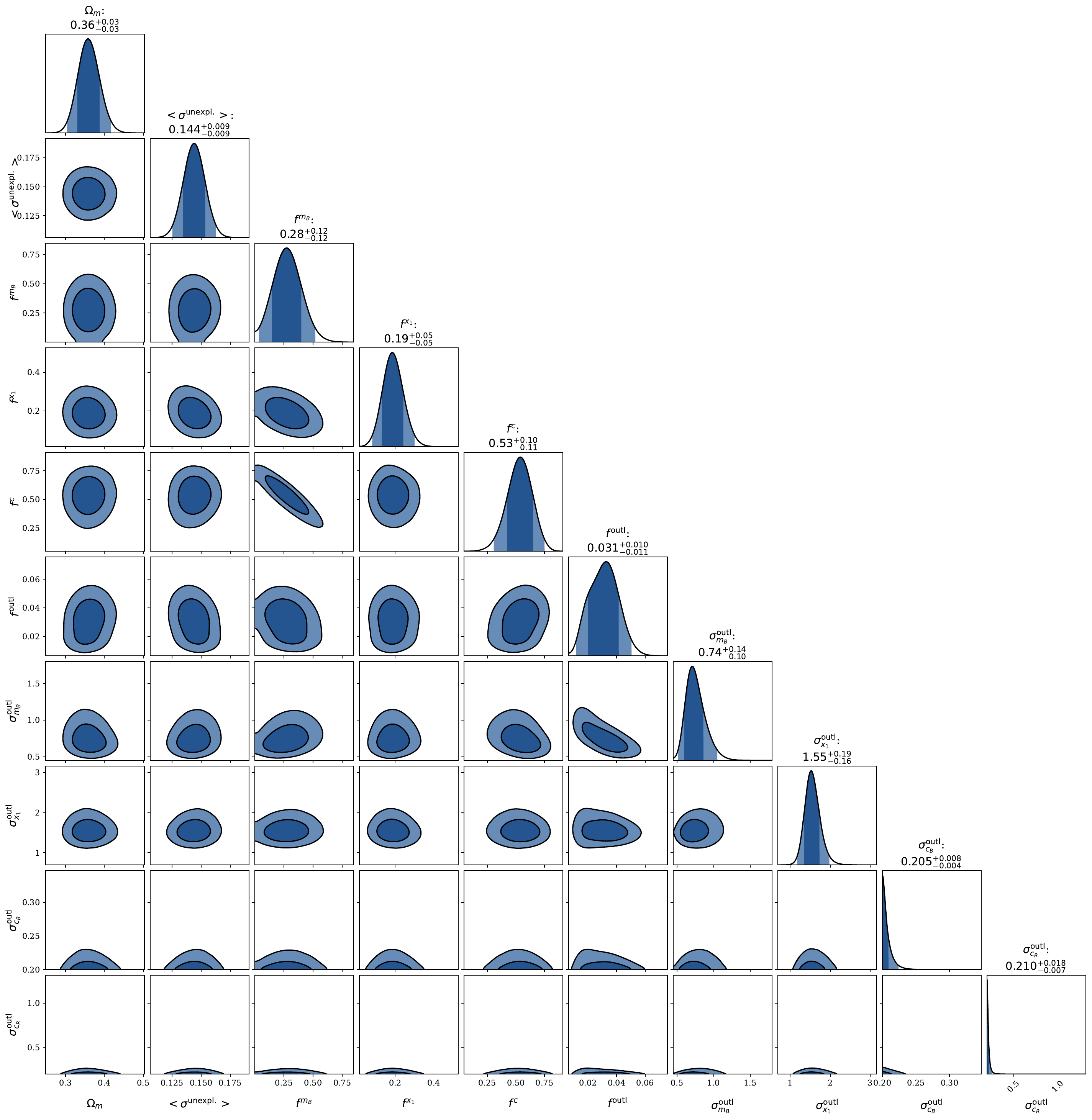}
\caption{Correlations between $\Omega_m$ and parameters relating to uncertainties and outliers. Each SN sample has its own $\sigma^{\text{unexplained}}$, so we show the average over all samples. We also show the fraction of the unexplained dispersion in $m_B$, $x_1$, and $c$. Finally, we show the outlier fraction and the estimated population dispersion of the outliers in $m_B$, $x_1$, and $c$. \uncertaintycorrelationsentence $c_B^{\mathrm{outl}}$ reaches its lower bound of $\ge 0.2$ in our default analysis; we note that changing the limit to $\ge 0.1$ has virtually no impact on any cosmological parameter. Changing the $c_R^{\mathrm{outl}}$ limit has a minor impact discussed in Appendix~\ref{sec:Priors}. \label{fig:corneruncertainties}}
\end{figure*}

Figure~\ref{fig:corneruncertainties} shows the correlations between $\Omega_m$ and the parameters relating to uncertainties and outliers. \uncertaintycorrelationsentence

We note that we have also investigated the non-cosmological parameters in Figures~\ref{fig:cornercoeffs}, \ref{fig:cornerstandardization}, and \ref{fig:cornerstandardization} when including external BAO + CMB constraints inside UNITY and using a flat $w_0$-$w_a$ cosmological model. None of these parameters change by any more than a modest fraction of their uncertainties.

\subsection{Cosmological Probe Likelihoods} \label{sec:CosmoProbeLike}

The following sections discuss the cosmological probes used in this analysis in more detail. As this work focuses on SN~Ia distance as a function of redshift, we only include other cosmology constraints that affect distances (and not, e.g., growth-of-structure constraints). We make our SN distances available for examining cosmological models not considered here or combining with other datasets at \url{https://doi.org/10.5281/zenodo.14090777}.

\subsubsection{SNe Ia}

\begin{figure}[h!tbp]
\centering
\includegraphics[width=0.5 \textwidth]{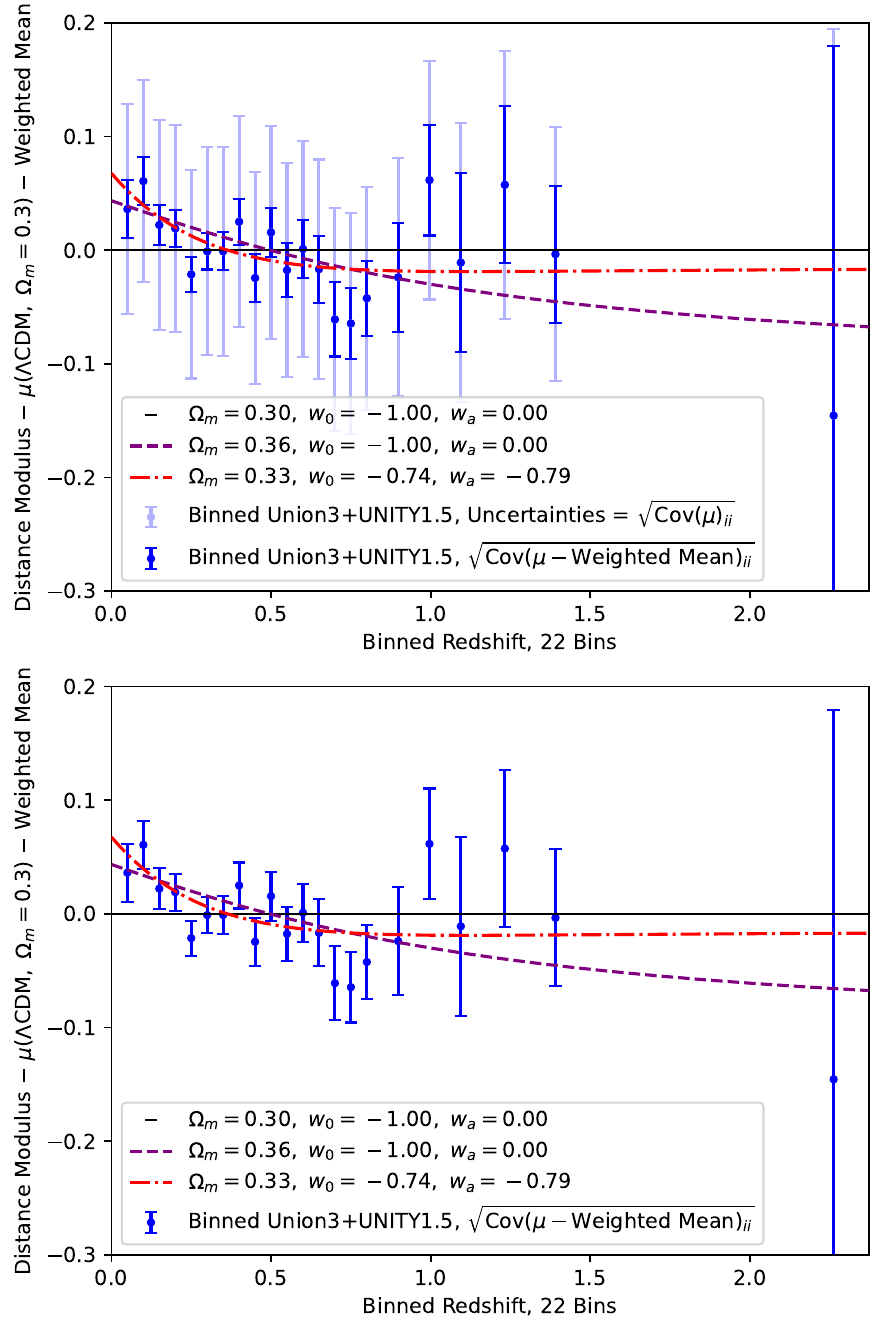}
\caption{The UNITY1.5-inferred spline-interpolated distances used in the frequentist portion of our cosmology analysis and released to the community. We show the uncertainties from the square root of the covariance of the UNITY1.5 posterior samples in light blue in the {\bf top panel} only. However, some of the covariance is in common among all the distance moduli; this covariance does not impact our cosmological analysis (as it is degenerate with \scriptM) but inflates the displayed uncertainties. Thus we also show the smaller uncertainties computed by subtracting the mean [distance modulus $-\ \mu(\Omega_m = 0.3)$] from each posterior sample in a darker blue ({\bf both panels}). The panels are otherwise identical. The deviation from $\Omega_m = 0.3$ is visible and we also show the distance moduli from two other cosmological models which match the data better (Table~\ref{tab:cosmoconstraints}): the best-fit SN-only flat-\LCDM model and the best-fit $w_0$-$w_a$ model using SNe+BAO+CMB. \label{fig:binneddistances}}
\end{figure}

The full UNITY1.5 analysis can infer cosmological parameters directly from the light-curve fits. However, most members of the community would prefer distance moduli as a function of redshift with a covariance matrix for frequentist analyses. We thus experiment with using UNITY1.5 to construct a smooth distance-modulus-as-a-function-of-redshift representation that can accurately reproduce the cosmological information in the data. For this frequentist purpose, we fit additive, interpolated perturbations as a function of redshift from a flat \LCDM model with $\Om=0.3$. We use redshift nodes in steps of 0.05 up to $z=0.8$, then redshift steps of 0.1 (or every 10 SNe, whichever is greater in redshift) for a total of 22 non-zero nodes. UNITY1.5 interpolates this perturbation with a second-order spline that is fixed to zero at $z=0$. After drawing 500,000 samples from the posterior, we compute the median posterior of the distance modulus and the covariance matrix. Figure~\ref{fig:binneddistances} shows these nodes and the distance-modulus perturbations. (Adding these perturbations to the $\Om = 0.3$ distance moduli produces the Union3+UNITY1.5 distances.)  Throughout the rest of this paper, we refer to these as ``spline-interpolated distance moduli'' and the analyses using them as ``frequentist analyses'' but they are based on Bayesian analyses underneath and as our cross check below shows, reproduce the full UNITY1.5 results.

As a cross-check that our spline captures the full UNITY analysis, we examine the SN-only \Om-$w$ contours as computed directly by putting $w$CDM in UNITY1.5, comparing with the frequentist contours from the spline-interpolated distance moduli. We draw 500,000 samples out of the $w$CDM posterior (400 chains with 1,250 discarded burn-in samples and 1,250 saved samples) and evaluate $\exp(-\chi^2/2)$ for each posterior sample, where $\chi^2$ is computed from the spline-interpolated distance moduli $\mu_i$ as:
\begin{eqnarray}
       & \chi^2_{\mathrm{SNe}} = \vec{r}^T \cdot C^{-1} \cdot \vec{r} \\
       & \vec{r} \equiv \mu(z_i,\ \Om,\ w) - \mu_i - \Delta\scriptM \nonumber
\end{eqnarray}
and $\Delta\scriptM$ absorbs the change in \scriptM with cosmological parameters. We find excellent agreement between the density of posterior samples and the $\exp(-\chi^2/2)$ values, even out into the tails. We thus conclude that our spline-interpolated distances and their covariance matrix are an accurate representation of the cosmological information in Union3+UNITY1.5. This is expected from the central limit theorem, as even though the posterior for one SN is not Gaussian, the posterior will become more Gaussian when averaging over many SNe.

\newcommand{\PPSentence}{The uncertainties especially in \OL are much smaller for Pantheon+, giving a contour with \UThreeVsPP of the area of Union3+UNITY1.5. We note that about 97\% of Pantheon+ SNe were considered for Union3 (although Union3 is about 1/3 larger), so the differences are not due to sample size.\xspace}
\newcommand{\Pantheonending}{to the Pantheon+ contour, indicating that UNITY1.5 assesses uncertainties much more conservatively than the Pantheon+ \BBC analysis. Future work should make a more direct comparison on the same input data or simulations to investigate this difference.\xspace}

\begin{figure}[h!tbp]
    \centering
    \includegraphics[width = 0.3 \textwidth]{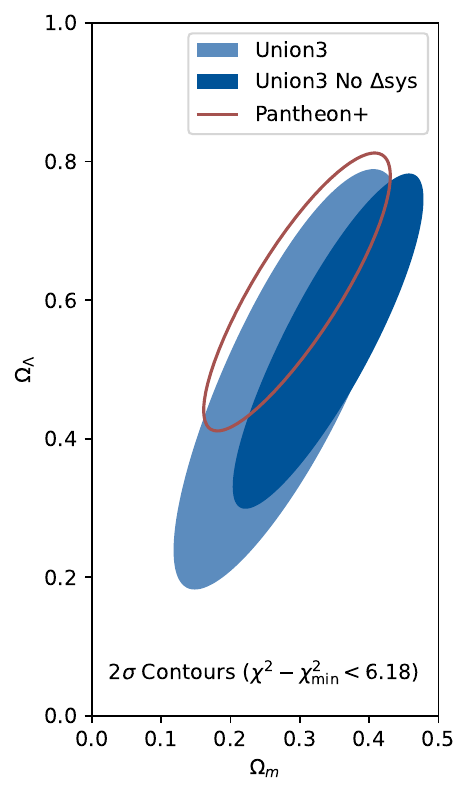}
    \caption{Comparison of Union3+UNITY1.5 (this work) and Pantheon+ \citep{Brout2022Cosmology} 2$\sigma$ confidence contours in the \Om-\OL plane. \PPSentence Also shown is the contour from Union3+UNITY1.5 with all small perturbative systematic terms (the \DZP terms in Section~\ref{sec:SmallPerturbations}) fixed to zero (dark blue area). Even this contour is of comparable size \Pantheonending}
    \label{fig:PPCompare}
\end{figure}

Figure~\ref{fig:PPCompare} shows a comparison of the Union3+UNITY1.5 (this work) and the Pantheon+ \citep{Brout2022Cosmology} 2$\sigma$ confidence contours in the \Om-\OL plane. \PPSentence Looking at our uncertainty analyses (Tables~\ref{tab:UncertainyAnalysis} and \ref{tab:UncertainyAnalysiswa}) shows some of the reasons for this change in assessed constraining power. Intergalactic extinction (Section~\ref{sec:intergalacticextinction}) and the redshift evolution of the host-galaxy correlations (Section~\ref{sec:hostgalaxycorrelations}) are some of our largest assessed uncertainties but are absent from Pantheon+. The Pantheon+ \BBC analysis propagates systematic uncertainties at first order, i.e., by finding the mean change in distance modulus as a function of redshift and adding it to the distance-modulus covariance matrix with appropriate scaling. It is not clear whether this is completely adequate; for example, interaction terms could be missed by this first-order procedure but would be naturally handled by our simultaneous analysis. We show a contour from Union3+UNITY1.5 with all small perturbative systematic terms (the \DZP terms in Section~\ref{sec:SmallPerturbations}, encompassing uncertainties from calibration, peculiar-velocity correlations, Milky-Way extinction, intergalactic extinction, and weak-lensing bias) fixed to zero (dark blue), and even this contour is comparable in size \Pantheonending

\subsubsection{Baryon Acoustic Oscillations}

BAO constraints measure distances relative to the sound horizon ($r_d$) at the redshift of the drag epoch ($z_d$). A recent summary of the SDSS/BOSS/eBOSS BAO measurements is \url{https://www.sdss4.org/science/final-bao-and-rsd-measurements}. In addition, we include 6dF, which extends to lower redshift than the SDSS measurements \citep{Beutler2011}. Appendix~\ref{sec:BAOCollection} shows the collected distances and their covariance matrix. 

Older BAO constraints use the fitting formula given in \citet{Eisenstein1998} for $z_d$. However, as noted by \citet{PlanckCollaboration2014}, Boltzmann codes (e.g., \texttt{CAMB} \citealt{Lewis2000}) return an $r_d$ scaled down by about 1/1.0275 compared to that fitting formula. We use the \citet{Eisenstein1998} fitting formula for our work, and must make sure that each BAO distance is quoted relative to that (after removing the differences in the mean, the fitting formulae match the same quantities from the Planck chains with an accuracy of parts per hundred thousand). In addition, BAO constraints are quoted assuming different fiducial cosmologies for $r_d$, and we must standardize here, too; Appendix~\ref{sec:BAOCollection} discusses these details as well.

\subsubsection{Big Bang Nucleosynthesis}

When computing SN + BAO constraints, there is no strong constraint on the early universe sound speed and thus the sound horizon size. Thus, we include an $\Omega_b h^2$ ($\omega_b$) constraint from Big Bang nucleosynthesis \citep{Cooke2016}. Following \citet{Abbott2018}, we average the $\omega_b$ values estimated from theoretical and observational $d(p,\gamma)^3$He reaction rates (taking one half the difference as the uncertainty) for a constraint of $\omega_b = 0.02208 \pm 0.00052$.

\subsubsection{Cosmic Microwave Background}

We include CMB constraints from the Planck baseline temperature and polarization power spectra \citep{PlanckCollaboration2020}. For simplicity, we compress the CMB to a low-dimensional set of parameters (see Appendix~\ref{sec:CMBCompression}).

\subsubsection{Hubble Constant}

In contrast with the other cosmological parameters, the value of the Hubble constant is more controversial (e.g., \citealt{DiValentino2021}). We therefore try comparing the results using two recent, high-precision distance-ladder-based Hubble-constant measurements. These measurements have some distance measures in common and are thus not independent, so we include them only one at a time. The \citet{Freedman2021} measurement is based on a distance ladder of geometric distances to Tip of the Red-Giant Branch (TRGB) distances to CSP SNe~Ia and finds $H_0 = 69.8 \pm 0.6$ (statistical) $\pm 1.6$ (systematic) km/s/Mpc. We combine the two uncertainties in quadrature to 1.7 km/s/Mpc. The other measurement we include is based on a distance ladder of geometric distances to Cepheid distances to SNe~Ia \citep{Riess2022}. We take the latest version which achieves slightly higher precision by better taking SN spectral variation into account \citep{Murakami2023}: $H_0 = 73.29 \pm 0.90$ km/s/Mpc. As many of the Hubble-flow SNe in our analysis are in common with these analyses, there should be a correlation between our SN distances and these values. The proper way to handle these correlations is to include the lower rungs of the distance ladder in UNITY. Fortunately, the cosmological impact of these correlations is small (e.g., \citealt{Riess2022} Figure~17), so we leave this for future work.

\subsection{SN+External Cosmological Constraints} \label{sec:cosmoexternalconstraints}

With our compressed likelihoods, we compute frequentist confidence intervals. It may seem strange for a Bayesian analysis to compute frequentist confidence intervals in the end. However, after compression, the distance constraints for all probes have essentially Gaussian uncertainties, and the cosmological models are fairly linear in their parameters (especially over the range of allowed parameters for the combined constraints), so a $\chi^2$-based analysis is a reasonable approximation and frequentist contours are visually smoother than contours constructed from MCMC samples. The frequentist spline-interpolated SN distances are also how we release our distances to the community, so these needed to be computed and checked in any case.

Our cosmological fits infer the Hubble constant ($H_0$), the fraction of the critical density in matter ($\Omega_m$), the fraction in curvature ($\Omega_k$), the dark energy equation-of-state parameter ($w$ for a constant equation of state, or $w_0$ for the value today), and the change in $w$ with scale factor ($w_a$, such that $w(a) = w_0 + (1 - a) w_a$). We do not consider any non-standard early universe physics, fixing $\Omega_{\gamma} h^2 = 2.4729\times 10^{-5}$ (CMB temperature of 2.72548~K, \citealt{Fixsen2009}), fixing $N_{\mathrm{eff}} = 3.04$, and thus fixing $\Omega_{r} h^2 = (1 + 0.2271 \, N_{\mathrm{eff}}) \, \Omega_{\gamma} h^2$ \citep{Mangano2005}. When including BAO or CMB constraints, we also fit for the fraction of the critical density in baryons ($\Omega_b$), as this affects the sound horizon in the early universe. The remaining portion of the critical density is the fraction of dark energy ($\Omega_{\mathrm{DE}}$, or $\Omega_{\Lambda}$ when dark energy is a cosmological constant).

Table~\ref{tab:cosmoconstraints} summarizes our cosmological constraints. We first solve for the best fit when varying each of the parameters specified in the rows. Then, we move each parameter in turn away from its best fit, find the best fit with that parameter fixed, and find the two points for that parameter (in the positive and negative direction) where $\chi^2$ increases by 1. This gives us the quoted plus and minus confidence intervals. We show constraints for SNe alone, SNe + CMB, BAO + CMB, SNe + BAO (plus an $\Omega_b h^2 \equiv \omega_b$ measurement to constrain the sound horizon), SNe + BAO + CMB, and SNe + BAO + CMB + $H_0$. The SNe + BAO + $\omega_b$ combination is sometimes known as the ``inverse distance ladder'' when constraining $H_0$. The Dark Energy Task Force Figure of Merit (DETF FoM) is the inverse area of the 95.4\% contour in the $w_0$-$w_a$ plane \citep{Albrecht2006} and we compute this number directly using the computed contours discussed below. We take the original definition, not the $\pi\, 6.18 = 19.4 \times$ higher definition sometimes quoted based on $\det(\mathrm{Cov}(w_0, \; w_a))^{-1/2}$.

\begin{deluxetable*}{lccccccc}
 \tablehead{
 \colhead{Probes} & \colhead{$\chi^2$ (DoF)}  & \colhead{$h$} & \colhead{$\Omega_m$} & \colhead{$\Omega_k$} & \colhead{$w$ or $w_0$} & \colhead{$w_a$} & \colhead{DETF FoM} }
 \startdata
\hline
 \multicolumn{8}{c}{Flat $\Lambda$CDM}\\ 
 \hline
SNe & 24.0 (20)  & \nodata  &  $0.356^{+0.028}_{-0.026}$  & \nodata  & \nodata  & \nodata  & \nodata  \\ 
SNe+CMB & 26.2 (21)  &  $0.671^{+0.006}_{-0.006}$  &  $0.319^{+0.008}_{-0.008}$  & \nodata  & \nodata  & \nodata  & \nodata  \\ 
BAO+CMB & 12.2 (13)  &  $0.677^{+0.004}_{-0.004}$  &  $0.311^{+0.006}_{-0.006}$  & \nodata  & \nodata  & \nodata  & \nodata  \\ 
SNe+BAO+$\omega_b$ & 38.9 (32)  &  $0.674^{+0.010}_{-0.010}$  &  $0.311^{+0.014}_{-0.013}$  & \nodata  & \nodata  & \nodata  & \nodata  \\ 
SNe+BAO+CMB & 39.0 (34)  &  $0.675^{+0.004}_{-0.004}$  &  $0.313^{+0.006}_{-0.006}$  & \nodata  & \nodata  & \nodata  & \nodata  \\ 
SNe+BAO+CMB+$H_0^{\mathrm{TRGB}}$ & 40.7 (35)  &  $0.677^{+0.004}_{-0.004}$  &  $0.311^{+0.006}_{-0.005}$  & \nodata  & \nodata  & \nodata  & \nodata  \\ 
SNe+BAO+CMB+$H_0^{\mathrm{Ceph.}}$ & 72.4 (35)  &  $0.686^{+0.004}_{-0.004}$  &  $0.299^{+0.005}_{-0.005}$  & \nodata  & \nodata  & \nodata  & \nodata  \\ 
\hline
 \multicolumn{8}{c}{Open $\Lambda$CDM}\\ 
 \hline
SNe & 22.6 (19)  & \nodata  &  $0.287^{+0.064}_{-0.066}$  &  $\phantom{-}0.203^{+0.183}_{-0.173}$  & \nodata  & \nodata  & \nodata  \\ 
SNe+CMB & 24.1 (20)  &  $0.632^{+0.027}_{-0.025}$  &  $0.358^{+0.030}_{-0.029}$  &  $-0.010^{+0.007}_{-0.007}$  & \nodata  & \nodata  & \nodata  \\ 
BAO+CMB & 12.0 (12)  &  $0.679^{+0.006}_{-0.006}$  &  $0.310^{+0.006}_{-0.006}$  &  $\phantom{-}0.001^{+0.002}_{-0.002}$  & \nodata  & \nodata  & \nodata  \\ 
SNe+BAO+$\omega_b$ & 36.4 (31)  &  $0.632^{+0.027}_{-0.026}$  &  $0.286^{+0.021}_{-0.021}$  &  $\phantom{-}0.114^{+0.076}_{-0.073}$  & \nodata  & \nodata  & \nodata  \\ 
SNe+BAO+CMB & 38.9 (33)  &  $0.677^{+0.006}_{-0.006}$  &  $0.312^{+0.006}_{-0.006}$  &  $\phantom{-}0.001^{+0.002}_{-0.002}$  & \nodata  & \nodata  & \nodata  \\ 
SNe+BAO+CMB+$H_0^{\mathrm{TRGB}}$ & 40.3 (34)  &  $0.679^{+0.006}_{-0.006}$  &  $0.310^{+0.006}_{-0.006}$  &  $\phantom{-}0.001^{+0.002}_{-0.002}$  & \nodata  & \nodata  & \nodata  \\ 
SNe+BAO+CMB+$H_0^{\mathrm{Ceph.}}$ & 65.3 (34)  &  $0.695^{+0.005}_{-0.005}$  &  $0.297^{+0.005}_{-0.005}$  &  $\phantom{-}0.004^{+0.002}_{-0.002}$  & \nodata  & \nodata  & \nodata  \\ 
\hline
 \multicolumn{8}{c}{Flat $w$CDM}\\ 
 \hline
SNe & 22.1 (19)  & \nodata  &  $0.244^{+0.092}_{-0.128}$  & \nodata  &  $-0.735^{+0.169}_{-0.191}$  & \nodata  & \nodata  \\ 
SNe+CMB & 23.2 (20)  &  $0.652^{+0.012}_{-0.012}$  &  $0.336^{+0.014}_{-0.013}$  & \nodata  &  $-0.924^{+0.044}_{-0.043}$  & \nodata  & \nodata  \\ 
BAO+CMB & 12.1 (12)  &  $0.681^{+0.014}_{-0.013}$  &  $0.308^{+0.012}_{-0.012}$  & \nodata  &  $-1.016^{+0.053}_{-0.057}$  & \nodata  & \nodata  \\ 
SNe+BAO+$\omega_b$ & 30.5 (31)  &  $0.619^{+0.022}_{-0.022}$  &  $0.288^{+0.017}_{-0.017}$  & \nodata  &  $-0.803^{+0.066}_{-0.067}$  & \nodata  & \nodata  \\ 
SNe+BAO+CMB & 37.5 (33)  &  $0.666^{+0.009}_{-0.008}$  &  $0.320^{+0.008}_{-0.008}$  & \nodata  &  $-0.957^{+0.034}_{-0.035}$  & \nodata  & \nodata  \\ 
SNe+BAO+CMB+$H_0^{\mathrm{TRGB}}$ & 40.3 (34)  &  $0.673^{+0.008}_{-0.008}$  &  $0.314^{+0.007}_{-0.007}$  & \nodata  &  $-0.980^{+0.032}_{-0.033}$  & \nodata  & \nodata  \\ 
SNe+BAO+CMB+$H_0^{\mathrm{Ceph.}}$ & 65.6 (34)  &  $0.699^{+0.007}_{-0.006}$  &  $0.291^{+0.006}_{-0.006}$  & \nodata  &  $-1.074^{+0.029}_{-0.029}$  & \nodata  & \nodata  \\ 
\hline
 \multicolumn{8}{c}{Flat $w_0$-$w_a$}\\ 
 \hline
SNe+CMB & 21.3 (19)  &  $0.665^{+0.014}_{-0.015}$  &  $0.323^{+0.015}_{-0.013}$  & \nodata  &  $-0.699^{+0.163}_{-0.170}$  &  $-1.05^{+0.78}_{-0.78}$  & 1.46  \\ 
BAO+CMB & 10.1 (11)  &  $0.650^{+0.025}_{-0.023}$  &  $0.339^{+0.026}_{-0.025}$  & \nodata  &  $-0.652^{+0.268}_{-0.261}$  &  $-1.00^{+0.71}_{-0.77}$  & 1.08  \\ 
SNe+BAO+$\omega_b$ $^a$ & 30.5 (30)  &  $0.629^{+0.029}_{-0.094}$  &  $0.299^{+0.032}_{-0.112}$  & \nodata  &  $-0.783^{+0.129}_{-0.095}$  &  $-0.22^{+0.83}_{-0.92}$  & 1.16  \\ 
SNe+BAO+CMB & 31.7 (32)  &  $0.660^{+0.009}_{-0.009}$  &  $0.329^{+0.009}_{-0.009}$  & \nodata  &  $-0.744^{+0.100}_{-0.097}$  &  $-0.79^{+0.35}_{-0.38}$  & 3.52  \\ 
SNe+BAO+CMB+$H_0^{\mathrm{TRGB}}$ & 35.7 (33)  &  $0.668^{+0.008}_{-0.008}$  &  $0.321^{+0.008}_{-0.008}$  & \nodata  &  $-0.796^{+0.096}_{-0.093}$  &  $-0.71^{+0.35}_{-0.38}$  & 3.88  \\ 
SNe+BAO+CMB+$H_0^{\mathrm{Ceph.}}$ & 64.1 (33)  &  $0.698^{+0.007}_{-0.007}$  &  $0.294^{+0.006}_{-0.006}$  & \nodata  &  $-0.973^{+0.089}_{-0.086}$  &  $-0.41^{+0.34}_{-0.37}$  & 4.67  \\ 
\hline
 \multicolumn{8}{c}{Open $w_0$-$w_a$}\\ 
 \hline
SNe+CMB & 20.5 (18)  &  $0.578^{+0.107}_{-0.048}$  &  $0.428^{+0.081}_{-0.123}$  &  $-0.037^{+0.044}_{-0.024}$  &  $-0.587^{+0.308}_{-0.236}$  &  $-3.29^{+2.60}_{-3.40}$  & \nodata  \\ 
BAO+CMB & \phantom{0}9.9 (10)  &  $0.647^{+0.026}_{-0.024}$  &  $0.341^{+0.027}_{-0.026}$  &  $-0.001^{+0.003}_{-0.003}$  &  $-0.610^{+0.302}_{-0.280}$  &  $-1.21^{+0.85}_{-1.02}$  & \nodata  \\ 
SNe+BAO+$\omega_b$ $^a$ & 30.3 (29)  &  $0.640^{+0.040}_{-0.102}$  &  $0.310^{+0.040}_{-0.119}$  &  $-0.044^{+0.097}_{-0.095}$  &  $-0.763^{+0.125}_{-0.107}$  &  $-0.29^{+0.87}_{-0.86}$  & 1.05  \\ 
SNe+BAO+CMB & 31.5 (31)  &  $0.658^{+0.009}_{-0.009}$  &  $0.330^{+0.010}_{-0.009}$  &  $-0.001^{+0.003}_{-0.003}$  &  $-0.715^{+0.118}_{-0.111}$  &  $-0.97^{+0.48}_{-0.54}$  & 2.23  \\ 
SNe+BAO+CMB+$H_0^{\mathrm{TRGB}}$ & 35.6 (32)  &  $0.667^{+0.008}_{-0.008}$  &  $0.321^{+0.008}_{-0.008}$  &  $-0.001^{+0.003}_{-0.003}$  &  $-0.785^{+0.109}_{-0.104}$  &  $-0.78^{+0.45}_{-0.51}$  & 2.52  \\ 
SNe+BAO+CMB+$H_0^{\mathrm{Ceph.}}$ & 63.5 (32)  &  $0.699^{+0.007}_{-0.007}$  &  $0.294^{+0.006}_{-0.006}$  &  $\phantom{-}0.002^{+0.002}_{-0.002}$  &  $-1.000^{+0.095}_{-0.091}$  &  $-0.22^{+0.40}_{-0.45}$  & 3.30  \\ 
 \enddata
 \caption{Constraints on cosmological parameters. The SN $\chi^2$ values are based on spline-interpolated distances (with 22 nodes, so SN DoF = $22- N_{\mathrm{fit}}$), so they are much smaller than the number of SNe (\totalNSNe). $^{a}$ For SNe + BAO with $w_0$-$w_a$, we enforce early matter domination to prevent the sound horizon changing and degrading the constraints.
 \label{tab:cosmoconstraints}}
\end{deluxetable*}

\begin{figure*}[h!tbp]
    \centering
    \includegraphics[width=0.7 \textwidth]{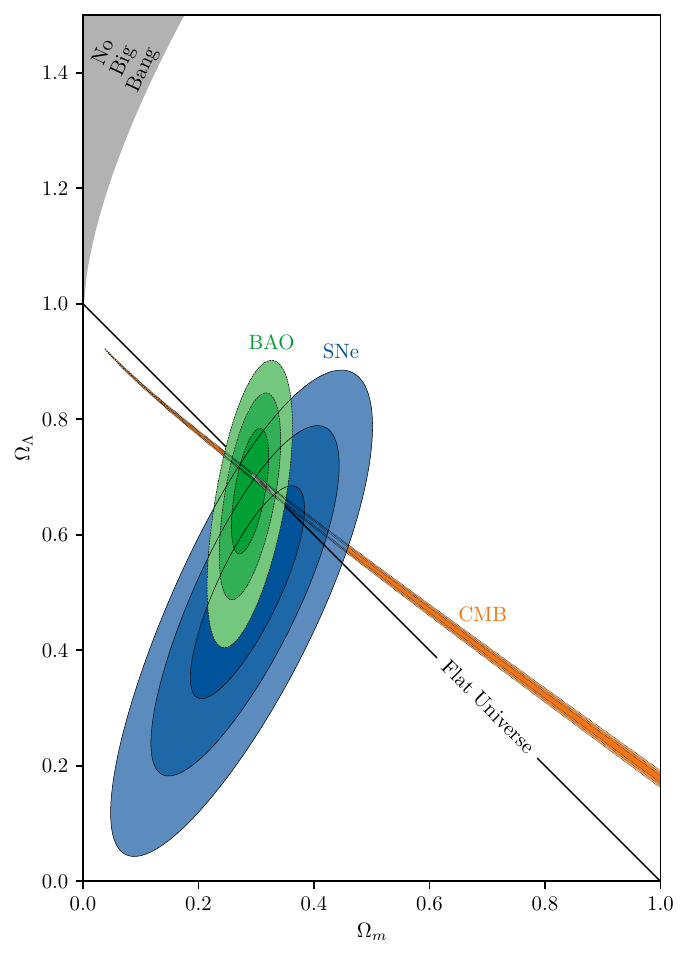}
    \caption{Constraints in the \Om-\OL plane. We show the constraints for SNe (in blue), BAO (in green), CMB (in orange), and combined (in gray).}
    \label{fig:OmOL}
\end{figure*}

\begin{figure*}[h!tbp]
    \centering
    \includegraphics[width=0.7 \textwidth]{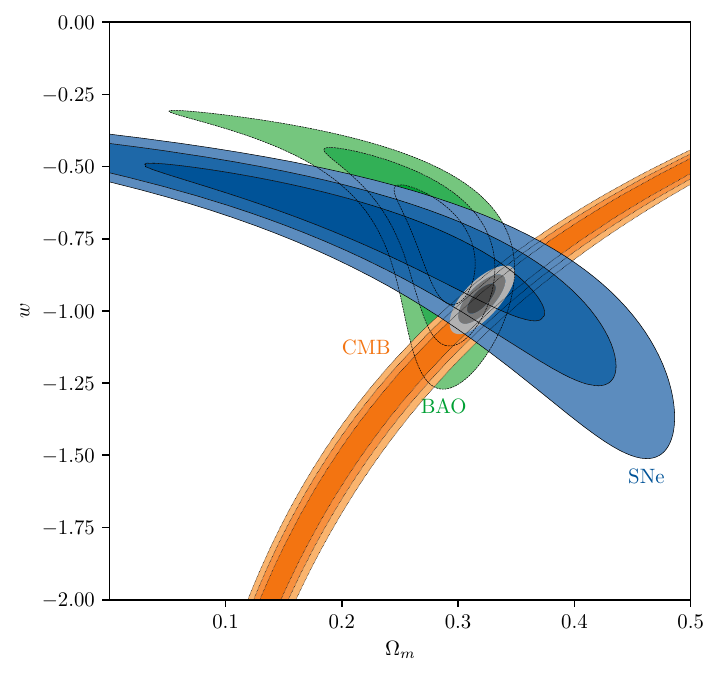}
    \caption{Constraints in the \Om-$w$ plane for flat-universe, constant equation-of-state parameter $w$ models. Again we show the constraints for SNe (in blue), BAO (in green), CMB (in orange), and combined (in gray). The probe constraints have different orientations and thus combine together for a much stronger measurement.}
    \label{fig:Omw}
\end{figure*}

\begin{figure*}[h!tbp]
    \centering
    \includegraphics[width=0.49 \textwidth]{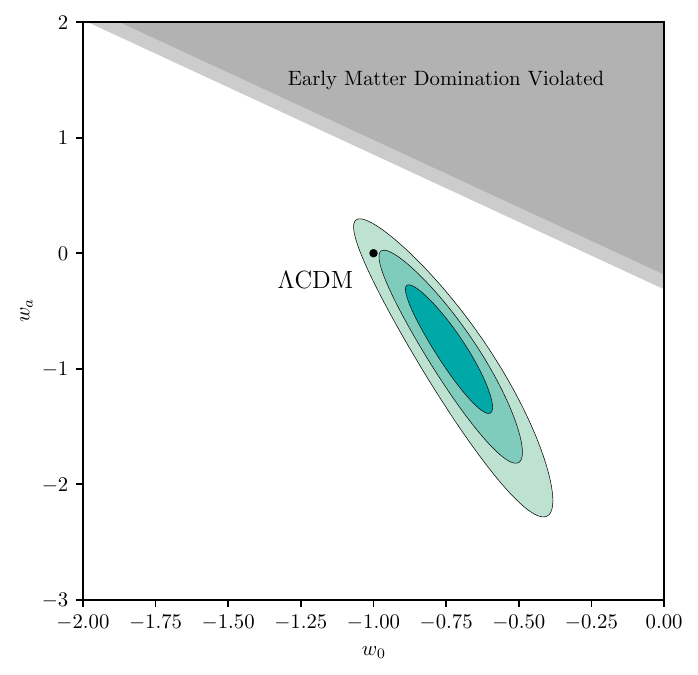}
    \includegraphics[width=0.49 \textwidth]{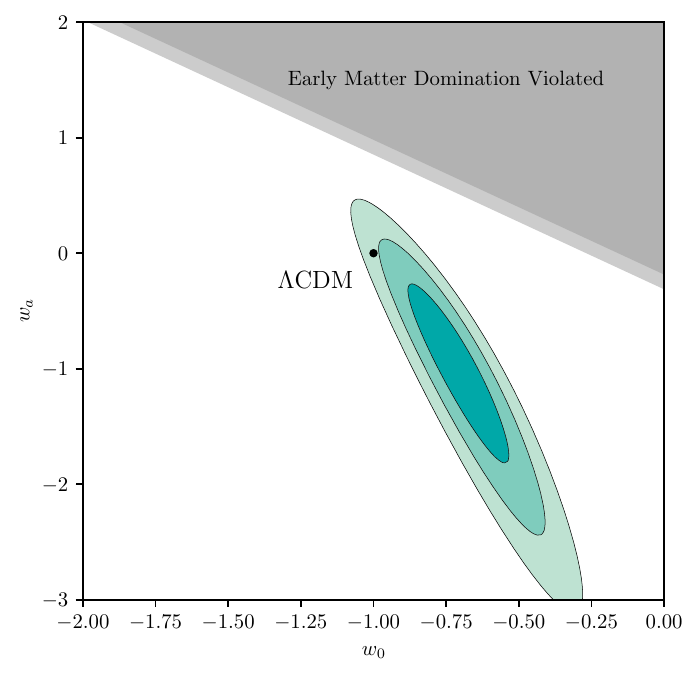}
    \caption{Constraints in the $w_0$-$w_a$ plane combining SNe, BAO, and CMB. The {\bf left panel} shows the 1, 2, and 3$\sigma$ contours assuming a flat universe, while the {\bf right panel} also fits for curvature. We also mark off the parameter space where early matter domination would begin to be violated (the shaded regions show 1\% and 10\% of the matter density at $z=1100$ assuming $\Omega_m = 0.3$). The contours show weak tension with \LCDM (indicated with a black dot).}
    \label{fig:w0wa}
\end{figure*}

\begin{figure*}[h!tbp]
    \centering
    \includegraphics[width=0.49 \textwidth]{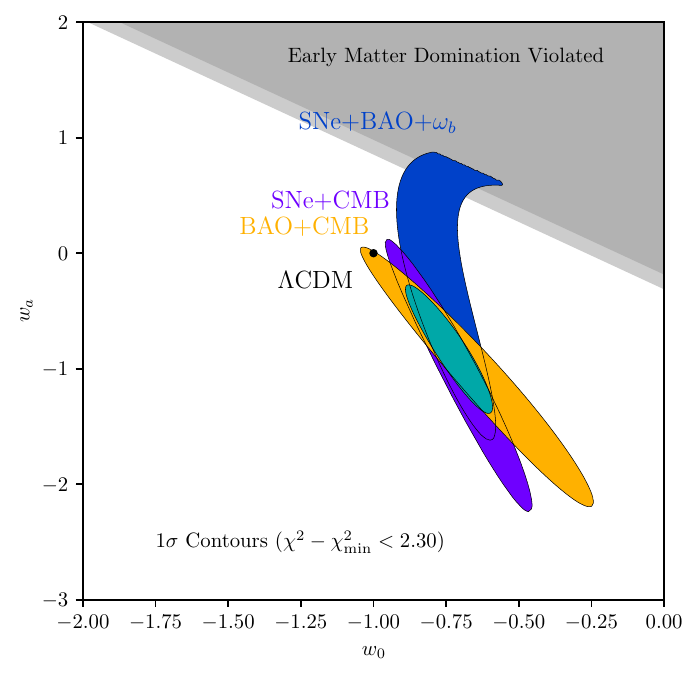}
    \includegraphics[width=0.49 \textwidth]{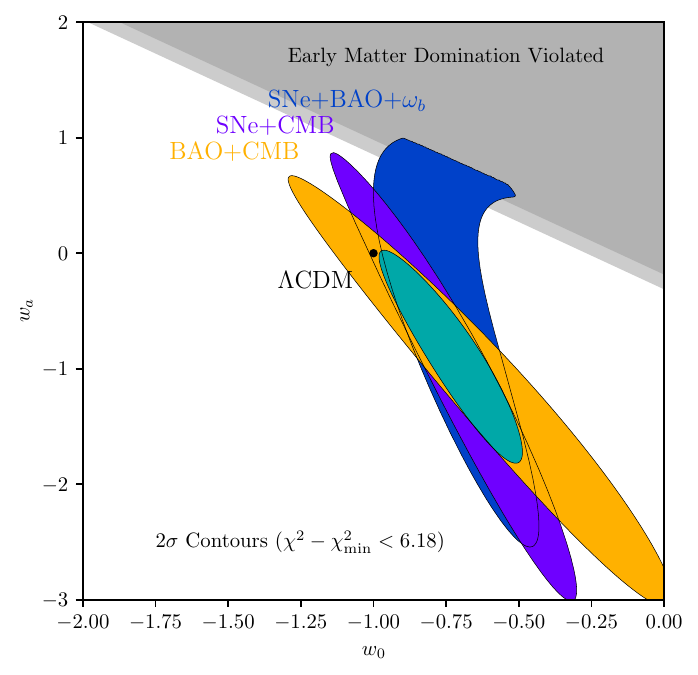}
    \caption{Constraints in the $w_0$-$w_a$ plane. The {\bf left panel} shows the $1 \sigma$ contours and the right panel shows $2 \sigma$ contours, both assuming a flat universe. The contours show constraints from BAO+CMB, SNe+CMB, and SNe+BAO (as well as the same BAO+CMB+SNe contour from Figure~\ref{fig:w0wa}). The contour for each pair of probes is angled differently, showing how the constraints get stronger when all three probes are combined together. The results from each pair of probes are compatible with the results that use all three.  
    \label{fig:w0waindiv}}
\end{figure*}

We compute frequentist contours ($\Delta \chi^2$ compared to the best fit of 2.296, 6.180, and 11.829 for 68.3\%, 95.4\%, and 99.7\% confidence) by fixing the two parameters shown in the plane and fitting for the others. We use an adaptive-refinement contour code that chooses points to evaluate.\footnote{\url{https://github.com/rubind/adaptive_contour}} Figure~\ref{fig:OmOL} shows our constraints in the \Om-\OL plane. Figure~\ref{fig:Omw} shows our constraints in the \Om-$w$ plane. Figure~\ref{fig:w0wa} shows our results in the $w_0$-$w_a$ plane (with and without allowing spatial curvature). Figure~\ref{fig:w0waindiv} investigates which individual data combinations drive our $w_0$-$w_a$ constraints.

Our cosmology analyses show us several findings, summarized below.
\begin{itemize}
    \item The strongest evidence for tension among all cosmological models and datasets considered is between the $H_0$ value referenced to the early universe and the $H_0$ from the local Cepheid+SN distance ladder. This tension persists for all cosmological models; adding the Cepheid $H_0$ to the SNe+BAO+CMB fit increases the $\chi^2$ by \HubbleCephChiInc for the addition of one degree of freedom, formally requiring us to reject either the measurement (i.e., its central value, its uncertainties, and/or its claimed Gaussianity might be in error) or reject the cosmological models \citep[e.g.,][]{DiValentino2021}. However, the SN+TRGB measurement shows no such strong tension due both to its lower central value and lower claimed precision.
    \item No evidence for curvature is seen.
    \item The $w_0$-$w_a$ models show weak tension with \LCDM, favoring ``thawing'' models with $w_0 > -1$ and $w_a < 0$ \citep{Caldwell2005}. The statistical significance of the tension ranges from \LCDMTension, depending on how one evaluates it. For example, adding $w_0$ and $w_a$ to flat \LCDM lowers the $\chi^2$ by \wOwaChiDec. Considering the same two data combinations, the statistical significance of the deviations of $w_0/w_a$ from \LCDM ranges from \wOwaSigmaRange. Interestingly, the tension is hinted at even without SNe, as the BAO+CMB $\chi^2$ drops by \BAOCMBChiDropwOwa when including $w_0$ and $w_a$ (and most of the drop is due to adding a single parameter: $w_a$, so this could be considered one degree of freedom).
    \item Taking the $H_0$ measurements at face value, including $H_0$ constraints increases the DETF FoM values by \wOwaTRGBFoMRange for the SN+TRGB measurement and by \wOwaCephFoMRange for the Cepheid+SN measurement when comparing to SNe+BAO+CMB alone. As noted by \citet{Riess2009}, this is a strong motivation for increasing the precision of the local $H_0$ measurements.
    \item However, the DETF FoM values are only $\sim 1$--5, so the constraints on any time variation of dark energy are still relatively weak. Future experiments should enable FoM values $\sim 10\times$ larger, e.g., SNe+CMB assuming a flat universe for the \RomanSpelled should reach FoM $\gtrsim 17$ \citep{Hounsell2018} with similarly large increases for future BAO experiments \citep{Kim2015}. Thus any hints of \LCDM tension seen in our analysis will be rapidly verified or excluded in the coming years. 
\end{itemize}

\section{Conclusions} \label{sec:Conclusions}

We present a new compilation of cosmological SNe~Ia (``Union3'') with different calibration paths and updated light-curve selection and fitting compared to previous compilations. We discuss the advantages of Bayesian regression for applications with non-Gaussian, incompletely understood, and multidimensional uncertainties, with a focus on supernova cosmology. We present the updated UNITY1.5 Bayesian analysis framework for inferring SN cosmological constraints while propagating astrophysical and observational uncertainties. The key improvements over earlier UNITY versions have been to the models of the population distributions, selection-effects, standardization, and some systematic uncertainties.

Our analysis suggests a stronger correlation between the color standardization coefficients of the reddest SNe and host-galaxy properties than between standardized luminosity and host-galaxy properties. Whether this correlation is due to a higher-$R_V$ reddening law in low-stellar-mass galaxies or some behavior of SNe themselves is not clear. We also find no evidence of redshift evolution of the host-galaxy correlations although our constraints are weak.

We demonstrate that UNITY1.5 can recover the unexplained dispersion distribution in $m_B$/$x_1$/$c$ on simulated data and then examine the real SNe. \unexplainedallocation We demonstrate a strong correlation between the amount of unexplained dispersion in color and the standardization coefficient for the bluest SNe. As with previous analyses, we find evidence of nonlinear color standardization.

We find virtually the same cosmological constraints allowing the $x_1$ and $c$ population distributions to vary with both redshift and host-galaxy stellar mass (our nominal analysis) as we do when these distributions only vary with host-galaxy stellar mass. This shows the advantage of modeling these distributions before selection effects, as the observed distributions change rapidly with redshift due to selection, as surveys cannot detect reddened SNe to as high a redshift on average as bluer SNe. Running a Bayesian analysis where the after-selection populations are modeled as redshift-independent introduces large biases \citep{Wood-Vasey2007, rubin16}.

We show cosmological constraints with Union3+UNITY1.5, including external constraints from BAO, CMB, BBN, and $H_0$. When considering cosmological models with a dark energy equation-of-state parameter that varies with time as $w_0 + (1 - a)w_a$, we find \LCDMTension tension with \LCDM, depending on which cosmological parameter one looks at and how many degrees of freedom one assumes (i.e., one degree of freedom for $w_0$ or $w_a$ and two degrees of freedom for both together).\footnote{Note added in refereeing: since the submission of this work, the evidence for time-varying dark energy has increased, \citealt{DESCollaboration2024, DESICollaboration2024}. None of our analyses have been changed in response to this new, semi-independent information.} We find no evidence of curvature, and none of the parameters we add to \LCDM ameliorate the $H_0$ tension from the Cepheid/SN~Ia distance ladder.

Our uncertainty analysis (summarized in Tables~\ref{tab:UncertainyAnalysis} and \ref{tab:UncertainyAnalysiswa}) suggests some paths forward. For Stage IV cosmological measurements, which target uncertainties several times smaller than current uncertainties \citep{Albrecht2006}, the entire top half of the list of uncertainties will need further effort. More and better calibrated nearby SN datasets with well understood selection functions will help to anchor the Hubble diagram more accurately and precisely. Observing more SNe further into the IR will help control host-galaxy, intergalactic, and Milky-way extinction. Two near-terms sources of such observations at high redshift are the \RomanSpelled \citep{Spergel2015, Rose2021} and Euclid \citep{Astier2014}.  The path to reducing SN~Ia astrophysical uncertainties is less clear. To some extent, more SNe will help, as SNe can be further subdivided in the analysis (e.g., by light-curve parameters or host-galaxy properties). It is clear that SALT misses SN diversity, as intrinsic SN variability depends on a few intrinsic parameters, certainly more parameters than just $x_1$ \citep[e.g.,][]{Branch2006, Rubin2020, Boone2021A, Stein2022}, and using additional SN information to standardize results in a smaller magnitude dispersion and smaller host-galaxy correlations \citep{Boone2021B}. For future high-redshift SN datasets, spectroscopy and well-sampled optical+NIR light curves may be a way to constrain additional SN variability parameters in large samples \citep{Swann2019, Rubin2022Prism} better than we can at the moment \citep[e.g.,][]{Sullivan2009, Foley2012}. On the whole, our analysis shows no insurmountable obstacles for the next large increase in precision possible with next-generation datasets.

\section*{Acknowledgments}

We thank David Jones and Kyle Barbary for help understanding \texttt{SNCosmo}'s SALT3 implementation. We also thank Michael Betancourt for suggesting tests with the posterior predictive distribution. We also thank the anonymous referee for thorough feedback. We thank Yungui Gong for catching a mislabeled parameter. The technical support and advanced computing resources from University of Hawaii Information Technology Services - Cyberinfrastructure, funded in part by the National Science Foundation MRI award \#1920304, are gratefully acknowledged. This research has made use of the NASA/IPAC Extragalactic Database (NED), which is funded by the National Aeronautics and Space Administration and operated by the California Institute of Technology. We acknowledge the usage of the HyperLeda database (\url{http://leda.univ-lyon1.fr}). This research has made use of the SIMBAD database,
operated at CDS, Strasbourg, France. This work was supported in part by the Director, Office of Science, Office of High Energy Physics of the U.S. Department of Energy under Contract No. DE- AC02-05CH11231. This work has made use of data from the European Space Agency (ESA) mission
{\it Gaia} (\url{https://www.cosmos.esa.int/gaia}), processed by the {\it Gaia}
Data Processing and Analysis Consortium (DPAC,
\url{https://www.cosmos.esa.int/web/gaia/dpac/consortium}). Funding for the DPAC
has been provided by national institutions, in particular the institutions
participating in the {\it Gaia} Multilateral Agreement. PR-L acknowledges support from grant PID2021-123528NB-I00, from the Ministerio de Ciencia e Innovaci\'on of Spain.

\software{
Astropy \citep{Astropy},
Extinction \citep{barbary_kyle_2016_804967},
Matplotlib \citep{matplotlib}, 
Numpy \citep{numpy},
PairV \citep{Davis2011},
PyStan \citep{allen_riddell_2018_1456206},
SciPy \citep{scipy},
SNCosmo \citep{sncosmo},
Stan \citep{Carpenter2017}
}

\clearpage

\restartappendixnumbering
\appendix

\section{Absolute Magnitude and Distance Modulii Parameterization} \label{sec:scriptM}

To remove the direct impact of $H_0$ on the distance moduli, we use a similar treatment as \citet{perlmutter97} for the last line of our Equation~\ref{eq:tripp}. We start with the absolute magnitude plus distance modulus:
\begin{eqnarray}
   & & M_B + \mu(z,\ \mathrm{cosmology}) \nonumber \\
   & = & M_B + 5 \log_{10}\left[\frac{(1 + \zhelio)}{10~\mathrm{pc}} \frac{c}{H_0} \frac{1}{\sqrt{|\Omega_k|}} \mathrm{sinn} \left\{ \sqrt{|\Omega_k|} \int_0^{\zCMB} \frac{d\zprime}{E(\zprime)} \right\} \right] \label{eq:MBmu}
\end{eqnarray}
where
\begin{equation}
    \mathrm{sinn}(x) \equiv
    \begin{cases}
    \begin{array}{ll}
    \sin(x) & \mathrm{for}\ \Omega_k < 0 \\
    x & \mathrm{for}\ \Omega_k = 0\\ 
    \sinh(x) & \mathrm{for}\ \Omega_k > 0 \;, \\
    \end{array}
    \end{cases}
\end{equation}
and
\begin{equation}
E(\zprime) \equiv \frac{H(\zprime)}{H_0}  = \sqrt{\Omega_m (1 + \zprime)^3 + \Omega_r (1 + \zprime)^4 + \Omega_{\mathrm{DE}} (1 + \zprime)^{3 (w_0+w_a+1)} e^{\frac{-3 w_a \zprime}{1 + \zprime}}  + \Omega_k (1 + \zprime)^2} \;,
\end{equation}
and $\Omega_m + \Omega_{\mathrm{DE}} + \Omega_r + \Omega_k = 1$ (the cosmic sum rule). If one is using a code that accommodates complex numbers and does not allow $\Omega_k$ to get too close to zero, sinn can be replaced with sinh with $\Omega_k$ written without taking the absolute value. \zhelio is the heliocentric redshift of the SN (or its host galaxy) and \zCMB is this redshift corrected for the peculiar motion of the solar system with respect to the CMB (e.g., \citealt{Davis2019}). By defining
\begin{equation} \label{eq:scriptM}
    \scriptM \equiv M_B - 5 \log_{10} (H_0/ 70~\mathrm{km/s/Mpc}) \;,
\end{equation}
and noting that
\begin{equation}
    5 \log_{10} \left[\frac{c}{(70~\mathrm{km/s/Mpc}) \, (10~\mathrm{pc})} \right] = 43.15861
\end{equation}
we can rewrite Equation~\ref{eq:MBmu} without $H_0$ made explicit:

\begin{eqnarray}
   & = & \scriptM + 43.15861 + 5 \log_{10}\left[(1 + \zhelio)  \frac{1}{\sqrt{|\Omega_k|}} \mathrm{sinn} \left\{ \sqrt{|\Omega_k|} \int_0^{\zCMB} \frac{d\zprime}{E(\zprime)} \right\} \right] \;.
\end{eqnarray}

\section{Parameters and Priors} \label{sec:Priors}

Table~\ref{tab:Priors} shows the priors we assume on the model parameters. Some priors are uninformative, e.g., the flat-in-angle priors on the standardization slopes. But most are weakly informative, i.e., broad compared to the posteriors, but enough to weakly constrain the parameters to the most plausible part of parameter space. Examples include the weak Gaussian priors on the population parameters. In some cases, we also impose parameter bounds.

We note that the lower bound of 0.2 on $\sigma_{c_R}^{\mathrm{outl}}$ was an oversight, and it should have been something more like 0.1--0.2$/\tau^c \approx 1$--2. We reran both flat \LCDM and flat $w_0$-$w_a$ cosmological inference as a test with this limit changed to 1 and see only minor effects: the inferred $\Omega_m$ decreases by about 0.007 or $1/4 \sigma$, and $w_0$ and $w_a$ have even less significant changes. We thus leave it for now, as it was only noticed after unblinding.

\newcommand{\highname}{H\xspace}
\newcommand{\lowname}{L\xspace}

\begin{deluxetable}{lr}
 \tablehead{
 \colhead{Prior} & \colhead{Parameter Description}  }
 \startdata
$\scriptM \sim \mathcal{N}(-19,\ 0.3^2)$ & Absolute Magnitude for $h = 0.7$ \\
$\tan^{-1}{\alpha} \sim \mathcal{U}(-0.2,\ 0.3)$ & $x_1$ Standardization Coefficient \\
$\tan^{-1}{\beta_B}  \sim \mathcal{U}(-1.4,\ 1.4)$ & $c_B$ Standardization Coefficient for Bluest SNe \\
$\tan^{-1}{\betaRL} \sim \mathcal{U}(-1.4,\ 1.4)$ & $c_R$ Standardization Coefficient for Reddest SNe in Low-Mass Galaxies \\
$\tan^{-1}{\betaRH}  \sim \mathcal{U}(-1.4,\ 1.4)$ & $c_R$ Standardization Coefficient for Reddest SNe in High-Mass Galaxies \\
$\delta(z = 0)  \sim \mathcal{N}(0,\ 0.2^2)$ & Host-Mass Standardization Coefficient, $z=0$ \\
$\delta(z = \infty)  \sim \mathcal{U}(0,\ 1)\, \delta(0)$ & Host-Mass Standardization Coefficient, $z=\infty$ \\
\hline
$\Omega_m  \sim \mathcal{U}(0,\ 1)$ & Fraction of Critical Density in Cold Matter \\
$w$ or $w_0 \sim \mathcal{U}(-2,\ 0)$ & Equation of State Parameter ($w$CDM or $w_0$-$w_a$ only) \\
$w_a  \sim \mathcal{U}(-4,\ 2)$ & Equation of State Parameter ($w_a$ only) \\
$\mu \sim  \mathcal{N}(0,\ 1^2) + \mu(\mathrm{Flat} \Lambda\mathrm{CDM},\ \Omega_m = 0.3)$ & Distance Modulus Nodes (spline-interpolated distance modulus only) \\
 & (These and other cosmology parameters: frequentist in Section~\ref{sec:CosmologyConstraints}) \\
\hline
$\sigma^{\text{unexplained}}_j  \sim \mathcal{U}(0.01,\ 0.3)$ & Unexplained Dispersion \\
$f^{m_B}, f^{x_1}, f^c  \sim $ 3D Unit Simplex & Fraction of Unexplained Dispersion in $\{m_B, x_1, c\}$ \\
\hline
$\xonetruei \sim (1 - f^{\mathrm{outl}}) \mathrm{ExpModNormal}(x_{1i}^*,\ (R^{x_1}_{i})^2,\ 1/\tau^{x_1}_i)$ & Modeled Latent $x_1$ Distribution \\
\hspace{0.25 in} + $f^{\mathrm{outl}} \mathcal{N}(0,\ (\sigma_{x_1}^{\mathrm{outl}})^2) $ & \\
$x_{1i}^* = \Phigh_i \, x^*_{1k(m = \highname)} + (1 - \Phigh_i) \, x^*_{1k(m = \lowname)}$ & Gaussian $x_1$ Mean\\
$R^{x_1}_i = \Phigh_i \, R^{x_1}_{k(m = \highname)} + (1 - \Phigh_i) \, R^{x_1}_{k(m = \lowname)}$ & Gaussian $x_1$ Dispersion \\
$\tau^{x_1}_{i} = \Phigh_i \, \tau^{x_1}_{k(m = \highname)} + (1 - \Phigh_i) \, \tau^{x_1}_{k(m = \lowname)}$ & Exponential $x_1$ Scale \\
$\cBtruei \sim (1 - f^{\mathrm{outl}}) \mathcal{N}(c_i^*,\ (R^c_i)^2) + f^{\mathrm{outl}} \mathcal{N}(0,\ (\sigma_{c_B}^{\mathrm{outl}})^2)$ & Modeled Latent $c_B$ Distribution \\
$c_i^* = \Phigh_i \, c^*_{k(m = \highname)} + (1 - \Phigh_i) \, c^*_{k(m = \lowname)}$ & Gaussian $c$ Mean\\
$R^c_i = \Phigh_i \, R^c_{k(m = \highname)} + (1 - \Phigh_i) \, R^c_{k(m = \lowname)}$ & Gaussian $c$ Dispersion \\
$\cRtruei \sim (1 - f^{\mathrm{outl}}) \mathrm{Exp}(1/\tau^c_{i}) + f^{\mathrm{outl}} \mathcal{N}(0,\ (\tau^c_{i} \, \sigma_{c_R}^{\mathrm{outl}})^2)$ & Modeled Latent $c_R$ Distribution \\
$\tau^c_{i} = \Phigh_i \, \tau^c_{k(m = \highname)} + (1 - \Phigh_i) \, \tau^c_{k(m = \lowname)}$ & Exponential $c$ Scale \\
$R^c_{km} \sim \mathcal{N}(0.1,\ 0.2)\, \mathcal{U}(0.01,\ 0.2)$ & Dispersion of Latent $c_B$ Distribution \\
$\tau^c_{km} \sim \mathcal{N}(0.1,\ 0.2^2) \, \mathcal{U}(-0.5,\ 0.5)$  & Exponential Scale of Latent $c_R$ Distribution \\
$c_{km}^* \sim \mathcal{N}(-0.1,\ 0.2^2)  \, \mathcal{U}(-0.5,\ 0.5) $ & Mean of Latent $c$  \\
$x_{1km}^* \sim \mathcal{N}(0,\ 2^2)  \, \mathcal{U}(-5,\ 5)$  & Gaussian Latent $x_1$ Distribution Central Value \\
$\tau^{x_1}_{km} \sim \mathcal{N}(-1,\ 2^2) \, \mathcal{U}(-5,\ 5)$  & Exponential Scale of Latent $x_1$ Distribution \\
$R_{km}^{x_1} \sim \mathcal{N}(1,\ 2^2) \, \mathcal{U}(0.1,\ 2) $ & Gaussian Dispersion of Latent $x_1$ Distribution \\
\hline
$f^{\mathrm{outl}} \sim \mathrm{LogNormal}(\log(0.02),\ 0.5^2) \, \mathcal{U}(0.001,\ 0.1)$ & Fraction of Outliers  \\
$\sigma_{m_B}^{\mathrm{outl}} \sim \mathcal{N}(0.5,\ 0.5^2) \, \mathcal{U}(0.2,\ 10)$ & Width of the Outlier Distribution in $m_B$  \\
$\sigma_{x_1}^{\mathrm{outl}}  \sim \mathcal{N}(3,\ 3^2) \, \mathcal{U}(0.2,\ 10)$ & Width of the Outlier Distribution in $x_1$  \\
$\sigma_{c_B}^{\mathrm{outl}} \sim \mathcal{N}(0.5,\ 0.5^2) \, \mathcal{U}(0.2,\ 10)$ & Width of the Outlier Distribution in $c_B$  \\
$\sigma_{c_R}^{\mathrm{outl}} \sim \mathcal{N}(10,\ 3^2) \, \mathcal{U}(0.2,\ 10)$ & Width of the Outlier Distribution in $c_R$ in units of $\tau^c_{i}$  \\
$ \DZP_l \sim \mathcal{N}(0, 1^2)$ & Normalized Systematics Perturbations \\
$m^{\mathrm{cut}}_j \sim \mathcal{N} (m^{\mathrm{nominal}}_j,\ 0.5^2) \, \mathcal{U}(14,\ 30)$ & Median Survey Depth (nominal in Table~\ref{tab:DatasetSummary}) \\
$\sigma^{\mathrm{cut}}_j \sim \mathcal{N} (\sigma^{\mathrm{depth}}_j,\ 0.25^2) \, \mathcal{U}(0.1,\ 3)$ & Survey Depth $1 \sigma$ Around Median (nominal in Table~\ref{tab:DatasetSummary}) \\
\enddata
 \caption{Priors used in this analysis. A subscript of $i$ indicates per-SN values, a subscript of $j$ indicates per-dataset values, a subscript of $k$ indicates redshift range (low-$z$, mid-$z$, or high-$z$), a subscript of $l$ indicates other (global) multiple values, and a subscript of $m$ indicates host-galaxy stellar mass ($\lowname$ for low or $\highname$ for high). \label{tab:Priors}}
\end{deluxetable}

\section{Light-Curve Shape and Color Population Distributions}\label{sec:colorpop}

\begin{figure}
    \centering
    \includegraphics[width = 0.9 \textwidth]{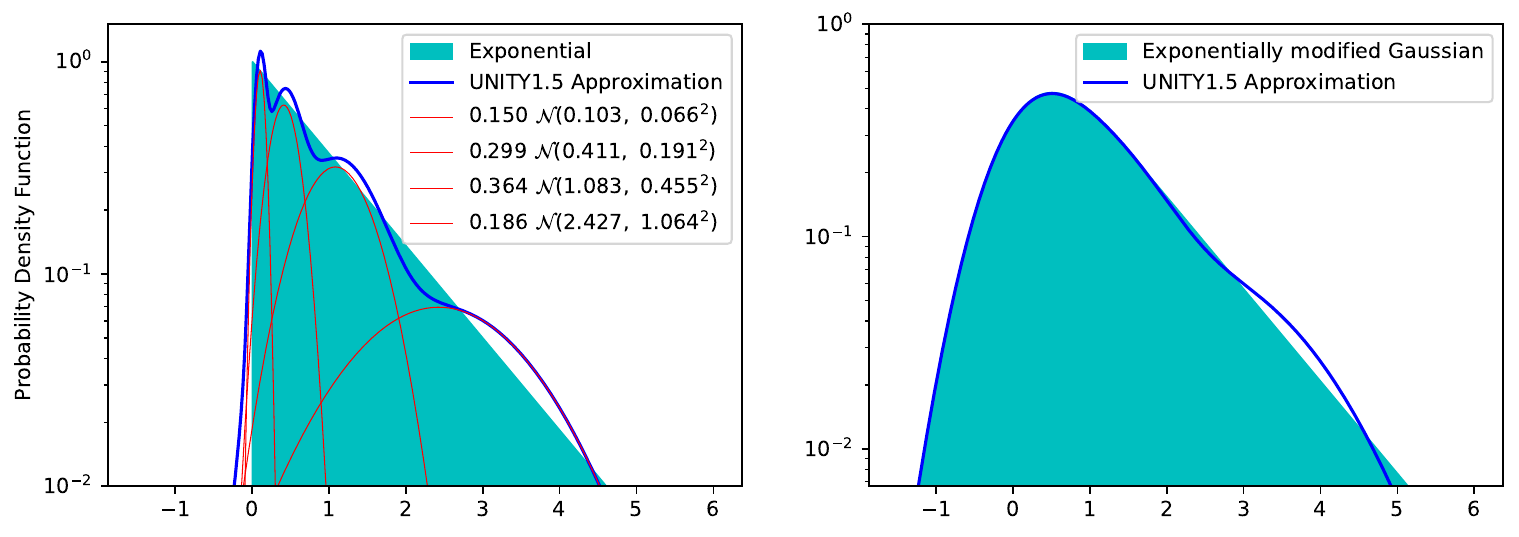}
    \caption{Illustration of the four-Gaussian UNITY1.5 approximation for the \xonetrue and \cRtrue populations. The {\bf left panel} shows the approximation of an exponential distribution with scale=1; the blue line shows the four-Gaussian approximation (each Gaussian is shown with a red curve) while the filled cyan region shows the exponential distribution. The approximation has some wiggles, but tracks the exponential well. The {\bf right panel} shows the same exponential convolved with a Gaussian with standard deviation=1/2. This convolution represents typical measurement uncertainties or the convolution of the exponential with the Gaussian part of the population distribution (both about the same size compared to the exponential scale). Here, the four-Gaussian approximation is a much better approximation than it is to the exponential alone. \label{fig:expapprox}}
\end{figure}

The distribution of colors (before any selection effects) for SNe Ia are skewed positive while the distribution of light-curve shapes ($x_1$) is skewed negative. New to this work, we use an approximation to the exponentially modified normal distribution for both. Treating selection effects requires an expression for the probability of detecting SNe based on the population parameters and the model of selection effects (e.g., \citealt{rubin15b}, \citealt{March2018}, \citealt{Hinton2017}). This requires integrating the SN population times the probability of selection as a function of magnitude. Prior work \citep{rubin15b, Hinton2019} has approximated these integrals. We choose instead to keep these integrals analytic by approximating the exponential distribution with a Gaussian mixture \citep{Kelly2007}. Four Gaussians is enough to give reasonable performance when fitting simulated data that was actually generated with exponential distributions (Section~\ref{sec:simulateddata}). Figure~\ref{fig:expapprox} compares this approximation to the analytic distribution and shows good agreement.

\section{Other Zeropoints Derived in This Work}\label{sec:MoreZeropoints}

\begin{deluxetable}{lrrrrrr}
 \tablehead{
 \colhead{Filter} & \multicolumn{6}{c}{MJD}  }
 \startdata
\nodata & 52334 & 53919 & 53920 & 54976 & 54977 & 59214 \\ 
\nodata & 2002.2 & 2006.5 & 2006.5 & 2009.4 & 2009.4 & 2021.0 \\ 
\hline
$F435W$  & 25.7932  & 25.7861  & 25.7607  & 25.7559  & 25.7676  & 25.7600 \\ 
$F475W$  & 26.1757  & 26.1684  & 26.1475  & 26.1421  & 26.1528  & 26.1451 \\ 
$F555W$  & 25.7373  & 25.7289  & 25.7127  & 25.7072  & 25.7177  & 25.7100 \\ 
$F606W$  & 26.4250  & 26.4163  & 26.4029  & 26.3969  & 26.4093  & 26.4015 \\ 
$F625W$  & 25.7530  & 25.7438  & 25.7319  & 25.7264  & 25.7392  & 25.7319 \\ 
$F775W$  & 25.2863  & 25.2806  & 25.2672  & 25.2629  & 25.2768  & 25.2691 \\ 
$F814W$  & 25.5324  & 25.5275  & 25.5108  & 25.5076  & 25.5209  & 25.5131 \\ 
$F850LP$  & 24.3554  & 24.3525  & 24.3274  & 24.3252  & 24.3374  & 24.3295 \\ 
\enddata
 \caption{Vega=0 zeropoints for ACS WFC as a function of time, computed from \citet{Bohlin2016} using alpha\_lyr\_stis\_008.ascii and averaged between the \bandpasses of the two CCD chips.\label{tab:ACSZP}}
\end{deluxetable}

In general, quoting instrumental sensitivity with respect to reference stars is easier to update when the SEDs of those reference stars are updated (as has happened with CALSPEC). Table~\ref{tab:ACSZP} shows the \citet{Bohlin2016} zeropoints converted into a Vega=0 system. We choose Vega=0 because of both convenience of historical comparisons and because Vega is near the mean color of the CALSPEC ACS calibrators. We linearly interpolate these zeropoints in time for the \citet{riess07}, \citet{suzuki12}, and \citet{rubin13} data.

Table~\ref{tab:PSZP} shows the Pan-STARRS1 aperture-photometry zeropoints updated for \CALSPEC. As the passbands for PS1 have been scanned, we only expect slow modifications with wavelength due to scattered light in the scans and variation in average image quality with wavelength. We fit for an exponential warp in wavelength for each filter, i.e., multiplying the passband by $\exp{(\alpha \, \lambda)}$, where $\alpha$ is a fit parameter). We only find a significant warping required in $g$ band; it is the broadest passband in wavelength, so this is plausible. We apply a prior that the modification over $1\mu$m is expected to be less than one $e$-folding, i.e., $\alpha \sim \mathcal{N}(0,\ (1$/$\mu$m)$^2$).

\begin{deluxetable}{ccccc}[h]
 \tablehead{
 \colhead{Filter} & \colhead{Zeropoint (PS1 $-$ AB)} & \colhead{Filter Warp (\ang)} & \colhead{Foundation - PS1} & \colhead{PSMD - PS1} }
 \startdata
$g$ & $0.0064 \pm 0.0049$ & $15.8 \pm 6.6$ & 0.0070 & -0.0066 \\
$r$ & $0.0065 \pm 0.0046$ & $1.6 \pm 7.5$ & 0.0017 & -0.0295 \\
$i$ & $0.0024 \pm 0.0048$ & $-2.2 \pm 7.3$ & 0.0000 & -0.0225 \\
$z$ & $0.0074 \pm 0.0036$ & $-0.1 \pm 8.5$ & -0.0011 & -0.0272 \\
\enddata
 \caption{AB offsets and filter shifts for PS aperture photometry. These are in the sense of mean PS aperture magnitudes minus synthetic \CALSPEC magnitudes.\label{tab:PSZP}}
\end{deluxetable}

\section{Baryon Acoustic Oscillation Distances} \label{sec:BAOCollection}

\newcommand{\rdours}{\ensuremath{r_d^{\mathrm{ours}}}\xspace}
\newcommand{\rdfidours}{\ensuremath{r_{d, \mathrm{fid}}^{\mathrm{ours}}}\xspace}
\newcommand{\rdfid}{\ensuremath{r_{d, \mathrm{fid}}}\xspace}

Here, we collect and standardize the BAO distances used in this analysis. BAO constraints are quoted using one of three distances (or their inverses): 1) Using the distribution of tracers perpendicular to the line of sight, BAO measures transverse comoving distance
\begin{equation}
    D_M(\zeff) \equiv (1 + \zeff) D_A(\zeff) \;.
\end{equation}
2) Using tracers along the line of sight, BAO measures the Hubble parameter
\begin{equation}
    D_H(\zeff) \equiv \frac{c}{H(\zeff)} \;.
\end{equation}
3) For smaller datasets, volume-averaged BAO distances measure 
\begin{equation}
    D_V(\zeff) \equiv \left[(1 + \zeff)^2 D_A^2(\zeff) \frac{c \, \zeff}{H(\zeff)} \right]^{1/3} \;.
\end{equation}
These distances are measured with respect to the sound horizon:
\begin{equation}
    r_d = \int_{z_d}^{\infty} \frac{c_s (\zprime) d \zprime}{H(\zprime)} \;.
\end{equation}
In other words, BAO measurements consist of constraints on $D_M/r_d$, $D_H/r_d$, and/or $D_V/r_d$. BAO measurements are frequently quoted with respect to a fiducial cosmology with a fiducial $r_d$, e.g., the measurement of \citet{Ross2015},
\begin{equation}
    D_V(\zeff = 0.15) \left[ \frac{\rdfid}{r_d} \right] = 664\pm25~\mathrm{Mpc} \;. \nonumber
\end{equation}
Unfortunately, a similar issue arises for defining $r_d$ as $r_s(\zstar)$ for the CMB: different approximations can scale $r_d$ by a cosmology-independent constant and different results are quoted relative to different fiducial cosmologies. Denoting \rdours to be $r_d$ using our approximations, we can rescale all measurements to be on a consistent scale. Taking the same example as above, we write
\begin{equation}
    D_V(\zeff = 0.15) \left[ \frac{\rdfidours}{\rdours} \right] = 664\pm25~\mathrm{Mpc} \;. \nonumber
\end{equation}
where \rdfidours has to be computed for the same cosmological parameters as each BAO analysis used to compute their \rdfid.

Table~\ref{tab:BAO} shows the BAO constraints that we include (and their covariance matrix) and \rdfidours for the fiducial parameters of each result. We take the 6dF $\zeff = 0.106$ constraints \citep{Beutler2011}, the $\zeff = 0.15$ SDSS constraints \citep{Ross2015}, the $\zeff = 0.38$ and 0.51 constraints from BOSS (\texttt{BAO\_consensus\_results\_dM\_Hz.txt} from \citealt{Alam2017}), the $\zeff=0.698$ constraints from BOSS and eBOSS (which supersede the $\zeff=0.61$ BOSS BAO point, providing a higher-redshift measurement uncorrelated with the lower BOSS bins, \texttt{sdss\_DR16\_LRG\_v12\_bao.txt} from \citealt{Gil-Marin2020, Bautista2021}), the eBOSS $\zeff=0.845$ measurement \citep{Raichoor2021, deMattia2021}, the eBOSS $\zeff=1.48$ measurement (\texttt{sdss\_DR16\_QSO\_BAO\_DMDH.txt} from \citealt{Neveux2020, Hou2021}), and the BOSS + eBOSS $\zeff=2.334$ measurement (\texttt{sdss\_DR16\_Combined\_BAO\_DMDHgrid.txt} from \citealt{duMasdesBourboux2020}, which we approximate with a Gaussian likelihood).

\begin{deluxetable}{lccccccccccccc}[h!tbp]
\tablehead{ \colhead{Type} & \colhead{$D_V$} & \colhead{$D_V$} & \colhead{$D_M$} & \colhead{$H$} & \colhead{$D_M$} & \colhead{$H$} & \colhead{$D_M$} & \colhead{$H$} & \colhead{$D_V$} & \colhead{$D_M$} & \colhead{$H$} & \colhead{$D_M$} & \colhead{$H$} }
\startdata
\zeff & 0.106 & 0.15 & 0.38 & 0.38 & 0.51 & 0.51 & 0.698 & 0.698 & 0.845 & 1.48 & 1.48 & 2.334 & 2.334\\
\rdfidours & $153.74$ & $152.42$ & $151.31$ & $151.31$ & $151.31$ & $151.31$ & $151.37$ & $151.37$ & $151.31$ & $151.37$ & $151.37$ & $150.87$ & $150.87$\\
Dist & $457.57$ & $664.00$ & $1512.4$ & $81.209$ & $1975.2$ & $90.903$ & $2639.1$ & $104.97$ & $2708.6$ & $4535.0$ & $152.98$ & $5523.1$ & $226.11$\\
\hline
 & $417.28$ & 0 & 0 & 0 & 0 & 0 & 0 & 0 & 0 & 0 & 0 & 0 & 0\\
 & 0 & $625.00$ & 0 & 0 & 0 & 0 & 0 & 0 & 0 & 0 & 0 & 0 & 0\\
 & 0 & 0 & $624.71$ & $23.729$ & $325.33$ & $8.3496$ & 0 & 0 & 0 & 0 & 0 & 0 & 0\\
 & 0 & 0 & $23.729$ & $5.6087$ & $11.643$ & $2.3400$ & 0 & 0 & 0 & 0 & 0 & 0 & 0\\
 & 0 & 0 & $325.33$ & $11.643$ & $905.78$ & $29.339$ & 0 & 0 & 0 & 0 & 0 & 0 & 0\\
 & 0 & 0 & $8.3496$ & $2.3400$ & $29.339$ & $5.4233$ & 0 & 0 & 0 & 0 & 0 & 0 & 0\\
 & 0 & 0 & 0 & 0 & 0 & 0 & $2351.3$ & $46.810$ & 0 & 0 & 0 & 0 & 0\\
 & 0 & 0 & 0 & 0 & 0 & 0 & $46.810$ & $8.3729$ & 0 & 0 & 0 & 0 & 0\\
 & 0 & 0 & 0 & 0 & 0 & 0 & 0 & 0 & $7716.8$ & 0 & 0 & 0 & 0\\
 & 0 & 0 & 0 & 0 & 0 & 0 & 0 & 0 & 0 & $13918$ & $-290.99$ & 0 & 0\\
 & 0 & 0 & 0 & 0 & 0 & 0 & 0 & 0 & 0 & $-290.99$ & $40.548$ & 0 & 0\\
 & 0 & 0 & 0 & 0 & 0 & 0 & 0 & 0 & 0 & 0 & 0 & $27389$ & $360.35$\\
 & 0 & 0 & 0 & 0 & 0 & 0 & 0 & 0 & 0 & 0 & 0 & $360.35$ & $22.669$\\
\enddata
\caption{Table of BAO distances. The first line specifies the type of distance, the second line shows the effective redshift, the third line is our fiducial sound horizon, the fourth line quotes the distance, and the remaining lines show the distance covariance matrix. \label{tab:BAO}}
\end{deluxetable}

\section{Cosmic Microwave Background Compression} \label{sec:CMBCompression}

We compress the \citet{PlanckCollaboration2020} MCMC samples down to three parameters: the shift parameter
\begin{equation}
    R \equiv \frac{\sqrt{\Omega_m H_0^2}}{c} (1 + \zstar) D_A(\zstar)
\end{equation}
\citep{Bond1997, Efstathiou1999}, the acoustic angular scale
\begin{equation}
\theta \equiv r_s(\zstar)/D_M(\zstar) \;,
\end{equation}
and the physical baryon density $\Omega_b h^2$ ($\omega_b$). We note that the WMAP5 cosmology paper \citep{Komatsu2009} is a great pedagogical resource for the relations we use. To ensure compatibility between our compression and the Planck cosmological parameters, we derive our own compression by reading in each chain, unpacking the repeated samples, computing our compressed parameters for each MCMC sample, and then computing the median and covariance matrix for the compressed parameters.

We use the baseline \texttt{plikHM\_TTTEEE\_lowl\_lowE} results and consider two cosmological models for deriving our compressed distances: \LCDM and flat $w$CDM. Curvature in the \LCDM chains is in mild tension with a flat universe: $\Omega_k = -0.043^{+0.016}_{-0.018}$, and this pulls the compressed CMB parameters about $1\sigma$ from their values in the flat $w$CDM chains. Restricting the range of curvature to be closer to zero (as would happen if external data were combined with the CMB) with rejection sampling gives similar results to the flat $w$CDM chains. In the end, we use the compressed parameters from the flat $w$CDM chains, as this gives better statistics than the \LCDM chains after rejection sampling (as we have to reject most of the samples from the \LCDM chains). Table~\ref{tab:CMBCompress} shows our compression.

\begin{deluxetable}{l|ccc}[h]
 \tablehead{
  & \colhead{$R$} & \colhead{$100 \, \theta$} & \colhead{$\omega_b$}  }
 \startdata
Values & 1.7492768568335353 & 1.039233410719115 & 0.02239245\\
\hline
$R$ & $92701.58172970748$ & $348041.8137694254$ & $1613445.8550364415$\\
$100\, \theta$ & $348041.8137694254$ & $13114681.644682042$ & $-3019007.1687636944$\\
$\omega_b$ & $1613445.8550364415$ & $-3019007.1687636944$ & $80842256.32398143$\\
\enddata
 \caption{Derived Planck CMB compression with inverse covariance matrix. \label{tab:CMBCompress}}
\end{deluxetable}

\section{Validation of Light-Curve Fitting} \label{sec:validLCfit}

\begin{figure*}[h!tbp]
    \centering
    \includegraphics[width = 0.98 \textwidth]{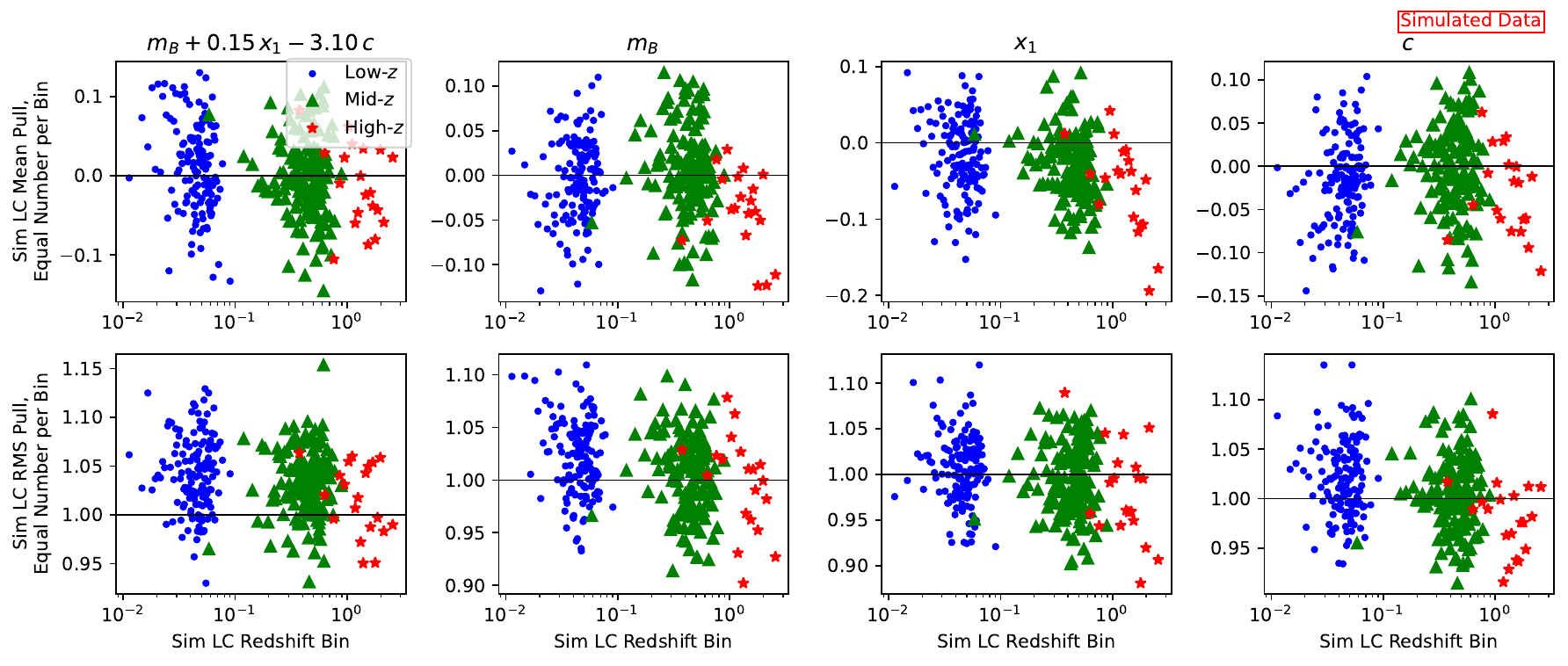}
    \caption{Validation of light-curve fitting using simulated data. The {\bf top panels} show the mean pulls in redshift bins with the low-$z$, mid-$z$, and high-$z$ simulated datasets separately plotted. In general, the mean pulls are consistent with zero (horizontal line), as expected for a unit normal but there is a small trend in redshift with cosmological implications explored in Figure~\ref{fig:simmeanresid}. The {\bf bottom panels} show the RMS pulls in redshift bins. \pullsaregood \label{fig:lcpulls}}
\end{figure*}

\begin{figure*}[h!tbp]
    \centering
    \includegraphics[width = 0.50\textwidth]{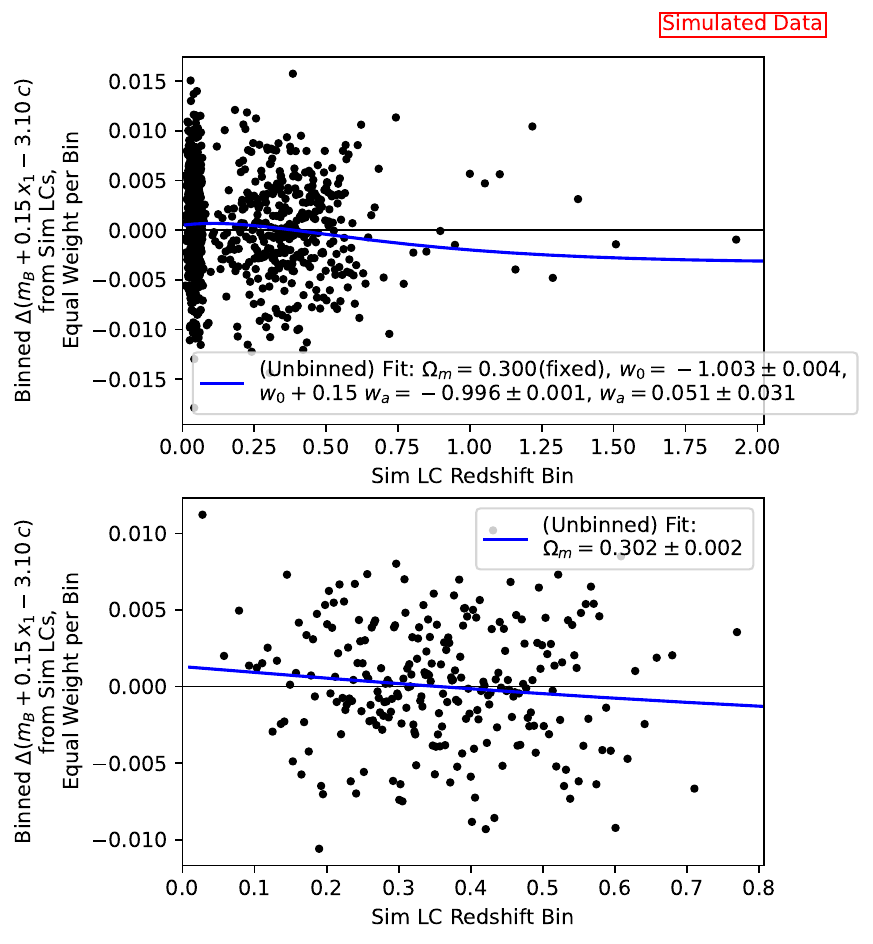}
    \caption{Evaluating the impact of the (tiny) observed light-curve-fitting biases on cosmological parameters, using our simulated-data testing. The {\bf top panel} shows the mean bias on distance moduli from all three sets of simulated samples (low-, mid-, and high-$z$). The {\bf bottom panel} shows the mean bias on distance moduli in redshift bins for the mid-redshift sample alone. Each bin contains the same amount of total inverse variance, so each has the same level of uncertainty in magnitudes. Our light-curve fitting shows small ($\lesssim$~1~mmag) biases which are strongly subdominant to the calibration uncertainties and are thus ignored in the cosmological analysis for Union3. However, the bias on $w_0 + 0.15\, w_a$ (which is chosen to be roughly uncorrelated with $w_a$) is visible when averaging all 100 sets of simulations (Table~\ref{tab:SimulatedSummaryLHV}).}
    \label{fig:simmeanresid}
\end{figure*}

This appendix uses our simulated light curves and their fits (more than 120,000 simulated SNe in total) to investigate the assumption in UNITY that light-curve fits have (reasonably correct) Gaussian uncertainties. Figure~\ref{fig:lcpulls} shows the mean and RMS pulls in equal-number-of-SNe bins. Figure~\ref{fig:simmeanresid} shows the mean Hubble residuals for both the mid-redshift simulated SNe and the low-, mid-, and high-redshift simulated SNe. Only very small biases are seen ($\lesssim$~1~mmag) which we fit with flat-$w_0$-$w_a$ and flat-\LCDM models to quantify the impact. The quoted uncertainties come from bootstrap resampling.

\end{document}